\documentclass[a4paper, 11pt]{article}
\usepackage[usenames]{xcolor}
\usepackage{xifthen}
\usepackage{algorithmic}
\usepackage{comment}
\usepackage{tcolorbox}
\usepackage{relsize}
\usepackage[font=footnotesize]{caption}
\usepackage[normalem]{ulem}
\usepackage{todonotes}

\usepackage{bm}
\usepackage{booktabs}
\usepackage{tabularx}
\usepackage{array}
\usepackage{makecell}
\usepackage{tikz}
\usetikzlibrary{shapes.geometric, arrows.meta, positioning, decorations.pathreplacing, calc}

\def\fontsettingup{2} 

\usepackage{amsmath, amsthm, amssymb}
\usepackage{thm-restate}
\usepackage{bbm}
\usepackage[T1]{fontenc}
\ifthenelse{\fontsettingup = 1}{ \usepackage{eulerpx, eucal, tgpagella}   }{}
\ifthenelse{\fontsettingup = 2}{ \usepackage{mathpazo, eufrak, tgpagella} }{}

\usepackage{hyperref}
\hypersetup{
  colorlinks=true,
  linkcolor=blue,
  filecolor=magenta,
  urlcolor=cyan,
  citecolor=blue
}

\usepackage[linesnumbered,boxed,ruled,vlined]{algorithm2e}
\usepackage{tikz}
\usepackage[capitalize]{cleveref}
\usepackage{makecell}

\usepackage[a4paper]{geometry}
\newgeometry{
textheight=9in,
textwidth=6.5in,
top=1in,
headheight=14pt,
headsep=25pt,
footskip=30pt
}

\newtheorem{theorem}{Theorem}

\newtheorem*{claim*}{Claim}

\newtheorem{lemma}[theorem]{Lemma}
\newtheorem{proposition}[theorem]{Proposition}
\newtheorem{corollary}[theorem]{Corollary}
\theoremstyle{definition}

\newtheorem{definition}[theorem]{Definition}
\newtheorem{remark}[theorem]{Remark}
\newtheorem*{remark*}{Remark}

\ifthenelse{\fontsettingup = 1}{
  \def\*#1{\mathbf{#1}} 
  \def\+#1{\mathcal{#1}} 
  \def\-#1{\mathrm{#1}} 
  \def\^#1{\mathbb{#1}} 
  \def\!#1{\mathfrak{#1}} 
}{}

\ifthenelse{\fontsettingup = 2}{
  \def\*#1{\boldsymbol{#1}} 
  \def\+#1{\mathcal{#1}} 
  \def\-#1{\mathrm{#1}} 
  \def\^#1{\mathbb{#1}} 
  \def\!#1{\mathfrak{#1}} 
}{}

\DeclareMathOperator*{\argmin}{arg\,min}

\def\oPr{\mathbf{Pr}}
\renewcommand{\Pr}[2][]{ \ifthenelse{\isempty{#1}}
  {\oPr\left[#2\right]}
  {\oPr_{#1}\left[#2\right]} }

\DeclareMathOperator*{\oVar}{\mathbf{Var}}
\newcommand{\Var}[2][]{ \ifthenelse{\isempty{#1}}
  {\oVar\left[#2\right]}
  {\oVar_{#1}\left[#2\right]} }

\def\oEnt{\mathbf{Ent}}
\newcommand{\Ent}[2][]{ \ifthenelse{\isempty{#1}}
  {\oEnt\left[#2\right]}
  {\oEnt_{#1}\left[#2\right]} }

\renewcommand{\epsilon}{\varepsilon}
\renewcommand{\emptyset}{\varnothing}
\newcommand{\norm}[1]{\left\Vert#1\right\Vert}

 \newcommand{\eps}{\varepsilon}

\newcommand{\cut}{\operatorname{cut}}
\newcommand{\diag}{\operatorname{diag}}
\newcommand{\KL}{D_{\mathrm{KL}}}
\newcommand{\tr}{\operatorname{tr}}
\newcommand{\E}{\mathbb{E}}

\newcommand{\ip}[2]{\left\langle #1,#2\right\rangle}

\let\epsilon=\varepsilon
\newcommand{\del}{\delta}

\newcommand{\one}{\mathbf{1}}
\newcommand{\ind}{\mathbb{1}}

\newcommand{\polylog}{\operatorname{polylog}}
\newcommand{\Lap}{\operatorname{Lap}}

\newcommand{\Prb}{\mathbf{Pr}}
\newcommand{\R}{\mathbb{R}}

\newcommand{\cM}{\mathcal{M}}
\newcommand{\bottom}{\bot}
\newcommand{\odotc}{\mathbin{\odot}}
\newcommand{\pdf}{\mathrm{pdf}}

\allowdisplaybreaks

\title{Polynomial-Time Differentially Private Synthetic Graph Release with Nearly-Linear Additive Error}
\author{
Marek Eliáš\thanks{Bocconi University. Email: \texttt{marek.elias@unibocconi.it}.}
\and Gautam Kamath\thanks{New York University. Email: \texttt{g@csail.mit.edu}. Supported by an Ontario Early Researcher Award, an NSERC Discovery Grant, and an NYU startup grant.}
\and Janardhan Kulkarni\thanks{Microsoft Research. Emails: \texttt{jakul@microsoft.com}, \texttt{smahabadi@microsoft.com}, \texttt{jatarnaw@microsoft.com}.}
\and George Z. Li\thanks{Carnegie Mellon University. Email: \texttt{gzli@cs.cmu.edu}.}
\and Jingcheng Liu\thanks{State Key Laboratory for Novel Software Technology, Nanjing University. Emails: \texttt{liu@nju.edu.cn}, \texttt{zou.zongrui@smail.nju.edu.cn}.}
\and Sepideh Mahabadi\footnotemark[3]
\and Pan Peng\thanks{University of Science and Technology of China. Emails: \texttt{ppeng@ustc.edu.cn}, \texttt{hangyuxu@mail.ustc.edu.cn}.}
\and Jakub Tarnawski\footnotemark[3]
\and Hangyu Xu\footnotemark[6]
\and Zongrui Zou\footnotemark[5]
}
\date{}

\begin{document}
\maketitle

\begin{abstract}
We give a polynomial-time \((\varepsilon, \delta)\)-edge-differentially private algorithm that releases a synthetic graph approximately preserving all cut values of a nonnegative weighted input graph \(G\).
Our algorithm outputs a nonnegative weighted graph $H$ such that, with probability at
least $1-\beta$, simultaneously for all $S\subseteq V$,
\[
 \left|w_G(S,V\setminus S)-w_H(S,V\setminus S)\right|
 \leq \gamma w_G(S,V\setminus S)
 + O\left(\frac{n}{\gamma\varepsilon}
 \log^7 \frac{n}{\delta\beta\gamma\varepsilon}\right).
\]
For fixed $\gamma, \varepsilon$ and $\delta = \beta = n^{-10}$, the additive error is $n \operatorname{polylog} n$, improving upon the previous efficient bound of $n^{13/12 + o(1)}$. Furthermore, we establish an $\Omega(n/\gamma\varepsilon)$ lower bound for the relevant parameter regime, showing that our result is tight up to polylogarithmic factors.

Our algorithm uses a log-det-regularized SDP to generate fractional
queries for multiplicative-weights updates and certify accuracy for all cuts.
The privacy analysis charges privacy loss only fractionally per mechanism call, based on the amount of mass on edge coordinates retired from subsequent iterations.

Our synthetic graph release yields improved error bounds for several downstream tasks in private graph analysis including correlation clustering, Max-Cut, Max-$k$-Cut, constrained cut problems and hierarchical clustering. Most notably, for unweighted complete-graph correlation clustering, we achieve the optimal $\widetilde O(n)$ additive error while reducing the best previously known approximation ratio for this additive error from several thousand to $1.387$.

We also provide a formalization of our results in Lean.

\end{abstract}
 \thispagestyle{empty}

\newpage
\pagenumbering{arabic}
\section{Introduction}

Graph data analysis is fundamental for theoretical computer science, providing the structural framework for routing, clustering, optimization, and network design. However, graphs frequently encode highly sensitive information including social relationships, transaction histories or communication logs. In such domains, directly releasing structural information or statistics can compromise individual privacy. Against this backdrop, differential privacy (DP) has emerged as the gold standard for privacy-preserving graph analysis since its introduction by Dwork, McSherry, Nissim, and Smith~\cite{DMNS06}. Given privacy budgets $\epsilon >0$ and $0\leq \delta<1$, a randomized algorithm $\mathcal{A}$ is $(\epsilon, \delta)$-differentially private if, for every pair of \emph{``neighboring''} datasets $D$ and $D'$ differing by at most one individual record, and for every set of possible outputs $\mathcal{S}$, we have that:
\[
    \Prb[\mathcal{A}(D) \in \mathcal{S}] \le e^\epsilon \cdot \Prb[\mathcal{A}(D') \in \mathcal{S}] + \delta.
\]
Intuitively, differential privacy provides an information-theoretic guarantee against privacy leakage by requiring that the output distributions remain close when a single individual's data is added or removed. Thus observing the output provides only limited additional evidence about that individual's participation. 

In terms of graph data, we follow the most fundamental and widely studied notion: \emph{edge differential privacy} (first proposed by Dwork, McSherry, Nissim, and Smith~\cite{DMNS06}). Here, two weighted graphs with edge-weight
vectors $w$ and $w'$ are considered \emph{``neighboring''} if $\|w-w'\|_1\leq1$. Edge level privacy protects the existence and intensity of relationships between individuals, assuming the public vertex set is known.

Within the rich landscape of private graph analysis, one of the most central and intensely studied problems is to release a synthetic graph under edge level privacy that preserves the values of all cuts~\cite{GRU12, blocki2012johnson, EKKL20, LUZ24, arora2019differentially, upadhyay2021differentially, zou2025almost, aamand2025breaking, fan2026private}. Formally, let $G = ([n],E,w)$ be a simple $n$-vertex weighted graph, and for any vertex subset $S \subseteq [n]$, let $w_G(S)$ denote the total weight of the edges crossing the cut $(S, [n]\setminus S)$. The goal is to design a computationally efficient, differentially private algorithm that outputs a synthetic weighted graph $\widetilde{G}$ on $[n]$ satisfying, simultaneously for all cuts $S \subseteq [n]$,
\begin{equation*}
    (1-\gamma)w_G(S) - \alpha \le w_{\widetilde{G}}(S) \le (1+\gamma)w_G(S) + \alpha,
\end{equation*}
where the relative error $\gamma \ge 0$ and the additive error $\alpha \ge 0$ are to be minimized. This problem is not only of theoretical interest in its own right, but also has direct applications to private correlation clustering~\cite{BEK21,liu2022better} and private optimization~\cite{aamand2025breaking, chandra2026differentially}.

The purely additive setting (where $\gamma = 0$) is by now well understood. As a straightforward baseline, adding i.i.d.\ Gaussian noise with variance ${O}(\log(1/\delta)/\epsilon^2)$ to the edge weight between every pair of vertices guarantees $(\epsilon, \delta)$-differential privacy; by a standard union bound, this incurs an error of $\widetilde{O}(|S|\sqrt{n}/\epsilon)$ simultaneously across all cuts $S\subseteq[n]$. This implies a worst-case purely additive error of $\widetilde{O}_{\delta}(n^{1.5}/\epsilon)$. A sequence of refined algorithms by Gupta, Roth, and Ullman~\cite{GRU12}, Eli{\'a}{\v s}, Kapralov, Kulkarni, and Lee~\cite{EKKL20}, and Liu, Upadhyay, and Zou~\cite{LUZ24} improved this additive error to $ \widetilde{O}_{\delta}(\sqrt{mn}/\epsilon)$ for graphs with $m$ weighted edges, while also proving that in the worst case where $m = \Theta(n^2)$, the $\widetilde{O}_{\delta}(n^{1.5}/\epsilon)$ scale is inherently unavoidable, if only purely additive error is allowed (i.e., $\gamma = 0$).

However, this $\widetilde{O}_{\delta}(n^{1.5}/\epsilon)$ scale of error is too large, and can completely hide sparse cuts, thereby undermining its usefulness for downstream applications, such as finding sparsest cuts or identifying expander components~\cite{nanongkai2017dynamic}. Consequently, it is natural to allow a relative error tolerance $\gamma > 0$, thereby securing meaningful accuracy even for small cuts. 

Yet, minimizing the additive error $\alpha$ while allowing a relative error $\gamma > 0$ is considered much more challenging, and has remained a long-standing open problem. As an information-theoretic benchmark, Blocki, Blum, Datta, and Sheffet~\cite{blocki2012johnson} proposed a private algorithm that achieves $(1\pm \gamma)$ relative error alongside a nearly linear additive error of $\alpha = \widetilde{O}(n/\gamma)$. Their approach leverages the \textit{existence} of graph cut sparsifiers, applying the exponential mechanism to select a valid candidate from the exponentially large space of all possible sparsifiers. Clearly the challenge is computational efficiency: there is no known polynomial-time implementation of such exponential mechanism over all sparse graphs.

\begin{sloppypar}
The elegant work of Aamand, Chen, Dalirrooyfard, Mitrovi{\'c}, Nevmyvaka, Silwal, and Xu~\cite{aamand2025breaking} made the first polynomial-time breakthrough beyond the $\widetilde{O}(n^{1.5})$ barrier in private all-cut estimation. They observe that if the graph is already a good expander, then it is highly connected so that any cut contains enough edges to absorb the $\widetilde{O}(|S|\sqrt{n})$ noise from basic Gaussian perturbation as a small relative error (i.e. $\gamma w_G(S)$).  Leveraging this property, they design a private expander decomposition using a private estimator of graph Laplacian matrices to find the sparsest cuts. Within each expander component, cuts are approximated with purely relative error; for the remaining inter-component residual edges (whose total count is shown to be bounded by $m_{\mathrm{res}} = O(n^{1.5 +o(1)}/\gamma)$), the algorithm falls back on the edge-sensitive additive error mechanism of~\cite{EKKL20}: substituting this residual edge count into the $\widetilde{O}_{\delta}(\sqrt{m_{\mathrm{res}}n})$ additive guarantee yields an overall additive error of only $\widetilde{O}_{\gamma}(n^{1.25+o(1)}) \ll \widetilde{O}(n^{1.5})$.
\end{sloppypar}

\begin{sloppypar}
More recently, Fan, Liu, Peng, Xu, and Zou~\cite{fan2026private} pushed this additive error to $\widetilde{O}_{\gamma}(n^{13/12+o(1)})$. They follow the similar high-level approach of~\cite{aamand2025breaking}, while they overcome key bottlenecks by introducing a set of new algorithmic tools. First, they developed a set of new degree-sensitive private spectral primitives to recursively shrink the residual graph down to only $m_{\mathrm{res}} = n^{5/4+o(1)}$ edges. Second, to release these leftover sparse edges without falling back to a purely additive baseline, they designed a relative-error terminal cut oracle with an additive error of $\widetilde{O}(n + (n^{2}m_{\mathrm{res}})^{1/3})$. Substituting this residual edge bound directly yields their $\widetilde{O}(n^{13/12+o(1)})$ additive error guarantee.
\end{sloppypar}

In contrast to these prior approaches~\cite{aamand2025breaking, fan2026private}, we depart from the recursive expander decomposition paradigm and address the problem purely through a continuous multiplicative weights framework. We design a \emph{fractional margin oracle} via a log-determinant regularized SDP, using Gaussian covariance samples to extract a witness that preserves the margin with constant probability. Furthermore, to control the adaptive privacy leakage across iterations without incurring standard composition penalties, we prove a \emph{success-charged composition} lemma (inspired by a number of prior works, including~\cite{gupta2010differentially,kaplan2021sparse,durfee2019practical,cohen2023target}), which charges the pair-local privacy loss of each edge directly to its telescoping fractional retirement, instead of the number of iterations. Integrating this oracle into an extension of the PREM framework \cite{PREM25} allows us to finally close the longstanding computational gap, achieving a polynomial-time release with a nearly-linear additive error of $\widetilde{O}_{\gamma}(n)$.
\subsection{Our Results}

Our central contribution is a polynomial-time edge-differentially private algorithm that resolves the longstanding computational bottleneck in relative-error cut release since Blocki, Blum, Datta, and Sheffet~\cite{blocki2012johnson}. As summarized in \Cref{tab:comparison}, our mechanism achieves a near-optimal additive error of $n\polylog n$, improving upon previous polynomial-time bounds. We formally state this guarantee in \Cref{thm:main}.

\begin{theorem}\label{thm:main}
Let
$n\geq2$, $0<\gamma\leq 1/4$, $0<\varepsilon\leq 1$, and
$0<\delta,\beta<1/10$.  There is an $(\varepsilon, \delta)$ edge-DP algorithm that, on every
nonnegative weighted $n$-vertex graph $G$, runs in time polynomial in
$n$, $1/\gamma$, $1/\varepsilon$, $\log(1/\delta)$, and $\log(1/\beta)$, and outputs a
nonnegative weighted graph $\widetilde{G}$ satisfying, with probability at least $1-\beta$, simultaneously
for every $S\subseteq V$,
\begin{align*}
 \left|w_G(S,V\setminus S)-w_{\widetilde{G}}(S,V\setminus S)\right|
 \leq \gamma w_G(S,V\setminus S)  + O\left(\frac{n}{\gamma\varepsilon}
 \left(\log(n/(\delta\beta\gamma\eps))\right)^{7}\right) .
\end{align*}
\end{theorem}
In particular, for fixed $\gamma$, $\varepsilon$ and $\delta=\beta=n^{-10}$, the additive error is
$n\polylog n$.
We provide a nearly-matching lower bound in \Cref{thm:lower_bound}. 
We build upon previous lower bounds for this problem in the purely additive (i.e., $\gamma=0$) case~\cite{EKKL20,LUZ24}, and show that the optimal additive error for this problem is $\Omega\left(\min\left\{\frac{n}{\varepsilon \gamma},\frac{n^{3/2}}{\varepsilon}\right\}\right)$. 

\begin{table}[t]
\centering
\caption{Comparison of error bounds for private all-cut release on an $n$-vertex, $m$-edge graph with weight vector $w$. The guarantee describes whether the synthetic graph preserves the true cut value $w_G(S)$ up to a purely additive error or a mixed relative-additive error. For simplicity, bounds are stated for constant privacy parameter $\epsilon = \Theta(1)$ and $\delta = n^{-c}$.}
\label{tab:comparison}
\begin{tabular}{@{}llll@{}}
\toprule
Result & Privacy & Running time &  Approximation guarantee \\
\midrule
\cite{EKKL20} & approximate & polynomial & $w_G(S) \pm \widetilde O(\sqrt{\|w\|_1n})$ \\
\cite{LUZ24} & approximate & polynomial & $w_G(S) \pm \widetilde O(\sqrt{mn})$ \\
\cite{blocki2012johnson} & pure & exponential$^*$ & $(1\pm\gamma)w_G(S) \pm  n\polylog n$ \\
\cite{aamand2025breaking} & approximate & polynomial & $(1\pm\gamma)w_G(S) \pm n^{5/4+o(1)}$ \\
\cite{fan2026private} & approximate & polynomial & $(1\pm\gamma)w_G(S) \pm n^{13/12+o(1)}$ \\
\textbf{This work} & {approximate} & {polynomial} & ${(1\pm\gamma)w_G(S) \pm n\polylog n}$\\\bottomrule
\multicolumn{4}{l}{\footnotesize $^*$For the information-theoretic exponential mechanism.}
\end{tabular}
\end{table}

\paragraph{Applications.}
The multiplicative-additive all-cut guarantee can be combined with existing nonprivate approximation algorithms to
give new polynomial-time private bounds for several important graph optimization and clustering
problems. In Appendix~\ref{app:applications}, we present a series of new bounds for correlation clustering, Max-Cut, Max-$k$-Cut, 
minimum bisection,
constant-balanced cuts, and hierarchical clustering, as summarized in Table~\ref{tab:applications}.
Here, we highlight the implications for two fundamental problems: correlation clustering and Max-Cut, for which we obtain additive error that is optimal up to polylogarithmic factors for constant $0<\varepsilon\leq1$.

\begin{table}[h!]
\centering
\caption{Comparison of guarantees for the graph optimization and clustering applications in \Cref{app:applications}. We write $(\rho,B)$ for a guarantee of at most $\rho\,\mathrm{OPT}+B$
in minimization problems and at least $\rho\,\mathrm{OPT}-B$
in maximization problems. The displayed multiplicative ratios are obtained by instantiating our reductions with specific nonprivate algorithms.
For the cut problems and correlation clustering, other nonprivate approximation ratios can likewise be transferred up to an arbitrarily small constant slack, retaining the shown additive bound and balance conditions.
We assume $\varepsilon=\Theta(1)$ and $\delta=n^{-c}$ for a fixed constant $c>0$; $\xi>0$ is an arbitrarily small fixed constant, and $\rho_k$ is defined in \Cref{thm:private-max-k-cut}. Entries citing \cite{fan2026private} are derived baselines, obtained by combining their cut-release guarantee with the indicated nonprivate algorithms; these applications are not stated in their paper. Their cited release theorem is for unweighted input graphs. The entries citing only \cite{aamand2025breaking} are stated in its applications section for weighted graphs. The lower bounds for correlation clustering and for Max-Cut hold for approximate DP,
establishing tightness of our approximate-DP guarantees. The balanced-cut lower bounds concern pure DP and weight privacy with the cited balance constraints and no multiplicative loss. For hierarchical clustering, the pure-DP lower bounds of \cite{ImolaEtAl23,DengEtAl25} hold for every multiplicative factor $o(n)$, while the approximate-DP lower bound of \cite{ImolaPaghRetschmeier26} assumes no multiplicative loss.
}
\label{tab:applications}
\small
\begin{tabularx}{\textwidth}{@{}>{\raggedright\arraybackslash}m{.27\textwidth}
  >{\centering\arraybackslash}m{.25\textwidth}
  >{\centering\arraybackslash}m{.22\textwidth}
  >{\centering\arraybackslash}X@{}}
\toprule
\textbf{Problem} & \textbf{Previous $(\rho,B)$} & \textbf{New $(\rho,B)$} & \textbf{Additive lower bound}\\
\midrule
Correlation clustering\par
\textit{(unweighted, min-disagree)}
& \makecell{$(1.387,\widetilde O(n^{3/2}))$\\
\cite{BEK21,liu2022better,garcia2026approximate}
\\$(O(1),\widetilde O(n))$\\\cite{CohenAddadEtAlPrivateCC}}
& $\textcolor{black}{(1.387,\widetilde O(n))}$
& \makecell{$\Omega(n)$\\\cite{CohenAddadEtAlPrivateCC}}\\
\addlinespace[.25em]
Correlation clustering\par
\textit{(unweighted,
min-disagree, fixed $k$ clusters)}
& \makecell{$(1+\xi,\widetilde O(kn^{3/2}))$\\pure DP \cite{BEK21}}
& $\textcolor{black}{(1+\xi,\widetilde O(n))}$
& ---\\
\addlinespace[.25em]
Max-Cut
& \makecell{$(0.878-\xi,n^{5/4+o(1)})$\\\cite{aamand2025breaking}\\$(0.878-\xi,n^{13/12+o(1)})$\\\cite{fan2026private,GoemansWilliamson95}}
& $\textcolor{black}{(0.878-\xi,\widetilde O(n))}$
& \makecell{$\Omega(n)$ \\\cite{dharangutte2026price},\\ Remark~\ref{rem:maxcut-tightness}}\\
\addlinespace[.25em]
Max-$k$-Cut
& \makecell{$(\rho_k-\xi,kn^{5/4+o(1)})$\\\cite{aamand2025breaking}\\$(\rho_k-\xi,n^{13/12+o(1)})$\\\cite{fan2026private,FriezeJerrum97}}
& $\textcolor{black}{(\rho_k-\xi,\widetilde O(n))}$
& ---\\
Minimum bisection
& \makecell{$(O(\log n),n^{5/4+o(1)})$\\\cite{aamand2025breaking}\\$(O(\log n),n^{13/12+o(1)})$\\\cite{fan2026private,Raecke08}}
& $\textcolor{black}{(O(\log n),\widetilde O(n))}$
& ---\\
\addlinespace[.25em]
Balanced cut\par
\textit{(constant balance, bicriteria)}
& \makecell{$(O(\sqrt{\log n}),n^{13/12+o(1)})$\\\cite{fan2026private,AroraRaoVazirani09}}
& $\textcolor{black}{(O(\sqrt{\log n}),\widetilde O(n))}$
& \makecell{$\Omega(\frac{n}{\eps\log^2 n})$\\\cite{DengEtAl25}}\\
\addlinespace[.25em]
Balanced sparsest cut\par
\textit{(edge expansion, constant balance, bicriteria)}
& \makecell{$(O(\sqrt{\log n}),n^{1/12+o(1)})$\\\cite{fan2026private,AroraRaoVazirani09}}
& $\textcolor{black}{(O(\sqrt{\log n}),\widetilde O(1))}$
& 
\makecell{$\Omega(\frac{1}{\eps\log^2 n})$\\\cite{DengEtAl25}}\\
\addlinespace[.25em]
Hierarchical clustering\par
\textit{(Dasgupta's cost)}
& \makecell{$(O(\sqrt{\log n}),\widetilde O(n^{5/2}))$\\\cite{ImolaEtAl23}}
& $\textcolor{black}{(O(\sqrt{\log n}),\widetilde O(n^2))}$
& \makecell{$\Omega(n^2)$\\\cite{ImolaEtAl23},\\\cite{DengEtAl25},\\\cite{ImolaPaghRetschmeier26}}\\
\bottomrule
\end{tabularx}
\end{table}

First, for correlation clustering~\cite{BansalBlumChawla04}, we consider the standard unweighted complete-graph min-disagree setting: given a complete graph where each edge is labeled as either positive ($+$) or negative ($-$), the objective is to partition the vertices into clusters to minimize the total number of disagreements (i.e., positive edges across different clusters plus negative edges within the same cluster). Following the synthetic-graph approach of Bun, Eli\'a\v{s}, and Kulkarni~\cite{BEK21}, we give a reduction that translates our mixed cut-error guarantee into a mixed clustering-error guarantee. This allows us to use \emph{any} non-private approximation algorithm as a black box, preserving its multiplicative approximation ratio (up to an arbitrarily small fixed slack) while adding only an $\widetilde{O}(n)$ additive error; it is known~\cite{CohenAddadEtAlPrivateCC} that $\Omega(n)$ error is necessary. Plugging in the current best $1.387$-approximation algorithm~\cite{garcia2026approximate} gives the following.

\begin{restatable}[Private correlation clustering]{corollary}{privatecorrelationclustering}
\label{cor:cc-intro}
For every constant $\varepsilon>0$ and $\delta=n^{-10}$, there is a
polynomial-time $(\varepsilon, \delta)$-edge-DP algorithm for unweighted complete-graph min-disagree
correlation clustering that, with high probability, outputs a partition with cost at most
\[
1.387\operatorname{OPT}_G+\widetilde O_{\varepsilon}(n).
\]
\end{restatable}
This matches the $\widetilde{O}(n)$ additive error of \cite{CohenAddadEtAlPrivateCC}
but improves the multiplicative factor from a constant factor of at least several thousand to $1.387$.
We note that this is \emph{\textbf{not}} a direct black-box application of \cref{thm:main},
as the nonprivate approximation algorithm expects an
unweighted complete signed instance whereas \cref{thm:main} outputs
a weighted graph (even on unweighted input). More importantly, the relative error on the total positive-edge
mass can be $\Theta(n^2)$ even when the optimal correlation-clustering cost is zero. To address these issues, instead of running our original algorithm for cut release (which builds \cref{thm:main}), we first adjust the algorithm in a non-black-box way by introducing a unit coordinate cap to ensure that the output graph has bounded weights. We then privately release the sparse \textit{disagreement set} of the private baseline~\cite{CohenAddadEtAlPrivateCC} (which has bounded positive-edge mass to prevent the relative error from exploding), rather than running on the original graph.

Second, for the Max-Cut problem, the goal is to partition the vertices into two sets to maximize the total weight of the crossing edges. Applying the celebrated Goemans-Williamson algorithm~\cite{GoemansWilliamson95} directly to our privately released synthetic graph yields the following.

\begin{restatable}[Private Max-Cut]{corollary}{privatemaxcut}
\label{cor:maxcut-intro}
For every constant $\xi>0, \varepsilon>0$, and $\delta=n^{-10}$, there is a polynomial-time $(\varepsilon, \delta)$-edge-DP algorithm for Max-Cut that, for any non-negative weighted graph $G$, with high probability outputs a bipartition with cut value at least
\[
(0.878-\xi)\operatorname{OPT}_G-\widetilde O_{\xi,\varepsilon}(n).
\]
Here, $\operatorname{OPT}_G$ is the value of the real max-cut in $G$.
\end{restatable}
Note that the $\widetilde{O}(n)$ lower bound on the error of private Max-Cut approximation stated by
~\cite{dharangutte2026price}, combined with Remark~\ref{rem:maxcut-tightness}, shows that the upper bound in Corollary~\ref{cor:maxcut-intro} is tight up to polylogarithmic factors. The best previously known polynomial-time private algorithms with the Goemans--Williamson factor had additive error $n^{5/4+o(1)}$ for weighted graphs~\cite{aamand2025breaking}, or $n^{13/12+o(1)}$ for unweighted graphs by combining the release of~\cite{fan2026private} with~\cite{GoemansWilliamson95}. Our result improves this additive error to nearly linear while retaining the Goemans--Williamson ratio up to arbitrarily small slack; this ratio is optimal under the Unique Games Conjecture~\cite{KKMO07,MOO10}.

\subsubsection{Lean formalization}
\Cref{thm:main}, including its polynomial running time, and \cref{thm:lower_bound} are formalized in Lean without any unproven assumptions.
For \cref{thm:main}, a single formal theorem establishes privacy, accuracy, and a worst-case polynomial running time for an implementation of our algorithm based on the ellipsoid method, as in \cref{sec:utility}.
Running time is measured in the real RAM model, with constant-cost arithmetic, comparisons, square roots, logarithms, exponentials, floor and ceiling, and exact sampling from the standard Gaussian and Laplace distributions; bit complexity is not analyzed.
For the applications in \cref{app:applications}, the formalization verifies the privacy and approximation guarantees, assuming the guarantees of the cited algorithms; their running times are not formally verified.
Some auxiliary lemmas, such as \cref{lem:ekkl-discrepancy,lem:hypercontractivity,lem:holder_interpolation}, are formalized only in the special cases needed for our proofs.
The code and a detailed description of the formalization are available at
\url{https://github.com/XhyDds/LEAN_DPCutRelease_NearlyLinear}.

\subsection{Technical Overview}
\label{sec:technical-overview}
\subsubsection{The Algorithm}
We start by introducing a few notational conventions.
Let $w\in\mathbb R_{\ge 0}^{\binom{n}{2}}$
be the vector of private edge weights. For any cut $S \subseteq V$, let
$c_S\in\{0,1\}^{\binom{n}{2}}$ denote its incidence vector. The cut value of $S$ on $w$ is then
\[\cut_w(S) = \langle w, c_S \rangle=\sum_{e\in\binom{n}{2}}w_e c_S(e).\]
We often consider fractional \emph{portions} of the graph.
A vector $a \in [0,1]^{\binom{n}{2}}$ represents a portion of the graph
containing a fraction $a_e$ of each edge $e\in \binom{n}{2}$.
We use $\odotc$ to denote coordinatewise multiplication, i.e., $(x\odotc y)_i=x_i y_i$ for every coordinate $i$.
Thus, $a\odotc w$ is the vector of edge weights in this portion,
and $\langle a, w\rangle$ is its total edge weight.

The main optimization procedure of our algorithm is built on the PREM framework of~\cite{PREM25}
for answering linear queries on histograms with relative-error guarantees, which in turn builds on foundational work on private multiplicative weights~\cite{HR10,hardt2012simple}.
It starts with the full graph (portion $a_1=\one_{\binom{n}2}$).
At the beginning of iteration $i$, let $w_i=a_i\odotc w$ denote
the current residual weights. Using multiplicative weight updates (MWU),
we find a portion $r_i\in[0,1]^{\binom n2}$ of this residual graph
and candidate weights $h_i$ satisfying the following properties:
\begin{itemize}
\item $\langle r_i\odotc h_i,c_S\rangle$ approximates
$\langle r_i\odotc w_i,c_S\rangle$ for each cut $S$,
with relative error $\gamma$ and additive error $B$;
\item $\langle r_i,w_i\rangle\geq C\|w_i\|_1$
for some constant $C>0$.
\end{itemize}
We release $r_i\odotc h_i$ and update the active portion to
$a_{i+1}=a_i\odotc(\one-r_i)$.
This decreases the remaining total edge weight by a constant factor,
leading to $O(\log M)$ iterations, where $M$ is the total edge weight $\|w\|_1$.
Moreover, $\tilde w = \sum_{i} r_i\odotc h_i$ approximates
$w$ with relative error $\gamma$ and additive error $O(B \log M)$.

Each iteration consists of $O(\log n/\gamma^2)$ steps of MWU,
each based on an \emph{aggregate witness}
$(g^+, g^-, r)$.
An aggregate witness is a fractional partition of the graph into three portions
satisfying $g^+ + g^- + r = \one_{\binom{n}2}$.
On the portion $r$, the current candidate $h_i$ already approximates
$w_i$ to the required precision.
On the portions $g^+$ and $g^-$, $h_i$ overestimates and underestimates
$w_i$, respectively.
To see why this partition is useful, consider the true mass of the
residual graph. One of the three portions must contain at least a third
of this mass. If this portion is $r$, we have an accurate approximation on a
substantial portion of the residual graph, which can be added to the publicly released synthetic graph. Otherwise, the corresponding
witness identifies substantial overestimation or underestimation,
providing an opportunity for significant progress in MWU.
We note that a direct implementation of PREM for cut queries would construct aggregate
witnesses by scanning the exponentially large family of cuts.
We instead construct these witnesses in polynomial time by aggregating
fractional witnesses obtained through the regularized SDP approach
as in \cite{EKKL20,fan2026private}.

Following the design of PREM, our algorithm requires only a polylogarithmic number
of aggregate witness computations. Each such computation accesses private
data and incurs privacy loss.
For comparison, the algorithm of~\cite{EKKL20} computes
$\tilde O(m/n)$ witnesses, each derived from a single fractional violating cut,
and bounds the privacy loss over these computations using advanced composition.
Our algorithm converges in far fewer iterations, but computing each
aggregate witness requires more work.
Indeed, we need to compute many fractional violating cuts to construct
a single aggregate witness. The key observation behind our improvement
over previous work is that the privacy loss incurred while constructing
a single aggregate witness can be bounded much more effectively
than with advanced composition.
In particular, we charge privacy loss to fractional progress towards
building the aggregate witness, so that the total privacy charge remains small
even when many violating cuts are computed.

\paragraph{Computing aggregate witnesses.}
We first describe how to compute, for a given public candidate $h$,
a single witness corresponding to a cut where $h$ does not approximate
the private weights $w$ to the desired precision. We then explain how to aggregate such witnesses.

As in~\cite{fan2026private}, we construct two error vectors:
\begin{align*}
	D^+
	&=
	h-(1+\gamma)w,
	\\
	D^-
	&=
	(1-\gamma)w-h.
\end{align*}
For a cut $S$, a positive value $\ip{D^+}{c_S}$ certifies overestimation,
whereas a positive value $\ip{D^-}{c_S}$ certifies underestimation. Ideally, to
identify \textit{where} the candidate $h$ violates the relative error bounds
the most, we would search for a cut $S$ that maximizes the
value $\langle D^\sigma, c_S \rangle$ for $\sigma \in \{+, -\}$.
This is a computationally hard problem resembling Max-Cut.\footnote{Unlike in Max-Cut, $D^+$ and $D^-$ may contain both positive and negative entries.}
Moreover, small changes in the private weights $w$ may cause very large changes in
the maximizer $S$, which makes it difficult to release.

We therefore adapt the approach of~\cite{EKKL20}, which uses a regularized SDP
to search for a violating cut $S$.
We optimize over the following elliptope, whose matrices represent
fractional cuts:
\[
 \mathcal X=
 \left\{
 X\in\R^{n\times n}:
 X=X^\top,\ \diag(X)=\one_n,\ X\succeq\frac1nI
 \right\}.
\]
This is the feasible region of the classical Max-Cut SDP of Goemans and Williamson~\cite{GoemansWilliamson95},
with the additional constraint $X\succeq \frac1n I$ to ensure that the regularized
objective remains bounded.
Each $X\in \mathcal{X}$ represents a fractional cut as follows.
For $e=\{u,v\}$, let $b_e=\mathbf e_u-\mathbf e_v$ and define
\begin{equation}\label{eq:qe}
 q_e(X)=\frac{1-X_{uv}}2=\frac14 b_e^\top Xb_e\in[0,1].
\end{equation}
This definition applies to every symmetric PSD matrix $X\in \mathbb{R}^{n\times n}$ with $\diag(X) = \mathbf{1}_n$.
In particular, if $s\in \{\pm 1\}^V$ encodes a cut, then $q_e(ss^\top) = 1$
if $e = \{u,v\}$ crosses that cut, and $q_e(ss^\top) = 0$ otherwise.
For a signed vector $D\in\R^{\binom{n}{2}}$ (here, $D$ is either $D^+$ or $D^-$),
the regularized separation objective used in the body of the paper is
\begin{equation}
 F(D)=
 \max_{X\in\mathcal X}
 \left\{
 \sum_{e\in\binom{n}{2}}D_e q_e(X)+\lambda\log\det X
 \right\}.
 \label{eq:overview-FD}
\end{equation}
The regularization term $\lambda\log\det X$, with $\lambda >0$, stabilizes the maximizer
and allows us to solve the program privately. Negating the strictly concave objective
gives a convex minimization problem. The log-determinant term makes the optimizer
$X_D$ unique and stable. Conceptually, $X_D$ is both a Gram matrix and a Gaussian
covariance matrix. The quantity $q_e(X)\in[0,1]$ measures how strongly
the two endpoints of edge $e$ disagree under this covariance.

The objective \eqref{eq:overview-FD} plays two distinct roles. First, its value provides guarantees for integral cuts.
For every cut $S$, we define the smoothed cut matrix
\[
 X_S=\left(1-\frac1n\right)ss^\top+\frac1nI,
 \qquad s\in\{\pm1\}^V.
\]
Then $X_S\in \mathcal{X}$, and \cref{lem:sdp_error} gives, for every cut $S$,
\begin{equation*}
 \ip{D}{c_S}
 \leq
 2\max\{F(D),0\}+O\left(\lambda n\log n+\frac{\norm{D}_1}{n}\right).
\end{equation*}
Consequently, small values of both $F(D^+)$ and $F(D^-)$ certify that no cut has a
large one-sided violation.

Second, when $F(D)$ is large, its optimizer yields a fractional direction for progress. Since
$\log\det X_D\leq0$,
\[
 \sum_eD_e q_e(X_D)\geq F(D).
\]
Thus $q(X_D)$ is already a bounded fractional edge query with a large
signed objective value. This query can then be used in the multiplicative weight updates.

\paragraph{From the SDP optimizers to an aggregate witness.}
Having computed the maximizer $X_D$ of \eqref{eq:overview-FD}, we
produce $k = O(\log n)$ noisy representations of it:
\[
Z^{(1)},\ldots,Z^{(k)}\sim N(0,X_D).
\]
Each of these vectors can be viewed as a noisy representation of $X_D$, since $\E[(Z^{(i)})(Z^{(i)})^T] = X_D$.
We average the resulting estimates of $q(X_D)$:
\[
\overline g_{\{u,v\}}
=
\frac1k\sum_{j=1}^k
\frac{(Z_u^{(j)}-Z_v^{(j)})^2}{4}.
\]
We clip $\overline g$, then mask and normalize the result to
obtain a candidate witness $g$ satisfying $0\leq g\leq r$
and $g_e=r_e>0$ for at least one
active edge $e$.
Construction of the witness may not be successful, e.g., if the resulting $g$ is too small.
In such case, we sample $Z^{(1)}, \dotsc, Z^{(k)}$ again and construct a new $g$.
These steps are new compared with prior work
(except for clipping, which is used in~\cite{fan2026private}).
They are important both for our privacy analysis and for ensuring
sufficient progress towards an aggregate witness.

The idea behind aggregation is to repeatedly extract witnesses from
the remaining portion $r$.
We start with $g^+ = 0$, $g^-=0$, and $r = \one_{\binom{n}{2}}$.
Whenever the construction yields a witness $g$ certifying overestimation
or underestimation, we add it to $g^+$ or $g^-$, respectively,
and subtract it from $r$.
We then repeat the construction with the error vectors restricted
to the updated portion:
$D^+ = r\odotc (h- (1+\gamma)w)$ and $D^- = r\odotc ((1-\gamma)w - h)$.
Consequently, aggregation terminates within $O(n^2)$ steps.
With high probability, both $F(D^+)$ and $F(D^-)$ are
sufficiently small at termination, certifying that
$(g^+,g^-,r)$ is a valid aggregate witness.
Each witness $g$ removes a fraction of each edge from further
consideration, and the total fraction removed from any edge $e$
is $g^+_e+g^-_e=1-r_e\leq1$.
In our privacy analysis (\Cref{sec:privacy-pwyg-comp}), we charge
the privacy cost of constructing each witness to this fractional
progress towards building the aggregate witness. This keeps the
total privacy charge small even when many witnesses are needed.

\subsubsection{The Analysis: Main Challenges}

\paragraph{The privacy accounting scheme.}
The main challenge is to construct an aggregate witness with a small privacy
loss, even though this may require many calls to FindWitness.
Applying advanced composition to all these calls would require stronger
regularization, increasing the additive error through the term
$\lambda n\log n$ in the cut approximation guarantee. 
We instead aim to charge each successful search in proportion to the
fraction of each edge that its witness removes from the current portion.
If $g^{(t)}$ is the witness returned by successful search $t$, then
\[
 \sum_t g^{(t)}(e)\leq1
 \qquad\text{for every edge }e.
\]
Thus, if privacy charges can be bounded by these fractions, their total
can remain small even when many witnesses are needed.
Making this scheme work requires a refinement of the SDP stability bounds
in~\cite{EKKL20,fan2026private}, control of the random witnesses, and an
accounting argument for unsuccessful searches.

\paragraph{A privacy bound that reflects fractional progress.}
For intuition, fix two neighboring inputs that differ only in the weight
of an edge $f$, by at most one unit, and consider a common history of the
aggregation procedure.
The current candidate $h$ and mask $r$ are then fixed.
A returned witness $g$ may remove only a tiny fraction $g(f)$ of this edge,
so a uniform privacy bound for each search is insufficient for this accounting.
We need the privacy charge to be small whenever this fraction is small.

The stability analyses of~\cite{EKKL20,fan2026private} already show how
log-det regularization makes Gaussian sampling from the SDP optimizer
private. Our refinement retains the dependence on the fractional-cut
value of the affected edge. If $X_0,X_1$ are the optimizers for the two
neighboring inputs and
$q_f^{\max}=\max\{q_f(X_0),q_f(X_1)\}$, then
\cref{lem:stability_of_optimizer} bounds their relative covariance distance by
$O(r_fq_f^{\max}/\lambda)$.
The corresponding uniform stability bound in~\cite{EKKL20} is $O(1/\lambda)$.
The resulting bound on the Gaussian privacy charge is, up to logarithmic
factors, $O(r_fq_f^{\max}/\lambda)$.
The factor $q_f^{\max}$ is essential: it allows a search to incur a small
charge when the SDP assigns little weight to the affected edge.
For general neighboring weight vectors, the analysis sums these contributions
with weights given by the absolute changes in the edge weights.

\paragraph{A random witness must remove enough to pay its charge.}
The refined stability bound depends on the SDP optimizers, whereas the
fraction removed from $r$ depends on the sampled witness. These quantities
must be related on the realized output: an unbiased estimate alone does not prevent
a sample from assigning very little weight to the affected edge.
This is why we average estimates obtained from $O(\log n)$ independent
Gaussian samples.
Together with clipping, averaging gives simultaneous constant-factor
control of all coordinates of the stochastic estimate.
The stability bound also ensures that $q_e(X_0)$ and $q_e(X_1)$ are
within a constant factor of each other. Consequently, on a common
high-probability event, every witness returned after masking and
normalization satisfies
\[
 r_fq_f^{\max}\leq O(g(f)).
\]
This connects the privacy charge to the fraction actually removed.

\paragraph{Unsuccessful searches still require privacy accounting.}
Not every attempt to construct a witness is successful.
If the constructed witness $g$ does not satisfy the requirements,
the algorithm samples new Gaussian vectors and tries again.
This process may be repeated several times.
The preceding argument bounds the sum of these charges only over
successful attempts.
An unsuccessful attempt removes no fraction of an edge, yet its failure
can reveal information and affect subsequent searches.
Moreover, even the number of successful searches can be large, so charging
a fixed amount per success would still be too expensive.

We address this using a type of success-charged composition,
building on several previous works, including~\cite{gupta2010differentially,kaplan2021sparse,durfee2019practical,cohen2023target}.
The composition lemma accounts for the full adaptive sequence of searches
using a bound on the sum of the charges for successful searches, together
with an overhead controlled by the largest charge and the logarithm of
the inverse failure probability. Here, the bound on the successful charges
follows from $\sum_tg^{(t)}(f)\leq1$.
The exceptional events in the Gaussian and coordinatewise bounds are
accounted for in the additive privacy parameter $\delta$.
This yields a privacy guarantee for the entire aggregate witness without
paying a separate full charge for every search.
Standard composition then applies to the polylogarithmically many
aggregate witnesses and the private mass estimates used by the outer algorithm.

\paragraph{Convergence despite opposing update directions.}
Once the aggregate witness guarantees are established, the outer analysis
follows the PREM framework~\cite{PREM25}. The updates must address both
overestimation and underestimation, measured by $F(D^+)$ and $F(D^-)$,
respectively.
An update aimed at reducing $F(D^+)$ may increase
$F(D^-)$, and conversely, so the two objectives need not
decrease together.
The usual KL potential between the normalized true and candidate weights
provides a common measure of progress. With high probability, each
selected aggregate witness has enough mass and relative
discrepancy to decrease this potential by $\Omega(\gamma^2)$, regardless
of the update direction.

\subsubsection{Comparison to Prior Work}
Our algorithm builds on the SDP-based private mirror-descent framework of
Eli{\'a}{\v{s}}, Kapralov, Kulkarni, and Lee~\cite{EKKL20}.
At each iteration, they solve a regularized SDP relaxation of the cut discrepancy
between the private input graph and the public synthetic candidate graph.
The log-det regularizer stabilizes the SDP optimizer, allowing a stochastic
gradient estimate to be released privately and used in the next mirror-descent
update. Advanced composition bounds the privacy loss across these releases.
This way, the privacy budget is spent within $\tilde O(m/n)$ iterations after which
the additive error remains
$\widetilde O\left(\sqrt{mn/\epsilon}\right)$
for a graph of total edge mass $m$.

Fan, Liu, Peng, Xu, and Zou~\cite{fan2026private} first use private spectral
methods and a recursive demand-aware expander decomposition to approximate parts of the graph, leaving
a sparse residual with at most $M=n^{5/4+o(1)}$ edges.
They approximate this residual using a private mirror-descent mechanism that
extends the SDP-based approach of~\cite{EKKL20} to allow relative error.
Each mirror-descent update uses a privately released stochastic gradient
estimate, and advanced composition bounds the privacy loss across iterations.
For fixed relative-error and privacy parameters, combining the convergence
analysis of mirror descent with the privacy bounds yields an additive error of
$\widetilde O\left(n+(n^2M)^{1/3}\right)$.
Applying this mechanism to the sparse residual gives an overall approximation
with relative error $\gamma$ and additive error
$\widetilde O(n^{13/12+o(1)})$.

In both~\cite{EKKL20} and~\cite{fan2026private}, the final error reflects a
tradeoff between convergence and privacy. More mirror-descent iterations
improve optimization accuracy, but composing the gradient releases under a
fixed privacy budget requires stronger regularization, which introduces
additional approximation error.

Our algorithm improves this tradeoff by constructing aggregate witnesses
that ensure substantial progress in each update. Although constructing a
single aggregate witness may require many individual gradient searches, we
show that, regardless of their number, the privacy loss over the entire aggregation
remains bounded.
In particular, this requires a more refined stability statement about the optimizers
of \eqref{eq:overview-FD} compared to \cite{EKKL20} and \cite{fan2026private}.
The resulting aggregate witnesses allow the PREM framework~\cite{PREM25}
for relative-error histogram release to converge in few multiplicative-weights
updates. Advanced composition therefore accounts for only a small number
of aggregation outputs. Combining this rapid convergence with effective
privacy accounting within each aggregation yields our improved
cut-approximation guarantee.

When applied to cut queries, the PREM framework \cite{PREM25} in fact
provides privacy and utility guarantees comparable to ours, up to
polylogarithmic factors. However, its direct implementation constructs
aggregate witnesses by scanning the exponentially large family of cuts.
In contrast, we construct aggregate witnesses in polynomial time from
fractional witnesses derived from a regularized SDP.
To accommodate these fractional witnesses, we maintain the residual graph
as a fractional portion of the original graph, whereas the original PREM
framework retains or removes each histogram coordinate in its entirety.

\paragraph{Organization.}
After the preliminaries in \cref{sec:prelim}, \cref{sec:CutRelease} presents CutRelease (\Cref{alg:main}).
\Cref{sec:witness} defines the regularized objective and describes FindWitness and AggregateWitnesses,
and \cref{sec:preprocess} presents ClipEdgeWeights. Section~\ref{sec:privacy} proves
optimizer stability, Gaussian covariance privacy, and the privacy guarantees of
\Cref{alg:gradient_aggregation} and \Cref{alg:main}.  Section~\ref{sec:utility} proves the
oracle certificate, the inner and outer iteration bounds, and the main utility theorem.  The
appendices contain the omitted proofs, the applications and the lower
bound.

\section{Preliminaries}\label{sec:prelim}

For $x,y\in\R^{n}$, $x\odotc y$ denotes coordinatewise multiplication,
i.e. $(x\odotc y)_i = x_iy_i$ for each coordinate $i=1, \dotsc, n$.
All logarithms in the paper are natural.

\paragraph{Differential privacy.}
We introduce basic definitions and mechanisms in differential privacy. In this paper, we consider the well-established \emph{edge-level} differential privacy (DP)~\cite{DMNS06}. Given a graph $G = (V,w)$, the public vertex set is fixed, and every unordered pair is an edge coordinate of the edge weights vector $w\in \mathbb{R}^{{n\choose 2}}$, including zero-weight non-edges. \begin{definition}
\label{def:DP}
Consider histograms $w,w' \in \R_{\geq0}^{\binom{n}{2}}$.
We say that histograms $w,w'$ are adjacent, written $w\sim w'$, if
$\norm{w-w'}_1\leq1$.
A randomized algorithm $\cM$ is $(\varepsilon, \delta)$ edge-DP if for
all adjacent $w,w'$ and measurable output events $A$,
\[
 \Prb[\cM(w)\in A]\leq e^\varepsilon\Prb[\cM(w')\in A]+\delta,
\]
and symmetrically with $w,w'$ exchanged.
\end{definition}

\begin{definition}[Laplace Distribution]
\label{def:laplace-dist}
The centered Laplace distribution with scale parameter $b > 0$, denoted as $\text{Lap}(b)$, is a continuous probability distribution with the probability density function
\[
    h(z) = \frac{1}{2b} \exp\left(-\frac{|z|}{b}\right), \quad z \in \mathbb{R}.
\]
If $Z \sim \text{Lap}(b)$, then $\Prb[|Z| \ge t] = \exp\left(-\frac{t}{b}\right)$ for $t\geq 0$.
\end{definition}

We use the Laplace mechanism, postprocessing, and basic and advanced adaptive composition in their standard
forms~\cite{DMNS06,DR14,DRV10}. 

\begin{lemma}[Laplace Mechanism]
\label{lem:laplace-mech}
Let $f \colon \mathcal{X} \to \mathbb{R}^k$ be a function with $\ell_1$-sensitivity $\Delta_1 f = \max_{x \sim x'} \|f(x) - f(x')\|_1$. For any $\varepsilon > 0$, the Laplace mechanism $\mathcal{M}_L$ defined by
\[
    \mathcal{M}_L(x) = f(x) + (Z_1, \dots, Z_k),
\]
where $Z_1, \dots, Z_k$ are drawn i.i.d. from $\text{Lap}(\Delta_1 f / \varepsilon)$, satisfies $\varepsilon$-differential privacy.
\end{lemma}

\begin{lemma}[Post-Processing]
\label{lem:post-processing}
Let $\mathcal{M} \colon \mathcal{X} \to \mathcal{Y}$ be a randomized mechanism that is $(\varepsilon, \delta)$-differentially private. Let $f \colon \mathcal{Y} \to \mathcal{Z}$ be an arbitrary randomized mapping whose internal randomness is independent of the private input data. Then the composite mechanism $f \circ \mathcal{M} \colon \mathcal{X} \to \mathcal{Z}$ remains $(\varepsilon, \delta)$-differentially private.
\end{lemma}

\begin{lemma}[Basic Adaptive Composition]
\label{lem:basic-composition}
Let $\mathcal{M}_1, \dots, \mathcal{M}_k$ be a sequence of adaptively chosen randomized mechanisms, where for each $i \in [k]$, the mechanism $\mathcal{M}_i$ is $(\varepsilon_i, \delta_i)$-differentially private. Then their adaptive composition is $\left( \sum_{i=1}^k \varepsilon_i, \sum_{i=1}^k \delta_i \right)$-differentially private.
\end{lemma}

\begin{lemma}[Advanced Adaptive Composition~\cite{DRV10,DR14}]
\label{lem:advanced-composition}
Let $\mathcal{M}_1, \dots, \mathcal{M}_k$ be a sequence of adaptively chosen randomized mechanisms, where, conditional on every fixed preceding transcript, each $\mathcal{M}_i$ is $(\varepsilon_0, \delta_0)$-differentially private for some $\varepsilon_0 > 0$ and $\delta_0 \ge 0$. Then for any $\delta'\in(0,1)$, the adaptive composition of these $k$ mechanisms is $(\varepsilon, k\delta_0 + \delta')$-differentially private, where
$$
   \varepsilon = \varepsilon_0 \sqrt{2k \ln(1/\delta')} + k \varepsilon_0 (e^{\varepsilon_0} - 1).
$$
\end{lemma}

\section{The Algorithm}

We now describe our algorithm. It proceeds in three main steps:
\begin{enumerate}
\item First, as a preprocessing step, we clip edge weights and privately release the excess weight of edges heavier than $U=O(\log n)$ using the Laplace mechanism (\Cref{lem:laplace-mech}). This is to ensure that the sum of the weights in the input graph is bounded, as it affects the number of iterations performed by the main algorithm.

\item The main optimization procedure, CutRelease, builds on the PREM framework of \cite{PREM25} for answering linear queries on histograms with relative-error guarantees. It proceeds in $O(\log nU)$ iterations, each operating at a progressively smaller scale.
In each iteration, it computes an approximation of a portion of the graph containing a constant fraction of the remaining edge weight, which is then removed from the graph.

Each partial approximation is computed using multiplicative weights updates (MWU). Let $M_i$ denote the total edge weight in the residual graph and let $h_i$ denote the edge-weight vector of the current candidate graph. Given high-quality \emph{aggregate witnesses} $g_1,\ldots,g_T$ satisfying, in particular, $\langle h_i,g_t\rangle \geq M_i/3$ at each step $t=1,\ldots,T$, the MWU procedure requires only a logarithmic number of updates. Privacy loss across the MWU steps is bounded using the advanced composition theorem (Lemma~\ref{lem:advanced-composition}).

\item The core of our algorithm is a separation oracle that produces \emph{aggregate witnesses} for the MWU procedure described above.
Each aggregate witness is derived from several fractional cuts computed using the regularized SDP \eqref{eq:overview-FD}, which certify errors in the approximation of the original weights $w$ by the current candidate weights $h$. This aggregation is necessary because individual fractional cut witnesses may have negligible mass and therefore provide insufficient progress in MWU. Up to $O(n^2)$ fractional cut witnesses may be needed to construct a single aggregate witness. Crucially, the privacy loss across these iterations can be bounded effectively.
We charge privacy loss to fractional progress towards
building the aggregate witness. The fractional contributions accumulated
on each edge sum to at most one, keeping the total privacy charge small
even when many violating cuts are computed.
\end{enumerate}

This section is organized as follows.
The technical but necessary preprocessing step is described in Section~\ref{sec:preprocess}.
Section~\ref{sec:CutRelease} introduces the main algorithm.
The computation and aggregation of fractional cut witnesses, which constitute the most challenging part of the algorithm, are described in Section~\ref{sec:witness}.

\subsection{CutRelease: The Main Optimization Procedure}
\label{sec:CutRelease}
We describe the CutRelease procedure, which guides the overall optimization process and builds on the PREM framework of \cite{PREM25}.
Here, we assume that every edge weight is bounded by $U=O(\log n)$.

CutRelease proceeds in iterations.
In each iteration, it finds an approximation of a fractional portion $r$ of the graph containing a constant fraction of the remaining total edge weight and removes this portion from further consideration.
The procedure performs $O(\log nU)$ iterations in total.
The portion $r$ and its approximation are found using an MWU algorithm, whose updates are based on the output of the oracle AggregateWitnesses.
The CutRelease procedure is summarized in Algorithm~\ref{alg:main}.

\begin{algorithm}[h]
\caption{$\mathsf{CutRelease}(w)$}
\label{alg:main}
\begin{algorithmic}[1]
\REQUIRE A preprocessed bounded graph $G=([n],E)$ with edge-weight vector $w$ (where $w_e \leq U$), a relative error tolerance $\gamma$, privacy parameters $\epsilon, \delta \in (0, 1)$, and a failure probability $\beta$.
\ENSURE A nonnegative weighted synthetic graph $\widetilde{G}$ with edge-weight vector $\widetilde{w}$.
\STATE Set parameters: $T_1=\Theta(\log (nU))$, $T_2=1+\left\lceil64(1+\gamma)^2\log n/\gamma^2\right\rceil$, $M_{\mathrm{stop}}=C(nU\log\frac{nU}{\delta\beta\gamma\eps}+\frac{n}{\gamma\varepsilon}(\log\frac{nU}{\delta\beta\gamma\eps})^6)$, $\eta=\gamma/[8(1+\gamma)]$, 
$\eps_2=\frac{\eps}{2T_1}$, $\beta_2=\frac{\beta}{2T_1}$, $\tau_2=\frac{\log(1/\beta_2)}{\eps_2}$,
$\eps_{\mathrm{GE}}=\frac{\eps}{4T_1\sqrt{2T_2\log(2T_1/\delta)}}$, $\delta_{\mathrm{GE}}=\frac{\delta}{2T_1T_2}$, $\beta_{\mathrm{GE}}=\frac{\beta}{2T_1T_2}$.
\STATE Let $a_1\gets \one_{\binom{n}{2}}$, $\widetilde{w}\gets \mathbf{0}$.
\FOR{$i=1,\ldots,T_1$}
  \STATE Let $w_i\gets a_i\odot w$.
  \STATE Privately release $\widetilde M_i$ for $M_i=\sum_{e}w_i(e)$ by $\widetilde M_i=\max(0,M_i+\Lap(1/\eps_2)+\tau_2).$
  \IF{$a_i=\mathbf{0}$ or $\widetilde M_i\le M_{\mathrm{stop}}$}
    \STATE \Return $\widetilde{w}$.
  \ENDIF
  \STATE Initialize $h_i$ by $h_i(e)=\widetilde M_i \frac{a_i(e)}{\sum_{e}a_i(e)}$ and $a_{i+1}$ by $a_{i+1}=a_i$.
  \FOR{$t=1,\ldots,T_{2}$}
    \STATE \COMMENT{Steps of MWU}
    \STATE $(g^+_i,g^-_i,r_i)\gets\mathsf{AggregateWitnesses}(w_i,h_i,\gamma)$.
    \IF{$\langle h_i,r_i\rangle \geq \max\{\langle h_i,g_i^+\rangle, \langle h_i,g_i^-\rangle\}$}
      \STATE Partial estimator collection: $\widetilde{w}\gets \widetilde{w}+h_i\odot r_i$.
      \STATE Mask update:  $a_{i+1}\gets a_i\odot(\one_{\binom{n}{2}}-r_i)$ and \textbf{break}.
    \ELSE
      \STATE Let $\theta=1, g\gets g^+_i$ if $\langle h_i,g^+_i\rangle\geq \langle h_i,g^-_i\rangle$, and $\theta=-1, g\gets g^-_i$ otherwise.
      \STATE Mirror descent update: $h_i(e)\gets\widetilde M_i\cdot\frac{h_i(e)\exp(-\theta\eta g(e))}{\sum_{e}h_i(e)\exp(-\theta\eta g(e))}$ for all $e$.
    \ENDIF
  \ENDFOR
\ENDFOR
\STATE \Return $\widetilde{w}$.
\end{algorithmic}
\end{algorithm}

\paragraph{Description of an iteration.}
Consider iteration $i$ of Algorithm~\ref{alg:main}.
Let $a_i$ denote the unreleased portion of the graph,
$w_i = a_i \odotc w$ the edge weights of this portion,
and $M_i = \sum_e w_i(e)$ its total weight.
Algorithm~\ref{alg:main} computes a private estimate $\tilde M_i$ of $M_i$
using the Laplace mechanism.
If $\tilde M_i \leq M_{\mathrm{stop}}$, Algorithm~\ref{alg:main} terminates.
Otherwise, we denote by
\[
\mathcal{W} := \{ h \in \mathbb{R}_{\geq 0}^{\binom{n}{2}} : \| h \|_1 = \tilde M_i\}
\]
the set of candidate edge-weight vectors.

The algorithm then uses MWU to compute a fractional portion $r_i$ containing a large mass of the
edge weights and its approximation $h_i$.
To this end, it uses an oracle that fractionally partitions the edge coordinates into three portions $g^+,g^-,r \in [0,1]^{\binom{n}2}$. The vectors $g^+$ and $g^-$ identify overestimation and underestimation, respectively, while $r$ is a portion on which the candidate is already sufficiently accurate.
We call $g^+$ and $g^-$ aggregate witnesses.

We assume that the oracle returns fractional vectors
$g^+,g^-,r\in[0,1]^{\binom{n}{2}}$ satisfying
\begin{align*}
g^++g^-+r &= \one_{\binom{n}{2}},\\
\langle h_i,g^+\rangle &\ge (1+\gamma)\langle w_i,g^+\rangle,\\
\langle h_i,g^-\rangle &\le (1-\gamma)\langle w_i,g^-\rangle.
\end{align*}
In addition, the oracle guarantees that $h_i$ approximates $w_i$
on the portion $r$, simultaneously for all cuts,
up to relative error $\gamma$ and additive error $B$.

Since these three vectors sum to the all-ones vector,
at least one of them carries a third of the mass of $h_i$.
If
$
\langle h_i,r\rangle
\geq \max\{\langle h_i,g^+\rangle,\langle h_i,g^-\rangle\},
$
we terminate MWU and release the approximation on $r$.
Otherwise, one of $g^+$ and $g^-$ carries at least a third
of the mass of $h_i$ and provides a direction for the next
MWU update.

With $\gamma$ denoting the multiplicative error tolerance,
$T_2=O(\log n/\gamma^2)$ MWU steps are enough
to obtain $r_i$ and $h_i$ such that
\[
\langle h_i,r_i\rangle
\geq \max\{\langle h_i,g^+\rangle,\langle h_i,g^-\rangle\}
\]
and therefore $\langle h_i,r_i\rangle \geq \frac13\|h_i\|_1$.
Since $h_i$ approximates $w_i$ on the portion $r_i$,
this implies
\[
\langle w_i,r_i\rangle\geq \frac{M_i}{20}.
\]
We can therefore release $h_i\odot r_i$ and remove the corresponding fractional
portion from the residual graph, decreasing its total weight by a constant
factor.
See Section~\ref{sec:mwu} for details.

\subsection{Producing aggregate witnesses}
\label{sec:witness}
Ideally, we would like to find $r_i$ and $h_i$
that minimize the violations of the constraints
\begin{itemize}
\item $\langle c_S, r_i\odotc h_i\rangle \leq (1+\gamma) \langle c_S, r_i\odotc w_i\rangle$ and
\item $\langle c_S, r_i\odotc h_i\rangle \geq (1-\gamma) \langle c_S, r_i\odotc w_i\rangle$
\end{itemize}
for each cut $S\subseteq V$ with incidence vector
$c_S \in \{0,1\}^{\binom{n}{2}}$.
This would require a separation oracle that finds cuts $S$
corresponding to violated constraints.
However, finding a maximally violating cut is computationally hard.
Moreover, such a cut may change drastically under a small change in $w_i$,
making it challenging to release privately.

We address these issues using an approach similar to that of~\cite{EKKL20}, optimizing a regularized objective over a set of fractional cuts represented by PSD matrices:
\begin{equation}
\label{eq:elliptope}
\mathcal{X}=\left\{X\in\mathbb{R}^{n\times n}\mid X = X^\top, \diag(X) = \one_n, X \succeq \frac{1}{n} I_n\right\}.
\end{equation}
Given $X\in \mathcal{X}$, let $q_{uv}(X) = \frac{1-X_{uv}}{2} \in [0,1]$ denote the fraction of edge $uv$
cut by $X$, and let $q(X) = (q_e(X))_e \in [0,1]^{\binom{n}{2}}$.
Let $D^+ = h - (1+\gamma)w$ and $D^- = (1-\gamma)w - h$ denote
the signed error vectors describing how well $h$ approximates $w$.
The violations of the constraints corresponding to $X$ are expressed as
$\langle D^+, q(X)\rangle$ and $\langle D^-, q(X)\rangle$, respectively.
We therefore consider the regularized objective
\begin{equation*}
\min_{h\in \mathcal{W}} \max \big\{F\big(h-(1+\gamma)w\big), F\big((1-\gamma)w-h\big)\big\},
\end{equation*}
where
\begin{equation}
\label{eq:F_D}
F(D) = \max_{X \in \mathcal{X}} \left\{\langle D, q(X)\rangle + \lambda \log \det(X) \right\}.
\end{equation}
Note that $\log\det(X)$ is bounded on $\mathcal{X}$
thanks to the constraint $X \succeq \frac1n I_n$ in \eqref{eq:elliptope}.
We compute the gradient of $F(D)$ using Danskin's theorem~\cite{danskin1966theory}. Since the strictly concave regularizer $\lambda \log\det(X)$ ensures that the maximizer $X_D\in\mathcal{X}$ is unique, Danskin's theorem implies that the gradient of $F(D)$ is the partial derivative of the inner objective with respect to $D$, evaluated at $X_D$. Therefore,
\[
\nabla_D F(D) = q(X_D).
\]
Applying the chain rule to $D^+$ and $D^-$, we obtain the gradients with respect to the candidate weights $h$:
\[
\nabla_h F(h-(1+\gamma)w)
    = q(X_{D^+}),
\]
\[
\nabla_h F((1-\gamma)w-h)
    = -q(X_{D^-}).
\]
These gradients provide natural directions for MWU.
In what follows, we describe how to transform them into aggregate witnesses
satisfying the requirements of Algorithm~\ref{alg:main}
with a small privacy loss.

\subsubsection{Computing a single witness}
\label{sec:GradEst}
First, we describe how to construct a witness corresponding to a single
gradient of $F(D)$.
Given a mask vector $r$ and an error vector $D$,
FindWitness first computes the gradient of $F(D)$.
As discussed, $\nabla_D F(D) = q(X_D)$, so the algorithm first computes the optimizer $X_D$ (line~\ref{line:optimize_XD}).
However, simply releasing $q(X_D)$ would not respect differential privacy.
Instead, following prior works~\cite{EKKL20,fan2026private}, we sample a vector $(Z_u)_{u\in [n]}$ from $\mathcal{N}(0,X_D)$, the Gaussian distribution with covariance matrix $X_D$. As shown by Fan, Liu, Peng, Xu, and Zou~\cite{fan2026private}, this preserves approximate differential privacy, and $\frac{(Z_u - Z_v)^2}{4}$ serves as an unbiased estimator of $q_{uv}(X_D)$ (see also Proposition~\ref{prop:unbiased_estimator}). 

However, we draw $k=O(\log n)$ samples from $\mathcal{N}(0,X_D)$ and
take the average of the resulting values of $\frac{(Z_u - Z_v)^2}{4}$ for each edge
to obtain a constant-factor estimate of $q_{uv}(X_D)$ for all pairs $\{u,v\}$ simultaneously (Lemma~\ref{lem:chi}).
To verify that the resulting estimate $\hat g$ indeed witnesses significant overestimation or underestimation of weights, we examine $\langle D, \hat g\rangle$, which, with constant probability, is at least $\frac12 \langle D, q(X_D)\rangle$ (Lemma~\ref{lem:chaos}).
Laplace noise is added to $\langle D, \hat g\rangle$ to produce a private estimate $Y$.
If $Y$ is small, this implies that we were unable to witness a significant error in $D$ over all relaxed cuts $X\in\mathcal{X}$.
Each trial computes these averages, clips the resulting vector, and internally computes its noisy score. On acceptance, it masks and normalizes the clipped vector to obtain $g$; otherwise, it tries again, up to $K$ trials.
If we are never able to witness significant error, we return $\bot$.
See Algorithm~\ref{alg:gradient_estimation} for details.
The final witness satisfies the following properties:
\begin{itemize}
\item $0\leq g_e \leq r_e$ for each $e\in \binom{n}2$ and
\item there exists $e\in \binom{n}2$ such that $g_e = r_e>0$.
\end{itemize}
This means that
we can see $g$ as a portion of the graph which
can be removed from $r$, eliminating at least one nonzero coordinate
from $r$.
See Section~\ref{sec:gradient_estimation} for details.

We note that
Algorithm~\ref{alg:gradient_estimation} does not release any information
on its own.
Its output will be released only as part of the aggregate
witnesses.

The following lemma shows that the estimator $\hat{g}$ in Algorithm~\ref{alg:gradient_estimation} is a constant-factor estimate of the true gradient, proved in Appendix~\ref{sec:Gaussian_approximate}.

\begin{algorithm}[h!]
\caption{$\mathsf{FindWitness}(D,r)$}
\label{alg:gradient_estimation}
\begin{algorithmic}[1]
\REQUIRE A signed objective vector $D$, a mask $r$, privacy parameters $\epsilon_{\mathrm{GE}}, \delta_{\mathrm{GE}} \in (0, 1)$, and a failure probability $\beta_{\mathrm{GE}}$.
\ENSURE A witness $g$ or the failure symbol $\bot$.
\STATE Set parameters: $k=\Theta(\log(\frac{n}{\beta_\chi\del_1}))$, $K=\Theta(\log(\frac{n}{\beta_{\mathrm{GE}}}))$, $\lambda=\Theta(\frac{1}{\eps_1}\log\frac{n}{\beta_{\mathrm{GE}}\del_{\mathrm{GE}}})$,  $\eps_1=\frac{\eps_{\mathrm{GE}}}{320K(1+\log(2/\del_{\mathrm{GE}}))}$, $\del_1=\frac{\del_{\mathrm{GE}}}{20n^2K}$, $\tau_1=\frac{\log(1/\beta_1)}{\eps_1}$, $\beta_1=\beta_{\chi}=\frac{\beta_{\mathrm{GE}}}{4n^2K}$.
\IF{$r=0$}
  \STATE \Return $\bot$
\ENDIF
\STATE Compute the optimizer $X_D$ of \eqref{eq:F_D}.\label{line:optimize_XD}
\FOR{$i=1,\dots,K$}
    \STATE Choose random vectors $Z^{(j)}\sim \mathcal{N}(0,X_D)$ for each $j=1,\dots,k$.
    \STATE Set $\overline g(u,v)=\frac1k\sum_{j=1}^k\frac{(Z_u^{(j)}-Z_v^{(j)})^2}{4},$ and $\hat g(u,v)=\min\{\overline{g}(u,v)/2,1\}$.
    \STATE Compute a private estimate $Y=\langle D, \hat{g} \rangle+L$, where $L\sim\Lap(1/\eps_{1})$.\STATE Set $\hat g_{\max}\gets\max\{\hat g(e):r(e)>0\}$.
    \IF{$Y<\tau_1$ or $\hat g_{\max}=0$} \label{line:accepted_trial}
    \STATE \textbf{Continue.}
  \ENDIF
  \STATE Set $g(e)\gets r(e)\cdot\hat g(e)/\hat g_{\max}$ for each $e$. \label{line:rescale_gradient}
  \RETURN $g$.
\ENDFOR
\STATE \Return $\bot$.
\end{algorithmic}
\end{algorithm}

\begin{lemma}\label{lem:chi}
For every fixed $X \in \mathcal{X}$ and $\beta_{\chi},\del_1\in(0,1)$, choose independent random vectors $Z^{(j)}\sim \mathcal{N}(0,X)$ for each $j=1,\dots,k$. Set $\overline g(u,v)=\frac1k\sum_{j=1}^k\frac{(Z_u^{(j)}-Z_v^{(j)})^2}{4},$ and $\hat g(u,v)=\min\{\overline{g}(u,v)/2,1\}$. For some universal constant $C > 0$, if
$k\geq C\log(n/(\beta_\chi\del_1))$, then with probability at least $1-\min(\beta_\chi, \del_1)$, for every pair of nodes $e=\{u,v\}$ simultaneously,
\[\frac{q_{e}(X)}{4}\leq \hat{g}(u,v)\leq q_{e}(X).\]
\end{lemma}

When $X = X_D$, the following lemma shows that, with constant probability, $\langle D, \hat{g} \rangle$ in Algorithm~\ref{alg:gradient_estimation} captures a constant fraction of the linear part of $F(D)$, also proved in Appendix~\ref{sec:Gaussian_approximate}.

\begin{lemma}\label{lem:chaos}
For every fixed $D \in \mathbb{R}^{\binom{n}{2}}$, $X \in \mathcal{X}$ and $\beta_{\chi},\del_1\in(0,1)$, choose independent random vectors $Z^{(j)}\sim \mathcal{N}(0,X)$ for each $j=1,\dots,k$. Set $\overline g(u,v)=\frac1k\sum_{j=1}^k\frac{(Z_u^{(j)}-Z_v^{(j)})^2}{4},$ and $\hat g(u,v)=\min\{\overline{g}(u,v)/2,1\}$.  For some universal constant $C > 0$, if
$k\geq C\log(n/(\beta_\chi\del_1))$, for a universal constant $c_0>0$, conditioning on the good event in Lemma~\ref{lem:chi},
\[
 \Prb\left[\langle D,\hat{g}\rangle\geq\frac{1}{2}\sum_{e=\{u,v\} \in \binom{n}{2}} D_{uv} q_{e}(X)\right]\geq c_0.
\]
\end{lemma}

\subsubsection{Aggregating the witnesses}
\label{sec:GradAgg}

A single invocation of FindWitness 
might return a highly localized witness. Feeding such a sparse witness directly to MWU in Algorithm~\ref{alg:main} would yield negligible progress while still incurring the full privacy cost of the oracle call.
Therefore, we aggregate multiple witnesses before they are released
and used in Algorithm~\ref{alg:main}.

AggregateWitnesses (Algorithm~\ref{alg:gradient_aggregation})
runs a loop in which it maintains a portion $r \in [0,1]^{\binom{n}{2}}$, initialized to the all-ones vector $\one_{\binom{n}{2}}$.
Applying this mask $r$ to both $D^+$ and $D^-$, the algorithm attempts to find witnesses for overestimation and underestimation with respect to the mask $r$ by calling Algorithm~\ref{alg:gradient_estimation}.
Algorithm~\ref{alg:gradient_aggregation} repeatedly calls Algorithm~\ref{alg:gradient_estimation} with different masks $r$ and aggregates the results into $g^+$ and $g^-$, which become aggregate witnesses of $h$ overestimating or underestimating weight on some set of edges, respectively.
Whenever an overestimating witness ($g^+$) or an underestimating witness ($g^-$) is successfully found (i.e., $g_{\mathrm{new}} \neq \bot$), the algorithm accumulates it, deducts it from the active residual mask $r$, and immediately loops back to search again on the updated mask.
With high probability, when Algorithm~\ref{alg:gradient_aggregation} terminates, we will have $\langle h, g^+ \rangle \geq (1+\gamma) \langle w, g^+\rangle$ and $\langle h, g^- \rangle \leq (1-\gamma) \langle w, g^-\rangle$ (Lemma~\ref{lem:gradient_estimation}).

\begin{algorithm}[ht]
\caption{$\mathsf{AggregateWitnesses}(w,h,\gamma)$}
\label{alg:gradient_aggregation}
\begin{algorithmic}[1]
\STATE $(g^+,g^-,r)\gets(0,0,\one_{\binom{n}{2}})$.
\LOOP
  \IF{$r=0$}
    \STATE Release and return $(g^+,g^-,0)$.
  \ENDIF
  \STATE Let $g_{\mathrm{new}}=\mathsf{FindWitness}(D^+,r)$ where $D^+=r\odotc(h-(1+\gamma)w)$.
  \IF{$g_{\mathrm{new}}\ne \bot$}
    \STATE $g^+\gets g^++g_{\mathrm{new}}$, $r\gets r-g_{\mathrm{new}}$ and \textbf{continue}.\label{line:aggregate-positive}
  \ENDIF
  \STATE Let $g_{\mathrm{new}}=\mathsf{FindWitness}(D^-,r)$ where $
 D^-=r\odotc((1-\gamma)w-h).$
  \IF{$g_{\mathrm{new}}\ne \bot$}
    \STATE $g^-\gets g^-+g_{\mathrm{new}}$, $r\gets r-g_{\mathrm{new}}$ and \textbf{continue}.\label{line:aggregate-negative}
  \ENDIF
\STATE Release and return $(g^+,g^-,r)$.
\ENDLOOP
\end{algorithmic}
\end{algorithm}

In Algorithm~\ref{alg:gradient_aggregation}, the loop reaches the final return statement if and only if both consecutive calls to \Cref{alg:gradient_estimation} output $\bot$.
With high probability, termination after both calls return $\bot$ certifies that $h$ approximates $w$ on the remaining mask $r$, simultaneously for all cuts, with relative error $\gamma$ and additive error $B$.

Observe that substituting Line~\ref{line:rescale_gradient} of Algorithm~\ref{alg:gradient_estimation} into the mask update rule (Line~\ref{line:aggregate-positive} or~\ref{line:aggregate-negative} in Algorithm~\ref{alg:gradient_aggregation}) yields
$r(e)\leftarrow  r(e) - r(e) \cdot \frac{\hat{g}(e)}{\hat{g}_{\max}} = r(e) \left( 1 - \frac{\hat{g}(e)}{\hat{g}_{\max}} \right)$ so that $r\in \mathbb{R}^{n\choose 2}$ remains nonnegative. Further, there is at least one $e\in{n\choose2}$ such that $\hat{g}(e) = \hat{g}_{\max}$, which ensures that every successful update zeroes out at least one active coordinate of $r$, guaranteeing that Algorithm~\ref{alg:gradient_aggregation} terminates after at most $\binom{n}{2} = O(n^2)$ successful mask updates.

Algorithm~\ref{alg:gradient_aggregation} performs a single release
once the final aggregate witness is found.
The privacy of this procedure is analyzed as a whole, including the calls to Algorithm~\ref{alg:gradient_estimation}, in \Cref{sec:privacy-pwyg-comp}.
The success-charged composition lemma
(Lemma~\ref{lem:pay-what-you-get}) is used to handle unsuccessful attempts in Algorithm~\ref{alg:gradient_estimation}.
See Section~\ref{sec:privacy-pwyg-comp}.

\subsection{Preprocessing: Clipping Edge Weights}\label{sec:preprocess}
Let the input $G=(V, E)$ be a simple weighted graph on $n$ vertices with edge-weight vector $w \in \mathbb{R}_{\ge 0}^{\binom{n}{2}}$. To ensure that the edge weights are properly bounded for the subsequent multiplicative weights updates, we handle the large-weight portion of each edge separately, allowing us to assume that all edge weights are bounded by $O(\log n)$. In particular, we decompose the weight $w_e$ of each edge into a bounded component $w_e'$ and an excess component $v_e$.
That is, $w_e = w_e' + v_e$.
We then privately release the excess component $v$ and run the rest of the
algorithm only on the bounded weights $w'$. Specifically, we define
\begin{equation}\label{eq: def_split_weight}
  v_e=(w_e-U)_+,\qquad w_e'=\min\{w_e,U\},
\end{equation}
where $U=\frac{8}{\eps_0\gamma}\log\frac{n}{\beta_0}$. Then, independently for each edge $e$, we draw $L_e\sim\Lap(1/\eps_0)$ and release
\begin{equation}\label{eq:noise_split_weight}
  Y_e=\max\left\{0,v_e+L_e-\frac{2}{\eps_0}\log\left(\frac{n}{\beta_0}\right)\right\},
\end{equation}
which serves as a private/noisy surrogate for the excess weight $v_e$. The algorithm is summarized in \Cref{alg:preprocess}.

\begin{algorithm}[h!]
\caption{$\mathsf{ClipEdgeWeights}(w)$}
\label{alg:preprocess}
\begin{algorithmic}[1]
\REQUIRE A weight vector $w \in \mathbb{R}_{\ge 0}^{\binom{n}{2}}$, a relative error tolerance $\gamma \in (0, 1)$, a privacy parameter $\eps_0 \in (0, 1)$, and a failure probability $\beta_0 \in (0, 1)$.
\ENSURE A bounded edge-weight vector $w'$ and a publicly released private estimate $Y$ of the excess weights.
\STATE Set the bounding threshold $U = \frac{8}{\eps_0\gamma}\log\left(\frac{n}{\beta_0}\right)$ and the noise shift $V = \frac{2}{\eps_0}\log\left(\frac{n}{\beta_0}\right)$.
\FOR{each edge $e \in \binom{n}{2}$}
  \STATE Decompose weights: $v_e \gets (w_e-U)_+$ and $w'_e \gets \min\{w_e, U\}$.
  \STATE Draw Laplace noise: $L_e \sim \Lap(1/\eps_0)$.
  \STATE Compute private excess: $Y_e \gets \max\left\{0, v_e + L_e - V \right\}$.
\ENDFOR
\RETURN $(w', Y)$.
\end{algorithmic}
\end{algorithm}

Algorithm~\ref{alg:preprocess} publishes $Y$ and leaves the bounded vector $w'$ for processing using Algorithm~\ref{alg:main}. The final output of the mechanism
is then $Y+\mathsf{CutRelease}(w')$.

This preprocessing step allows us to privately release the excess component of all edges with weight $\gg U$, which induces only additional relative error. 

\begin{lemma}\label{lem:preprocess}
Fix any $\eps_0, \beta_0, \gamma \in (0, 1)$, and $n$-vertex input graph $G$ with edge-weight vector $w$. Decompose the weight vector $w=w'+v$, where $w'$ and $v$ are defined in \eqref{eq: def_split_weight}. 
Then releasing $Y$ (as defined in \eqref{eq:noise_split_weight}) is $\eps_0$-DP, and, with probability at least $1-\beta_0$, for every cut $S$,
\[
 |Y(S,V\setminus S)+w'(S,V\setminus S)-w(S,V\setminus S)|\leq(\gamma/2)w(S,V\setminus S).
\]
\end{lemma}

\begin{proof}
Since the $\ell_1$-sensitivity of $v$ is $1$, by the Laplace Mechanism (Lemma~\ref{lem:laplace-mech}), the release $Y$ is $\eps_0$-DP. 
By tail bounds on the Laplace distribution and a union bound, with probability at least $1-\beta_0$,  for every edge $e$, $|L_e|\leq\frac{1}{\eps_0}\log\left(\frac{\binom{n}{2}}{\beta_0}\right)\leq\frac{2}{\eps_0}\log(\frac{n}{\beta_0})$. In the remainder of the proof, we condition on this good event.

Recall that $Y_e=\max\left\{0,v_e+L_e-V\right\}$ where $V :=\frac{2}{\eps_0}\log(\frac{n}{\beta_0})$. With the aforementioned conditioning on $L_e$, we have that $L_e - V \in [-2V, 0]$.
Thus $v_e + L_e - V$ decreases $v_e$ by at most $2V$, and clipping this quantity at $0$ cannot increase the error.
Therefore $|Y_e - v_e| \leq 2V$.
If $v_e=0$, then $Y_e=0$.
Otherwise, if $v_e>0$, the definition of $v_e$ implies that $w_e > U = 4V/\gamma$.
Therefore, $|Y_e - v_e| \leq 2V \leq (\gamma/2) w_e$.

Finally, since $w=w'+v$, $|Y_e+w_e'-w_e|=|Y_e-v_e|\leq\gamma w_e/2$. For every cut $S$, summing over all edges crossing $(S, V\setminus S)$, we see that $|Y(S,V\setminus S)+w'(S,V\setminus S)-w(S,V\setminus S)|\leq(\gamma/2)w(S,V\setminus S)$ as desired.
\end{proof}

\section{Privacy Analysis}\label{sec:privacy}

We prove the privacy guarantee in two main steps: first, in \Cref{sec:privacy-sdp-stability}, we establish the stability of the log-det SDP optimizer to obtain a pair-local privacy bound for each Gaussian attempt in Algorithm~\ref{alg:gradient_estimation}; then, in \Cref{sec:privacy-pwyg-comp}, we employ success-charged composition (Lemma~\ref{lem:pay-what-you-get}) to amortize these local privacy losses over the adaptive sequence of attempts within each invocation of Algorithm~\ref{alg:gradient_aggregation}. Finally, the overall privacy guarantee for our main mechanism, Algorithm~\ref{alg:main}, follows via standard sequential composition across all $\text{polylog}(n)$ outer and inner rounds.

\subsection{Stability of SDP Optimizer}\label{sec:privacy-sdp-stability}

We first analyze the stability of the optimizer \eqref{eq:F_D} over any pair of neighboring graphs.
For any signed vector $D \in \mathbb{R}^{\binom{n}{2}}$, we let
\begin{equation}\label{eq: A_D}
    A_D=\frac14\sum_{e\in\binom{[n]}{2}}D_eb_eb_e^\top.
\end{equation} Under the standard trace inner product $\langle A, B \rangle = \tr(A^\top B)$ and since $A_D$ is symmetric, $\langle A_D, X \rangle = \tr(A_D X)$.
We therefore have 
\[\langle A_D, X \rangle = \frac{1}{4}\tr\left(\sum_{e\in\binom{[n]}{2}}D_eb_eb_e^\top X\right)=\sum_{e\in\binom{[n]}{2}}D_eq_e(X) = \langle D, q(X)\rangle.\] Thus we define $X_D$ as the unique maximizer of the regularized objective in \eqref{eq:F_D}, namely, $X_D = \arg\max_{X \in \mathcal{X}} \left\{\langle A_D, X \rangle + \lambda \log\det X \right\}$.
Consider another signed vector $D'$ that is a perturbation of $D$, in the sense that $D' = D + \xi$ for some vector $\xi \in \mathbb{R}^{\binom{n}{2}}$.
(Later, in the privacy analysis, $\xi$ will be induced by a pair of neighboring graph weight vectors $w$ and $w'$.)
Consequently, $A_{D'}=A_{D}+\sum_e\xi_e b_eb_e^\top/4$. For simplicity, we denote $A_0:=A_{D}$ and $A_1:=A_{D'}$. By a similar but more delicate proof than \cite[Lemma 4.9]{EKKL20} and \cite[Lemma 44]{fan2026private}, we can bound the difference between the corresponding optimizers, which we denote by $X_0$ and $X_1$.
Specifically, we bound $\norm{X_0^{-1/2}(X_1-X_0)X_1^{-1/2}}_F$: while this type of ``relative Frobenius error'' may seem mysterious, it arises naturally when bounding the likelihood ratio between Gaussian distributions with covariances $X_0$ and $X_1$ (see, e.g., \Cref{lem:gaussian-privacy}).

\begin{lemma}\label{lem:stability_of_optimizer}
Let $A_1=A_0+\sum_e\xi_e b_eb_e^\top/4$, and let
\[
 X_i=\arg\max_{X\in\mathcal X}\{\ip{A_i}{X}+\lambda\log\det X\},
 \quad i\in\{0,1\}.
\]
Then
\[\norm{X_0^{-1/2}(X_1-X_0)X_1^{-1/2}}_F\leq \frac{\sum_e|\xi_e|\sqrt{q_e(X_0)q_e(X_1)}}{\lambda}.\]
Additionally, if $\norm{X_0^{-1/2}(X_1-X_0)X_1^{-1/2}}_F\leq1/2$, then for every $e$ it holds that
\[\frac{1}{2}q_e(X_1)\leq q_e(X_0)\leq 2q_e(X_1).\]
\end{lemma}

\begin{proof}
Since the function $\ip{A_i}{X}+\lambda\log\det X$ is strictly concave on the convex set
$\mathcal X$ for $\lambda > 0$, its maximizer is unique. 
Observe that $\nabla_X (\langle A_i, X \rangle + \lambda \log \det X) = A_i + \lambda X^{-1}$.
Therefore, by first-order optimality, we have that
\[
 \ip{A_0+\lambda X_0^{-1}}{X_1-X_0}\leq0,
 \qquad
 \ip{A_1+\lambda X_1^{-1}}{X_0-X_1}\leq0.
\]
Adding the two inequalities gives
\begin{equation}\label{eq:foo_ineq}
\lambda\ip{X_0^{-1}-X_1^{-1}}{X_1-X_0}\leq\ip{A_1-A_0}{X_1-X_0}.
\end{equation}
We start by rewriting the left-hand side of \eqref{eq:foo_ineq}.
Observe that $X_0^{-1} - X_1^{-1} = X_0^{-1}(X_1 - X_0)X_1^{-1}$.
Then we can write:
\begin{align*}
    \lambda\ip{X_0^{-1}-X_1^{-1}}{X_1-X_0} 
    &=\lambda\ip{X_0^{-1}(X_1 - X_0)X_1^{-1}}{X_1-X_0} \\
    &=\lambda\tr\left(X_0^{-1}(X_1-X_0)X_1^{-1}(X_1-X_0)\right)\\
    &=\lambda\tr\left(\left(X_0^{-1/2}(X_1-X_0)X_1^{-1/2}\right)\left(X_0^{-1/2}(X_1-X_0)X_1^{-1/2}\right)^\top\right)\\
    &=\lambda\norm{X_0^{-1/2}(X_1-X_0)X_1^{-1/2}}_F^2.
\end{align*}
The third equality uses the cyclic property of trace, and the fourth is since $\norm{B}_F^2 = \tr(BB^\top)$.

We next upper bound the right-hand side of \eqref{eq:foo_ineq}. Recall that $A_1-A_0=\sum_e\xi_e b_eb_e^\top/4$, thus
\begin{align*}
\ip{A_1-A_0}{X_1-X_0}
    &=\sum_e \frac{\xi_e}{4}
      \ip{b_eb_e^\top}{X_1-X_0}\\
    &=\sum_e \frac{\xi_e}{4}
      b_e^\top(X_1-X_0)b_e\\
    &=\sum_e \xi_e\bigl(q_e(X_1)-q_e(X_0)\bigr)\\
    &\le \sum_e |\xi_e|\,|q_e(X_1)-q_e(X_0)|.
\end{align*}
Putting this together, we now have 
\begin{equation}\label{eq:frob-sq-ineq}
\lambda\norm{X_0^{-1/2}(X_1-X_0)X_1^{-1/2}}_F^2\le \sum_e |\xi_e|\,|q_e(X_1)-q_e(X_0)|.
\end{equation}

It remains to bound the quantity $|q_e(X_1)-q_e(X_0)|$.
For each $e$, we let $u = X_0^{1/2}b_e/2$ and $v=X_1^{1/2}b_e/2$.  Then
\[
 q_e(X_1)-q_e(X_0)=\frac{1}{4}b_e^\top(X_1-X_0)b_e
 =u^\top X_0^{-1/2}(X_1-X_0) X_1^{-1/2}v.
\]
Since $\norm u_2=\sqrt{q_e(X_0)}$ and $\norm v_2=\sqrt{q_e(X_1)}$, by the Cauchy-Schwarz inequality, we have
\begin{equation}\label{eq:q-endpoint-difference}
 |q_e(X_1)-q_e(X_0)|\leq\sqrt{q_e(X_0)q_e(X_1)}\norm{X_0^{-1/2}(X_1-X_0) X_1^{-1/2}}_F.
\end{equation}
Combining 
\eqref{eq:frob-sq-ineq} and \eqref{eq:q-endpoint-difference} gives
\[\norm{X_0^{-1/2}(X_1-X_0)X_1^{-1/2}}_F\leq \frac{\sum_e|\xi_e|\sqrt{q_e(X_0)q_e(X_1)}}{\lambda}.\]
Additionally, if $\norm{X_0^{-1/2}(X_1-X_0) X_1^{-1/2}}_F\leq1/2$, by \eqref{eq:q-endpoint-difference}, we have,\[|q_e(X_1)-q_e(X_0)|\leq\frac{1}{2}\sqrt{q_e(X_0)q_e(X_1)}.\]
Dividing both sides by $q_e(X_1)$ and letting $y = q_e(X_0)/q_e(X_1)$, squaring the inequality yields $(y-1)^2 \leq y/4$. The roots of this quadratic lie within the interval $[1/2, 2]$. Thus we have $q_e(X_0)/q_e(X_1) \in [1/2, 2]$, which concludes the proof.
\end{proof}

The following lemma together with Lemma~\ref{lem:stability_of_optimizer} implies the privacy guarantee for the release of $(Z^{(1)}, \dots, Z^{(k)})$ in \Cref{alg:gradient_estimation}. We use a different analysis from \cite{EKKL20} to absorb the influence of $k$ samples.
A similar lemma appears as Theorem 5.1 of~\cite{alabi2023privately}, which also provides a DP-style guarantee for data sampled from Gaussians with similar covariance matrices.

\begin{lemma}\label{lem:gaussian-privacy}
Let $\del_1 \in(0,1/2)$ be a fixed parameter and $X_0,X_1\in\mathbb{R}^{n\times n}$ be symmetric positive definite matrices. Denote $E=X_0^{-1/2}(X_1-X_0)X_1^{-1/2}$. We use $\pdf_0$ and $\pdf_1$ to denote the probability density functions of $\mathcal{N}(0,X_0)$ and $\mathcal{N}(0,X_1)$ respectively. Suppose $\norm{E}_F<1/2$. Then for \[\eps_1=O\left(k\norm{E}_F^2+\sqrt{k\log\frac{1}{\del_1}}\norm{E}_F+\log\frac{1}{\del_1}\norm{E}_F\right),\] we have \[e^{-\eps_1}\prod_{j=1}^k\pdf_1(Z^{(j)})\leq\prod_{j=1}^k\pdf_0(Z^{(j)})\leq e^{\eps_1}\prod_{j=1}^k\pdf_1(Z^{(j)})\] with probability at least $1-\del_1$ over the joint distribution of the batch   $(Z^{(1)}, \dots, Z^{(k)})$, where each $Z^{(j)} \sim \mathcal{N}(0, X_0)$ is drawn independently.
\end{lemma}

The proof uses the following concentration bound for weighted chi-squared random variables.
The following signed version of the Laurent--Massart inequality follows from
Verzelen and Gassiat~\cite[Lemma~A.1]{verzelen2017adaptive}, applied to
$A=\operatorname{diag}(a_1,\ldots,a_n)$ and $-A$, together with a union bound.
\begin{lemma}[\cite{laurent2000adaptive}]\label{lem:concentration_chi-squares}
For a fixed parameter $t>0$, given independent standard Gaussian variables $\Sigma_i\sim\mathcal{N}(0,1)$ and weight $a_i$ for $i=1,\dots,n$, we have
\[
 \Prb\left[\left|\sum_{i=1}^n a_i(\Sigma_i^2-1)\right|
 >2\bigl(\norm a_2\sqrt t+\norm a_\infty t\bigr)\right]\leq2e^{-t}.
\]
\end{lemma}

\begin{proof}[Proof of Lemma~\ref{lem:gaussian-privacy}]
For a symmetric positive definite matrix $X$, the density function of $\mathcal{N}(0,X)$ is\[\pdf_X(Z) = \frac{1}{\sqrt{(2\pi)^n \det(X)}} \exp\left(-\frac{1}{2} Z^\top X^{-1} Z\right).\]

Denote $M=X_0^{-1/2}X_1X_0^{-1/2}$. Let the eigenvalues of $M$ be $e^{\theta_1},\dots,e^{\theta_n}$. Since $Z^{(j)}\sim \mathcal{N}(0,X_0)$, we denote $Z^{(j)}=X_0^{1/2}\Sigma^{(j)}$ where $\Sigma^{(j)}_i\sim\mathcal{N}(0,1)$. Then we have,

\begin{align}
    2\ln\frac{\prod_{j=1}^k\pdf_0(Z^{(j)})}{\prod_{j=1}^k\pdf_1(Z^{(j)})}&=k\ln\left(\frac{\det(X_1)}{\det(X_0)}\right)+\sum_{j=1}^k(Z^{(j)})^\top (X_1^{-1}-X_0^{-1}) Z^{(j)}\notag\\
    &=k\ln(\det(M))+\sum_{j=1}^k(\Sigma^{(j)})^\top (M^{-1}-I) \Sigma^{(j)}\notag\\
    &=k\sum_{i=1}^n{\theta_i}+\sum_{j=1}^k\sum_{i=1}^n(e^{-\theta_i}-1)(\Sigma^{(j)}_i)^2\notag\\
    &=k\sum_{i=1}^n({\theta_i}+e^{-\theta_i}-1)+\sum_{j=1}^k\sum_{i=1}^n(e^{-\theta_i}-1)((\Sigma^{(j)}_i)^2-1).\label{eq:logquotient}
\end{align}

By the inequality that $e^{x} + e^{-x}-2\geq x^2$ for all $x \in \mathbb{R}$, we have
\begin{align*}
    \norm{E}_F^2&=\tr\left(X_0^{-1}(X_1-X_0)X_1^{-1}(X_1-X_0)\right)\\
    &=\tr(X_0^{-1}X_1+X_1^{-1}X_0-2I) \\
    &= \sum_{i=1}^n (e^{\theta_i} + e^{-\theta_i} - 2)\geq \sum_{i=1}^n\theta_i^2.
\end{align*}
 Therefore, since $\norm{E}_F<1/2$, we have $|\theta_i|<1/2$ for every $i$. Since the inequality $0 \leq x + e^{-x} - 1 \leq x^2$ holds for all $|x|<1/2$, we can bound the first term in \Cref{eq:logquotient}:
 \begin{equation}\label{eq:part_A}
     0\leq k\sum_{i=1}^n({\theta_i}+e^{-\theta_i}-1)\leq k\sum_{i=1}^n\theta_i^2\leq k\norm{E}_F^2.
 \end{equation}

To bound the second term in \Cref{eq:logquotient}, we use Lemma~\ref{lem:concentration_chi-squares} and the inequality that $|e^{x} - 1| \leq 2|x|$ holds for all $|x|<1/2$ in combination with $\sum_{i=1}^n \theta_i^2 \leq \norm{E}_F^2$ which we established earlier.
\begin{align}
    \left|\sum_{j=1}^k\sum_{i=1}^n(e^{-\theta_i}-1)((\Sigma^{(j)}_i)^2-1)\right|&\leq2\sqrt{t\sum_{j=1}^k\sum_{i=1}^n(e^{-\theta_i}-1)^2}+2t\max_{i=1}^n|e^{-\theta_i}-1|\notag\\
    &\leq4\sqrt{tk\norm{E}_F^2}+4t\norm{E}_F,\label{eq:part_B}
\end{align}
with probability at least $1-2e^{-t}$. By choosing $t=\log(2/\del_1)$, combining \eqref{eq:logquotient}, \eqref{eq:part_A} and \eqref{eq:part_B} gives the desired conclusion.
\end{proof}

\subsection{Privacy via Success-Charged Composition}\label{sec:privacy-pwyg-comp}

In this section, we analyze the privacy cost of all calls to \Cref{alg:gradient_estimation}, performed as a subroutine within \Cref{alg:gradient_aggregation}. 
Naively, this procedure is invoked up to $O(n^2)$ times, and a crude calculation (using advanced composition) would incur privacy cost roughly $O(nK\varepsilon_1)$.
Instead, we will perform a more careful privacy accounting, where we only pay proportional to the mass $r$ retired in each call to \Cref{alg:gradient_estimation}, allowing us to bound the total privacy cost by something that resembles $O(K\varepsilon_1)$ -- entirely avoiding a dependence on $n$.
This privacy analysis strategy where we try to ``only charge for success'' appears in many places in the literature~\cite{gupta2010differentially,kaplan2021sparse}, including, e.g., pay-what-you-get composition~\cite{durfee2019practical}, target charging~\cite{cohen2023target}
and PREM~\cite{PREM25} which combines target charging with individual privacy accounting based on removing histogram coordinates. Our analysis extends this principle to fractional removal, using refined SDP stability bounds to relate each privacy charge to the fraction removed.
To make things self-contained and avoid unnecessary generality, we derive a bound for our particular situation from first principles.

We start with the following lemma, which analyzes a sequence of private mechanisms, each of which either outputs a result or a special failure symbol $\bot$. 
These private mechanisms and their privacy parameters may be chosen adaptively, i.e., based on the outputs of previous mechanisms. 
The key property is that we require that we have a worst-case privacy bound over all sequences of invocations, counting \emph{only successful} (i.e., non-$\bot$) invocations of the mechanisms.
This is in contrast to standard composition, when we would need such a bound over all sequences, counting both successes and failures. 
The resulting privacy bound is roughly equivalent to this bound that only counts successes, plus the worst-case cost of any particular call. 
For technical reasons, we consider an \emph{auxiliary} sequence of probability distributions $N$ rather than the actual distribution of outputs of the mechanisms $P$, which are close in total variation distance -- this is because \Cref{lem:gaussian-privacy} only holds with high probability, so we must account for this failure event.
The proof of the following lemma appears in Appendix~\ref{sec:pay-what-you-get}.

\begin{lemma}[Success-Charged Composition]
\label{lem:pay-what-you-get} 
Suppose that the following conditions hold for every pair of adjacent inputs $w$ and $w'$. Let $\mathcal{M}_1, \dots, \mathcal{M}_k$ be a sequence of adaptively chosen randomized mechanisms, where each mechanism outputs either a valid result or a designated failure symbol $\bottom$. 
For each step $t \in [k]$ and every common transcript prefix $y_{<t} = (y_1, \dots, y_{t-1})$, let $P_t^0$ and $P_t^1$ denote the conditional output distributions of $\mathcal{M}_t(w)$ and $\mathcal{M}_t(w')$, respectively. Suppose there exist auxiliary probability measures $N_t^0$ and $N_t^1$, a privacy loss parameter $\ell_t = \ell_t(y_{<t}) \ge 0$, and global constants $\ell^* < 1/10$ and $\delta' \ge 0$ such that, for any measurable set $A$ of return values and $b \in \{0, 1\}$:
\begin{itemize}
    \item If $P_t^b(A)=0$, then $N_t^b(A)=0;$
    \item $|P_t^b(A) - N_t^b(A)| \le \delta';$
    \item $e^{-\ell_t} N_t^1(A) \le N_t^0(A) \le e^{\ell_t} N_t^1(A),$
\end{itemize}
where $\ell_t(y_{<t}) \le \ell^*$.
Furthermore, suppose there exists a deterministic bound  $L$
such that for any sequence of outputs $y_1, \dots, y_k$ sampled sequentially according to
$y_t\sim N_t^b(\cdot\mid y_{<t}),$ for $t=1,\ldots,k$, 
the accumulated privacy loss over successful attempts is bounded by $L$:
\begin{equation*}
    \sum_{t=1}^k \ell_t(y_{<t}) \cdot \ind[y_t \ne \bottom] \le L.
\end{equation*}

Then, for any fixed $\delta'' \in (0,1)$, the adaptive composition $(\mathcal{M}_1, \dots, \mathcal{M}_k)$ is $(\eps, \del)$-differentially private, where 
\begin{equation*}
    \eps = 4L + 4\ell^*\log(1/\del''), \qquad \del = \del'' + (1+e^{\eps})k\del'.
\end{equation*}
\end{lemma}

Note that the bound $\ell_t$ may depend on the fixed input pair and the common
history, but is fixed before the next output is drawn. 

Given neighboring graphs $G$ and $G'$ with weight vectors $w$ and $w'$, such that $\norm{w-w'}_1\leq1,$ the overall privacy cost of each invocation of \Cref{alg:gradient_aggregation} is given below.

\begin{lemma}[Privacy of \Cref{alg:gradient_aggregation}]
\label{lem:gradient_privacy}
Fix any $\eps_{\mathrm{GE}}, \delta_{\mathrm{GE}}$ and $\beta_{\mathrm{GE}}\in(0,1)$, let $\eps_1=\frac{\eps_{\mathrm{GE}}}{320K(1+\log(2/\del_{\mathrm{GE}}))}, \del_1=\frac{\del_{\mathrm{GE}}}{20n^2K}, k=\Theta(\log\frac{n}{\beta_{\chi}\del_1}), \beta_{\chi}=\beta_{\mathrm{GE}}/4n^2K$ and $\lambda=\Theta(\frac{1}{\eps_1}\log\frac{n}{\beta_{\mathrm{GE}}\del_{\mathrm{GE}}}).$ Then each invocation of
\Cref{alg:gradient_aggregation} in \Cref{alg:main} with input $(w_i,h_i,\gamma)$ is
$(\eps_{\mathrm{GE}}, \delta_{\mathrm{GE}})$-differentially private.
\end{lemma}

\begin{proof}
Fix neighboring graphs $G$ and $G'$ with weight vectors $w$ and $w'$,
such that $\norm{w-w'}_1\leq1$.
At a common outer history in \Cref{alg:main}, write
$w_i=a_i\odotc w$ and $w_i'=a_i\odotc w'$.
Since each coordinate of $a_i$ is at most $1$, the vector
$\zeta:=w_i-w_i'$ satisfies $\norm{\zeta}_1\leq1$.

\par\smallskip\noindent\textit{Setup for privacy of one invocation of \Cref{alg:gradient_estimation}.}
We first consider a fixed invocation of \Cref{alg:gradient_estimation} after an arbitrary common history, so that the current candidate $h_i$, residual mask $r$, and choice of sign $\sigma\in\{\pm\}$ are fixed. Let $D$ and $D'$ be the signed objective vectors $D^\sigma$ evaluated at $w_i$ and $w_i'$, respectively. Recall that $D^+ = r \odot (h_i - (1+\gamma)w_i)$, $D^- = r \odot ((1-\gamma)w_i - h_i)$, and $\xi = D' - D$. Depending on the chosen sign $\sigma$, this objective discrepancy $\xi$ evaluates to:
\begin{itemize}
    \item For $\sigma = +$: $\xi = r \odot (h_i - (1+\gamma)w'_i) - r \odot (h_i - (1+\gamma)w_i) = (1+\gamma) r \odot \zeta$.
    \item For $\sigma = -$: $\xi = r \odot ((1-\gamma)w'_i - h_i) - r \odot ((1-\gamma)w_i - h_i) = -(1-\gamma) r \odot \zeta$.
\end{itemize}
Note that $1+\gamma < 2$ as $\gamma\in (0,1)$. This allows us to bound the magnitude of $\xi_e$ as:
\begin{equation}
\label{eq:xi-leq-r-zeta}
    |\xi_e| \leq (1+\gamma)|r_e \zeta_e| \leq 2|r_e \zeta_e|.
\end{equation}

Denote $A_0:=A_D$ and $A_1:=A_{D'}$. By the definition in \eqref{eq: A_D}, we can express the perturbed matrix as $A_1 = A_0 + \sum_e\xi_e b_eb_e^\top/4$. Since the residual mask satisfies $r_e \in [0, 1]$ for all $e$, and we have already established that the input discrepancy is bounded as $\norm{\zeta}_1 \leq 1$, it immediately follows that $\norm{\xi}_1 \leq 2 \norm{r \odot \zeta}_1 \leq 2\norm{\zeta}_1 \leq 2$.

Let
$X_i=\arg\max_{X\in\mathcal X}\{\ip{A_i}{X}+\lambda\log\det X\}$ for $i\in\{0,1\}.$
Let $E=X_0^{-1/2}(X_1-X_0)X_1^{-1/2}$.
Set $q_e^{\max}=\max(q_e(X_0),q_e(X_1))$,
$q_e^{\min}=\min(q_e(X_0),q_e(X_1))$.
By
Lemma~\ref{lem:stability_of_optimizer},
\begin{equation}\label{eq:direct-covariance-distance}
 \norm{E}_F=\norm{X_0^{-1/2}(X_1-X_0)X_1^{-1/2}}_F
 \leq\frac{\sum_e|\xi_e|\sqrt{q_e(X_0)q_e(X_1)}}{\lambda}
 \leq\frac{2\sum_e|\zeta_e|r_eq_e^{\max}}{\lambda}.
\end{equation}
Additionally,  since $q_e(X_0),q_e(X_1)\leq1$ and $\norm{\xi}_1\leq2$, by choosing $\lambda=\Theta(\frac{1}{\eps_1}\log\frac{n}{\beta_{\mathrm{GE}}\del_{\mathrm{GE}}})\geq8$,
\[\norm{E}_F\leq\frac{\sum_e|\xi_e|q_e^{\max}}{\lambda}\leq\frac{\norm{\xi}_1}{\lambda}<\frac{1}{2}.\]

We proceed to bound the privacy of a single trial within \Cref{alg:gradient_estimation}, followed by the privacy of all $K$ trials and thus the entire invocation of \Cref{alg:gradient_estimation}.

\par\smallskip\noindent\textit{Privacy of one trial.}
Denote $\pdf_0$ and $\pdf_1$ as the probability density functions of $\mathcal{N}(0,X_0)$ and $\mathcal{N}(0,X_1)$ respectively. Then by Lemma~\ref{lem:gaussian-privacy}, with probability at least $1-\del_1$ over the joint distribution of the batch   $Z=(Z^{(1)}, \dots, Z^{(k)})$, where each $Z^{(j)} \sim \mathcal{N}(0, X_0)$ is drawn independently, 
\begin{equation}\label{eq:direct-gaussian-loss}
-\eps'\leq\sum_{i=1}^k\ln\frac{\pdf_0(Z^{(i)})}{\pdf_1(Z^{(i)})}\leq \eps',
\end{equation}
where
\[\eps'=O\left(k\norm{E}_F^2+\sqrt{k\log\frac{1}{\del_1}}\norm{E}_F+\log\frac{1}{\del_1}\norm{E}_F\right)=O\left(\norm{E}_F\log\frac{n}{\beta_{\mathrm{GE}}\del_{\mathrm{GE}}}\right).\]
Combining with \eqref{eq:direct-covariance-distance} and the choice of $\lambda=\Theta(\frac{1}{\eps_1}\log\frac{n}{\beta_{\mathrm{GE}}\del_{\mathrm{GE}}})$ gives, 
\begin{equation}
    \label{eq:eps_prime}\eps'
=O\left(\frac{\sum_e|\zeta_e|r_eq_e^{\max}}{\lambda}\log\frac{n}{\beta_{\mathrm{GE}}\del_{\mathrm{GE}}}\right)
\leq\sum_e|\zeta_e|r_eq_e^{\max}\cdot\eps_1.
\end{equation}

By Lemma~\ref{lem:stability_of_optimizer} and $\norm{E}_F\leq1/2$, we have \[\frac{1}{2}q_e(X_1)\leq q_e(X_0)\leq 2q_e(X_1),\]
which further gives $q_e^{\max}\leq2q_e^{\min}$.

By Lemma~\ref{lem:chi} and the choice of $k$, with an additional failure probability at
most $\del_1$, the estimator $\hat{g}$ in the same attempt satisfies, for every $e$, \begin{equation}\label{eq:direct-constant-approximation}
 \frac{q_e^{\max}}8\leq\hat g(e)\leq2q_e^{\min}\leq2q_e^{\max}.
\end{equation}

The bounds derived in \Cref{eq:direct-gaussian-loss} and \Cref{eq:direct-constant-approximation} were with respect to Gaussian samples drawn from $\mathcal{N}(0, X_0)$, but observe the same derivation holds if $X_0$ and $X_1$ were swapped.
Let $\mathcal G$ be the common event that
\eqref{eq:direct-gaussian-loss} and
\eqref{eq:direct-constant-approximation} both hold.
It has probability at least $1-2\del_1$ whether samples are drawn from the Gaussian with covariance matrix $X_0$ or $X_1$.

An accepted trial (i.e., one that does not satisfy the if-statement in Line~\ref{line:accepted_trial} of Algorithm~\ref{alg:gradient_estimation}) in $\mathcal G$ returns
$g(e)=r(e)\hat g(e)/\hat g_{\max}$.
Since $\hat g_{\max}\leq1$, 
\begin{equation}
    \label{eq:qmax-leq-g}
    r(e) q_e^{\max}\leq8r(e)\hat{g}(e)=8\hat{g}_{\max}g(e)\leq8g(e).
\end{equation}

It remains to account for the noisy score
$Y=\langle D,\hat g\rangle+L$, where $L\sim\Lap(1/\eps_1)$.  For a fixed batch $Z\in\mathcal G$, by \eqref{eq:xi-leq-r-zeta} and \eqref{eq:direct-constant-approximation}, its two Laplace centers differ by
\[
 \left|\langle D-D',\hat g\rangle\right|
 \leq\sum_e|\xi_e|\hat g(e)
 \leq4\sum_e|\zeta_e|r_eq_e^{\max}.
\]

Thus, by the Laplace Mechanism (Lemma~\ref{lem:laplace-mech}) and \eqref{eq:eps_prime}, for every  $Z \in \mathcal{G}$, the joint densities $p_0,p_1$ of $(Z,Y)$ (i.e., not the densities conditioned on $\mathcal{G}$) obey
\begin{equation}\label{eq:loss_in_one_sample}
    \left|\log\frac{p_0(Z,Y)}{p_1(Z,Y)}\right|
 \leq\eps'+\eps_1\cdot4\sum_e|\zeta_e|r_eq_e^{\max}=5\sum_e|\zeta_e|r_eq_e^{\max}\cdot\eps_1.
\end{equation}

\par\smallskip\noindent\textit{Privacy of all $K$ trials.}
To handle all $K$ trials in one invocation of \Cref{alg:gradient_estimation}, equivalently draw their $K$
independent pairs $(Z,Y)$ in advance and return the first accepted
witness, or $\bot$ if none is accepted.
Let $\mathcal G^{[K]}$ be the event that $\mathcal G$ holds for every sampled $(Z, Y)$ pair.

Set $\ell=10K\sum_e|\zeta_e|r_eq_e^{\max}\cdot\eps_1$. By \eqref{eq:loss_in_one_sample}, for any set of samples when $\mathcal{G}^{[K]}$ holds, we have that the likelihood ratio (over the full sequence of $(Z,Y)$ pairs) lies in $[e^{-\ell/2}, e^{\ell/2}]$.
(Note that $\ell$ is fixed
before the next output is drawn.)

The preceding statement is for the unconditional densities. We examine the effect on the likelihood ratio when we condition on $\mathcal{G}^{[K]}$.
Indeed, using the definition of conditional probability on both densities, the likelihood ratio of any event under the conditional distribution can only be multiplicatively shifted by the likelihood ratio of the conditioning event itself.
Integrating the already-derived likelihood ratio bound over the event $\mathcal{G}^{[K]}$, we can place the likelihood ratio for $\mathcal{G}^{[K]}$ in $[e^{-\ell/2}, e^{\ell/2}]$, resulting in the likelihood ratio for the densities conditional on $\mathcal{G}^{[K]}$ falling in the range $[e^{-\ell}, e^{\ell}]$.

Let $P_b$ be the actual output distribution and $N_b$ be the distribution conditioned on $\mathcal{G}^{[K]}$ where $b = 0,1$ (corresponding to input $G$ and $G'$, respectively). 
Since these outputs follow the same deterministic post-processing of the samples, we thus have the same property as before for these distributions. 
Specifically, for any event $A$, we thus have that
\begin{equation}
 e^{-\ell}N_1(A)\leq N_0(A)\leq e^{\ell}N_1(A).
\end{equation}

Furthermore, note that the probability of $\mathcal{G}^{[K]}$ is at least $1-2K\del_1$ under input graph $G$ or $G'$. 
Thus, we have that 
\begin{equation}
 |P_b(A)-N_b(A)|\leq2K\del_1\quad(b=0,1).
\end{equation}

Finally, note that $P_b(A) = 0$ implies that $N_b(A) = 0$, since conditioning the underlying samples can not give non-zero probability to a previously impossible event. 
We have now verified the three distributional conditions required by \Cref{lem:pay-what-you-get}.

\par\smallskip\noindent\textit{Privacy of \Cref{alg:gradient_aggregation}.}
We now compose all invocations of \Cref{alg:gradient_estimation} in \Cref{alg:gradient_aggregation}.
Denote the $t$-th output by $g^{(t)}$.
At each history before call $t$, take $N_t^b$ and $\ell_t$ to be the
conditional law $N_b$ and bound $\ell$ constructed above. $r_e^{(t)}$ and $(q_e^{\max})^{(t)}$ are defined similarly.

The update rule in \Cref{alg:gradient_aggregation} gives
$\sum_{t:g^{(t)}\ne\bot}g^{(t)}\leq\one_{\binom n2}$ on every transcript, including those
of the auxiliary process. 
Thus, by \eqref{eq:qmax-leq-g},
\[
 \sum_{t:g^{(t)}\ne\bot}\ell_t
 \leq10K\eps_1\sum_e|\zeta_e|\sum_{t:g^{(t)}\ne\bot}r_e^{(t)}(q^{\max}_e)^{(t)}
 \leq80K\eps_1\sum_e|\zeta_e|\sum_{t:g^{(t)}\ne\bot}g^{(t)}(e)
 \leq 80K\eps_1\norm{\zeta}_1
 \leq 80K\eps_1.
\]
Thus, the total cost of successful calls is at most $80K\varepsilon_1$ along every execution of the auxiliary process. 

Since $\norm{\zeta}_1\leq1$ and $r_e,q_e^{\max}\leq1$, each local charge satisfies $\ell_t\leq10K\eps_1$.
Apply \Cref{lem:pay-what-you-get} with $k=n^2$ (padding shorter executions with zero-cost dummy calls)$, \del'=2K\del_1,
 L=80K\eps_1$, $\ell^*=10K\eps_1$, and $\del''=\del_{\mathrm{GE}}/2$. Then
\begin{align*}
 \ell^*&=\frac{\eps_{\mathrm{GE}}}{32(1+\log(2/\del_{\mathrm{GE}}))}<\frac1{32}<\frac1{10},\\
 4L+4\ell^*\log(1/\del'')
 &=320K\eps_1+40K\eps_1\log(2/\del_{\mathrm{GE}})
 \leq\eps_{\mathrm{GE}}.
\end{align*}
Since $\eps_{\mathrm{GE}}<1$ and $\del_1=\frac{\del_{\mathrm{GE}}}{20n^2K}$, the additive privacy parameter is at most
\[
 \del''+(1+e^{\eps_{\mathrm{GE}}})k\del'
 \leq\frac{\del_{\mathrm{GE}}}{2}
 +(1+e)\,2n^2K\del_1
 <\del_{\mathrm{GE}}.
\]
Therefore, each invocation of
\Cref{alg:gradient_aggregation} in \Cref{alg:main} is
$(\eps_{\mathrm{GE}}, \delta_{\mathrm{GE}})$-differentially private.

\end{proof}

Finally, we apply advanced adaptive composition (Lemma~\ref{lem:advanced-composition}) to account for the total privacy cost of our core mechanism Algorithm~\ref{alg:main}.

\begin{lemma}[Privacy of \Cref{alg:main}]\label{lem:main-privacy}
For $0<\varepsilon\leq1$ and $0<\delta<1$, fix $\eps_2=\frac{\eps}{2T_1}$, $\eps_{\mathrm{GE}}=\frac{\eps}{4T_1\sqrt{2T_2\log(2T_1/\delta)}}$, $\delta_{\mathrm{GE}}=\frac{\delta}{2T_1T_2}.$
Then \Cref{alg:main} is
$(\eps, \del)$-differentially private.
\end{lemma}

\begin{proof}
Fix neighboring graphs $G$ and $G'$ with weight vectors $w$ and $w'$, such that $\norm{w-w'}_1\leq1.$ Fix a common transcript before outer round $i$. The mask $a_i$ is then fixed, and
$w_i=a_i\odotc w$, $w_i'=a_i\odotc w'$ satisfy
$\norm{w_i-w_i'}_1\leq\norm{w-w'}_1\leq1$.
Consequently $|M_i(w)-M_i(w')|\leq1$, so the release
$M_i+\Lap(1/\eps_2)+\tau_2$ is $\eps_2$-DP by Laplace mechanism (\Cref{lem:laplace-mech}).

Conditioned on each subsequent common transcript, the candidate $h_i$ is fixed and 
\Cref{lem:gradient_privacy} gives $(\eps_{\mathrm{GE}},\delta_{\mathrm{GE}})$-DP for the
next call to AggregateWitnesses. There are at most $T_2$ such calls in this round. Apply advanced composition theorem (\Cref{lem:advanced-composition}). Since $\eps_{\mathrm{GE}}\leq1$,
$e^{\eps_{\mathrm{GE}}}-1\leq2\eps_{\mathrm{GE}}$, and hence their composed privacy
parameter satisfies
\begin{align*}
 \eps_{\mathrm{GE}}\sqrt{2T_2\log(2T_1/\delta)}
 +T_2\eps_{\mathrm{GE}}(e^{\eps_{\mathrm{GE}}}-1)
 &\leq \frac{\eps}{4T_1}
 +\frac{\eps^2}{16T_1^2\log(2T_1/\delta)}
 \leq\frac{\eps}{2T_1},\\
 T_2\delta_{\mathrm{GE}}+\delta/2T_1
 &=\frac{\delta}{T_1}.
\end{align*}
The last inequality uses $\eps\leq1$, $T_1\geq1$, and $\log(2T_1/\delta)\geq\log2$.
Including the mass release, one outer round is therefore
$(\eps/T_1,\delta/T_1)$-DP. Basic adaptive composition (Lemma~\ref{lem:basic-composition}) over at most $T_1$ outer rounds
gives $(\eps,\delta)$-DP. The candidate and mask updates and the released synthetic graph
are postprocessing of this transcript.
\end{proof}

\section{Utility Analysis}\label{sec:utility}
We first establish the utility guarantees of the fractional-witness oracle, showing that it either identifies a valid over- or underestimation direction or certifies that the current estimate is accurate on the remaining mask (\Cref{sec:gradient_estimation}).  We then use these guarantees to analyze the signed multiplicative-weights updates and the geometric progress of the outer fractional decomposition, culminating in the utility and running-time guarantees of \Cref{alg:main} and finalizing the proof of our main theorem (\Cref{sec:mwu}).

\subsection{The Witness Aggregation Oracle}\label{sec:gradient_estimation}
For any cut $S \subseteq V$, the Boolean cut query $c_S\in\{0,1\}^{\binom{n}{2}}$ is defined as:
\[
 c_S(\{u,v\})=\ind[\{u,v\}\in(S,V\setminus S)],
\] hence the cut value of $S$ on $w$ equals $\cut_w(S) = \langle w, c_S \rangle$. Given a vector $D\in\mathbb{R}^{\binom{n}{2}}$, the regularized separation objective $F(D)$ is defined in \eqref{eq:F_D}. We first bound the gap between the regularized $F(D)$ and the true maximum cut value with respect to $D$. The proof is similar to Lemma 4.3 in \cite{EKKL20}.
\begin{lemma}\label{lem:sdp_error}
For every signed vector $D \in \mathbb{R}^{\binom{n}{2}}$ and every cut $S \subseteq [n]$,
\[\langle D, c_S\rangle\leq 2\max\{F(D),0\}+O(\lambda n\log n+\norm{D}_1/n).\]
Moreover, $\langle D,\one\rangle\leq2F(D)$.
\end{lemma}

\begin{proof}
For every cut $S$, let $s\in\{\pm1\}^V$ encode $S$, i.e., $s_u=1$ for all $u\in S$ and $s_u=-1$ for all $u\in V\setminus S$. Let $X_S=(1-\frac{1}{n})ss^\top+\frac{1}{n} I_n$. Since $X_S$ is a symmetric matrix with unit
diagonal and $X_S\succeq\frac{1}{n}I_n$, we have $X_S\in\mathcal{X}$. Therefore, \[
F(D)\geq\sum_{e=\{u,v\} \in \binom{[n]}{2}} D_{uv} q_{e}(X_S) + \lambda \log\det X_S.
\]
Note that for any $e = \{u,v\}$, $q_{e}(X_S)=\frac{1-(X_S)_{uv}}{2}=(1-\frac{1}{n})c_S(\{u,v\})+\frac{1}{2n}$. Additionally, the eigenvalue of $X_S$ in direction $s$ is $(1-\frac{1}{n})n+\frac{1}{n}$, and its other $n-1$ eigenvalues are $\frac{1}{n}$.
Consequently, $\det X_S = \left(\left(1 - \frac1n\right)n + \frac1n\right)\frac{1}{n^{n-1}}$ and $0 \leq - \log \det X_S = (n-1) \log n - \log \left(n-1 +\frac1n\right) = O(n \log n)$. Therefore,
\[
 F(D)\geq \left(1-\frac{1}{n}\right)\langle D,c_S\rangle+\frac{1}{2n}\sum_{\{u,v\}\in\binom{[n]}{2}}D_{uv}+\lambda\log\det X_S.
\]
Using
$(1-1/n)^{-1}\leq2$ we have $\langle D, c_S\rangle\leq 2\max\{F(D),0\}+O(\lambda n\log n+\norm{D}_1/n).$ Finally, $I_n\in\mathcal X$, $q_e(I_n)=1/2$, and $\log\det I_n=0$, so $F(D)\geq\langle D,\one\rangle/2$.
\end{proof}

In the inner iteration of \Cref{alg:main}, we recursively refine the partial synthetic estimate $h_i$ of the weights at the current iteration $w_i$. Given the output $(g^+,g^-,r)$ of $\mathsf{AggregateWitnesses}(w_i,h_i,\gamma)$, the lemma proved below shows that, when $\langle h_i,r\rangle\geq \langle h_i,g^+\rangle$ and $\langle h_i,r\rangle \geq \langle h_i,g^-\rangle$, the partial estimate $h_i$ already serves as a good approximation of $w_i$ under the mask $r$.

\begin{lemma}\label{lem:gradient_estimation}
Let $0<\gamma<1$, $K=\Theta(\log(n/\beta_{\mathrm{GE}}))$, $\tau_1=\log(1/\beta_1)/\eps_1$ and $\beta_1=\beta_{\chi}=\beta_{\mathrm{GE}}/4n^2K$. Given nonnegative edge weights $w$ and a nonnegative partial estimate $h$, suppose $w_e\leq U$ for every $e$. Let $(g^+,g^-,r^*)=\mathsf{AggregateWitnesses}(w,h,\gamma)$ (\Cref{alg:gradient_aggregation}).
Then with probability at least $1-\beta_{\mathrm{GE}}$:
\begin{itemize}
    \item $\langle h,g^+\rangle\geq(1+\gamma)\langle w,g^+\rangle;$
    \item $\langle h,g^-\rangle\leq(1-\gamma)\langle w,g^-\rangle;$
    \item For every cut $S$ simultaneously, \[(1-\gamma)\langle w,c_S\odotc r^*\rangle-B\leq \langle h,c_S\odotc r^*\rangle\leq(1+\gamma)\langle w,c_S\odotc r^*\rangle+B,\]
where $B=\Theta\left(\tau_1+\lambda n\log n+nU\right).$ 
\end{itemize}
\end{lemma}

\begin{proof}
Observe that whenever $g_{\mathrm{new}} \neq \bot$,
at least one coordinate of $r$ becomes 0, because
we have $g_{\mathrm{new}}(e) = r(e) > 0$ for at least one edge $e$.
Hence \Cref{alg:gradient_aggregation} makes at most $\binom{n}{2}$ successful calls to \Cref{alg:gradient_estimation}.
Accounting for failures, there are at most $2\cdot\binom{n}{2} \leq n^2$ total calls to \Cref{alg:gradient_estimation}.
Therefore, by the tail bound of Laplace distribution, Lemma~\ref{lem:chi} and union bound, $|L|\leq\tau_1$ and $q_{uv}(X_D)/4\leq \hat{g}(u,v)\leq q_{uv}(X_D)$ hold simultaneously across all calls of \Cref{alg:gradient_estimation} with probability at least $1-n^2K(\beta_1+\beta_{\chi})$. Denote this good event by $\mathcal E$.

On $\mathcal E$, for each call of $g_{\mathrm{new}}=\mathsf{FindWitness}(D^+,r)$ in \Cref{alg:gradient_aggregation}, if $g_{\mathrm{new}}\ne\bot$, then in at least one of the $K$ iterates, $Y=\langle D^+,\hat{g}\rangle+L\geq\tau_1$. Thus $\langle D^+,\hat{g}\rangle\geq\tau_1-\tau_1\geq0$. Since $D^+=r\odotc(h-(1+\gamma)w)$ and $\hat g_{\max} > 0$, this gives \[0\le\frac{1}{\hat g_{\max}}\langle r\odotc(h-(1+\gamma)w), \hat{g}\rangle=\left\langle h-(1+\gamma)w, r\odotc\frac{\hat{g}}{\hat g_{\max}}\right\rangle=\langle h,g_{\mathrm{new}}\rangle-(1+\gamma)\langle w,g_{\mathrm{new}}\rangle.\]
Summing over all successful calls of $\mathsf{FindWitness}(D^+,r)$ gives $\langle h,g^+\rangle\geq(1+\gamma)\langle w,g^+\rangle.$
By a symmetric argument, $\langle h,g^-\rangle\leq(1-\gamma)\langle w,g^-\rangle$ also holds.

To bound the probability of an unsuccessful search, condition only on the history before the call $\mathsf{FindWitness}(D,r)$, with either sign of the error vector. Its inputs $D,r$ are then fixed.
Denote $X_D$ as the SDP optimizer of \eqref{eq:F_D} for $D$.
Due to the condition $\diag(X)=\one_n$ for all $X \in \mathcal{X}$, the sum of $X_D$'s eigenvalues is $n$ and AM-GM implies their product is $\leq 1$, and therefore $\log\det X_D\leq0$.
We then have
\[\sum_{e=\{u,v\} \in \binom{[n]}{2}} D_{uv} q_{e}(X_D)\geq F(D).\]
Suppose $F(D)\geq4\tau_1$.
By Lemmas~\ref{lem:chi} and~\ref{lem:chaos}, with probability at least $c_0(1-\beta_\chi)$ in each trial,
\[\langle D,\hat{g}\rangle\geq\frac{1}{2}\sum_{e=\{u,v\} \in \binom{[n]}{2}} D_{uv} q_{e}(X_D) \geq \frac12 F(D) \geq2\tau_1.\]
Independently, $\Prb[L\geq-\tau_1]\geq1/2$, so with probability at least $c_0(1-\beta_\chi)/2\geq c_0/4$, both $\langle D,\hat{g}\rangle\geq2\tau_1$ and $L\geq-\tau_1$ hold.
On this event, $Y\geq\tau_1$ and $\hat g_{\max}>0$, since $\langle D,\hat{g}\rangle>0$ and $D$ is supported on the active coordinates of $r$. Thus the trial returns a nonzero witness.
Since the trials use fresh independent randomness, the probability that this call returns $\bot$ is at most $(1-c_0/4)^K$.
This bound holds conditional on every pre-call history with $F(D)\geq4\tau_1$. A union bound over at most $n^2$ adaptive calls therefore shows that, with probability at least $1-n^2(1-c_0/4)^K$, every call returning $\bot$ has $F(D)<4\tau_1$.
Let's call this event $\mathcal{E}'$.

For the remainder of the proof, work on the intersection $\mathcal{E}' \cap \mathcal E$.
If $r^* = \mathbf{0}$, then the third clause is immediate.
Therefore, assume that $r^* \neq \mathbf{0}$.
In the last iteration of \Cref{alg:gradient_aggregation}, the output of $\mathsf{FindWitness}(D^+,r^*)$ and $\mathsf{FindWitness}(D^-,r^*)$ are both $\bot$, where
\[D^+=r^*\odotc(h-(1+\gamma)w),\qquad D^-=r^*\odotc((1-\gamma)w-h).\]
Consequently, $F(D^+)<4\tau_1$ and $F(D^-)<4\tau_1$.

Since $I_n\in\mathcal X$, $q_e(I_n)=1/2$, and $\log\det I_n=0$, the bound on $F(D^+)$ gives
\[
 \frac12 \big(\langle h,r^*\rangle-(1+\gamma)\langle w,r^*\rangle\big)
 = \langle D^+, q(I_n)\rangle + \lambda \log\det I_n
 \leq F(D^+)<4\tau_1.
\]
Thus, for $\sigma\in\{+,-\}$, nonnegativity and $\norm{w}_1\leq n^2U$ imply
\begin{align*}
 \norm{D^\sigma}_1
 &\leq\langle h,r^*\rangle+(1+\gamma)\langle w,r^*\rangle\\*
 &\leq  8\tau_1 + 2(1+\gamma)\norm{w}_1
 =O(\tau_1 + n^2U).
\end{align*}
Combining with Lemma~\ref{lem:sdp_error} and absorbing $\tau_1/n$ into $\tau_1$ gives,
for every cut $S$ and $\sigma\in\{+,-\}$,
\[
 \langle D^\sigma,c_S\rangle\leq O(\tau_1+\lambda n\log n+nU)=B.
\]
Rearranging this gives \[(1-\gamma)\langle w,c_S\odotc r^*\rangle-B\leq \langle h,c_S\odotc r^*\rangle\leq(1+\gamma)\langle w,c_S\odotc r^*\rangle+B,\]
with probability at least
\[
\Prb(\mathcal{E} \cap \mathcal{E}') \geq
1-n^2(1-c_0/4)^K-n^2K(\beta_1+\beta_{\chi}).
\]
By setting $K=\Theta(\log(n/\beta_{\mathrm{GE}}))$ with a sufficiently large constant and $\beta_1=\beta_{\chi}=\beta_{\mathrm{GE}}/4n^2K$, this happens with probability at least $1-\beta_{\mathrm{GE}}$.
\end{proof}

\subsection{Multiplicative-Weight Updates}\label{sec:mwu}
The following lemma computes the number of inner iterations required by \Cref{alg:main}, using a potential function argument.

\begin{lemma}\label{lem:inner_round}
Let $0 < \gamma \leq 1/4$, $\eta=\gamma/[8(1+\gamma)]$, $\tau_2=\log(1/\beta_2)/\eps_2$,
$B=O(\tau_1+\lambda n\log n+nU)$ as in \Cref{alg:main}, and
$M_{\mathrm{stop}}=C(nU\log\frac{nU}{\delta\beta\gamma\eps}+\frac{n}{\gamma\varepsilon}(\log\frac{nU}{\delta\beta\gamma\eps})^6)\geq10(B+\tau_2/\gamma)$, for a sufficiently large universal constant $C$.
For any fixed outer iteration $i$, with probability at least $1-T_2\beta_{\mathrm{GE}}-\beta_2$, if $\widetilde M_i>M_{\rm stop}$, then within at most $T_2=1+\left\lceil64(1+\gamma)^2\log n/\gamma^2\right\rceil=O(\log n/\gamma^2)$ inner iterations, $\mathsf{AggregateWitnesses}(w_i,h_i,\gamma)$ returns $(g^+_i,g^-_i,r_i)$ satisfying
\[
\langle h_i,r_i\rangle\geq\max\{\langle h_i,g^+_i\rangle,\langle h_i,g^-_i\rangle\}.
\]
\end{lemma}

\begin{proof}
We condition on the conclusion of Lemma~\ref{lem:gradient_estimation} and $M_i\leq\widetilde{M}_i\leq M_i+2\tau_2$ holding simultaneously across all $T_2$ calls of \Cref{alg:gradient_aggregation}. By union bound, this joint event occurs with probability at least $1 - T_2\beta_{\mathrm{GE}}-\beta_2$. Since $\widetilde M_i>M_{\mathrm{stop}}\geq10(B+\tau_2/\gamma)$ and $\gamma\leq1/4$,
\begin{equation}\label{eq:continuing-mass-bounds}
 M_i\geq\widetilde M_i-2\tau_2
 >10B+\left(\frac{10}{\gamma}-2\right)\tau_2
 \geq10B+\frac{8\tau_2}{\gamma}.
\end{equation}
Consequently,
\[
 1\leq\frac{\widetilde M_i}{M_i}
 \leq1+\frac{2\tau_2}{M_i}
 \leq1+\frac{\gamma}{4}.
\]
Denote $h_i^{(j)}$ as the estimate at the $j$-th inner iteration in \Cref{alg:main}. We consider \[\Phi(j):= M_i \cdot D_{\mathrm{KL}}\left(\frac{w_i}{M_i}\middle\|\frac{h_i^{(j)}}{\tilde M_i}\right)=\sum_e w_i(e)\log\frac{w_i(e)}{h_i^{(j)}(e)} + M_i \log \frac{\tilde M_i}{M_i}\] as a potential function, measuring discrepancy between the true weights $w_i$ for iteration $i$  and our candidate synthetic weights $h_i^{(j)}$. 
The remainder of the proof will quantify how $\Phi(j)$ decreases as $j$ increases.

The initialization is $h_i^{(1)}(e)=\widetilde M_i \frac{a_i(e)}{\sum_{e}a_i(e)}$.
For every coordinate with $w_i(e)>0$, using $w(e)\leq U$, $a_i(e)\leq1$,
$B=\Theta\left(\tau_1+\lambda n\log n+nU\right)\geq nU$, and \eqref{eq:continuing-mass-bounds}, we obtain
\[
 \frac{w_i(e)/M_i}{h_i^{(1)}(e)/\widetilde M_i}
 =\frac{a_i(e)w(e)/M_i}{a_i(e)/\sum_{e}a_i(e)}
 =\frac{w(e)\sum_{e'}a_i(e')}{M_i}
 \leq\frac{n^2U}{10B}\leq n.
\]
It follows that
\[\Phi(1)=\sum_e w_i(e)\log\frac{w_i(e)/M_i}{h_i^{(1)}(e)/\widetilde{M}_i}\leq \sum_e w_i(e)\log n=M_i\log n.\]

In the $j$-th inner iteration, denote the output of \Cref{alg:gradient_aggregation} as $(g^+_i,g^-_i,r_i)$. Suppose that $\langle h_i^{(j)},g^+_i\rangle\geq  \max\{\langle h_i^{(j)},g^-_i\rangle,\langle h_i^{(j)},r_i\rangle\}$, we have $\langle h_i^{(j)},g^+_i\rangle\geq\norm{h_i^{(j)}}_1/3=\widetilde{M}_i/3$. By Lemma~\ref{lem:gradient_estimation}, we also have $\langle h_i^{(j)},g^+_i\rangle\geq(1+\gamma)\langle w_i,g^+_i\rangle.$ Using further that $\tilde M_i/M_i \leq 1 + \gamma/4$,
\[\left\langle \frac{h_i^{(j)}}{\widetilde{M}_i}- \frac{w_i}{M_i},g^+_i\right\rangle\geq \left(1-\frac{\tilde M_i/M_i}{1+\gamma}\right)\left\langle \frac{h_i^{(j)}}{\widetilde{M}_i},g^+_i\right\rangle\geq\left(1-\frac{1+\gamma/4}{1+\gamma}\right)\cdot \frac13=\frac{\gamma}{4(1+\gamma)}.\]

By a symmetric argument, suppose that $\langle h_i^{(j)},g^-_i\rangle\geq \max\{\langle h_i^{(j)},g^+_i\rangle,\langle h_i^{(j)},r_i\rangle\}$, we have \[\left\langle \frac{w_i}{M_i}-\frac{h_i^{(j)}}{\widetilde{M}_i},g^-_i\right\rangle\geq\left(\frac{\tilde M_i/M_i}{1-\gamma}-1\right)\left\langle \frac{h_i^{(j)}}{\widetilde{M}_i},g^-_i\right\rangle\geq \left(\frac{1}{1-\gamma}-1\right)\frac13\geq\gamma/3.\]

Therefore, for either chosen pair $(\theta, g) \in \{(1, g^+_i), (-1, g^-_i)\}$, we have
\begin{equation}\label{eq:diff_in_j_iterate}
    \theta\left\langle \frac{h_i^{(j)}}{\widetilde{M}_i} - \frac{w_i}{M_i}, g \right\rangle\geq\frac{\gamma}{4(1+\gamma)}.
\end{equation}

The remaining argument is similar to \cite{fan2026private}. Denote $W_j=\sum_{e}h_i^{(j)}(e)\exp(-\theta\eta g(e)).$ By the update rule in \Cref{alg:main}, 
\begin{align*}
    \Phi(j)-\Phi(j+1)&=\sum_ew_i(e)\log\frac{h_i^{(j+1)}(e)}{h_i^{(j)}(e)}\\
    &=\sum_ew_i(e)\log\frac{\widetilde M_i\cdot h_i^{(j)}(e)\exp(-\theta\eta g(e))/W_j}{h_i^{(j)}(e)}\\
    &=-\theta\eta \sum_ew_i(e)g(e)+M_i\log\frac{\widetilde{M}_i}{W_j}.
\end{align*}

Since $e^{x} \leq 1 + x + x^2$ holds for every $x\leq1/4$ and $\eta g(e)\leq\gamma/[8(1+\gamma)]\leq\frac{1}{4}$, together with $\log(1+y) \leq y$, we have,
\begin{align*}
    \log\frac{W_j}{\widetilde{M}_i}
    &=\log \left( \sum_{e} \frac{h_i^{(j)}(e)}{\widetilde{M}_i} \exp(-\theta\eta g(e))\right)\\
    &\leq\log \left( \sum_{e} \frac{h_i^{(j)}(e)}{\widetilde{M}_i} (1-\theta\eta g(e)+\eta^2g^2(e))\right)\\
    &\leq\sum_{e} \frac{h_i^{(j)}(e)}{\widetilde{M}_i} (-\theta\eta g(e)+\eta^2 g^2(e))\\
    &\leq-\frac{1}{\widetilde{M}_i}\sum_{e}\theta\eta h_i^{(j)}(e)g(e)+\eta^2,
\end{align*}
where the last inequality comes from $g^2(e)\leq1$. Therefore, by $\eta=\gamma/[8(1+\gamma)]$ and \eqref{eq:diff_in_j_iterate}, 
\begin{align*}
    \Phi(j) - \Phi(j+1) &\geq \eta M_i \theta \left \langle \frac{h_i^{(j)}}{\widetilde{M}_i} - \frac{w_i}{M_i}, g \right\rangle - \eta^2M_i\\
    &\geq\frac{\eta M_i\gamma}{4(1+\gamma)}-\eta^2M_i=\frac{\gamma^2 M_i}{64(1+\gamma)^2}.
\end{align*}
Since KL divergence is nonnegative, combining with $\Phi(1)\leq M_i\log n$, there can be at most
$64(1+\gamma)^2\log n/\gamma^2=O(\log n/\gamma^2)$ update steps in each fixed outer iteration in \Cref{alg:main}.

\end{proof}

The next lemma bounds the number of outer iterations required by \Cref{alg:main}.

\begin{lemma}\label{lem:outer_round}
With probability at least $1-T_1T_2\beta_{\mathrm{GE}}-T_1\beta_2$, \Cref{alg:main} terminates with $a_i=0$ or $\widetilde{M}_i\leq M_{\mathsf{stop}}$ in at most $T_1 = O(\log(nU))$ outer rounds.
\end{lemma}

\begin{proof}
Let $\mathcal E$ be the event that every released mass estimate satisfies $M_i\leq\widetilde M_i\leq M_i+2\tau_2$, including in the stopping round, and every invocation of $\mathsf{AggregateWitnesses}$ satisfies the conclusions of Lemma~\ref{lem:gradient_estimation}.
Applying these guarantees conditional on the preceding history and taking a union bound over at most $T_1$ mass releases and $T_1T_2$ oracle calls gives $\Prb[\mathcal E]\geq1-T_1\beta_2-T_1T_2\beta_{\mathrm{GE}}$.
Assume that $\mathcal E$ holds. By the argument of Lemma~\ref{lem:inner_round}, in each continuing outer round, $\mathsf{AggregateWitnesses}(w_i,h_i,\gamma)$ returns $(g^+_i,g^-_i,r_i)$ satisfying $\langle h_i,r_i\rangle\geq\max\{\langle h_i,g^+_i\rangle,\langle h_i,g^-_i\rangle\}$ within at most $T_2=O(\log n/\gamma^2)$ inner iterations.
Fix such an outer round $i$ and observe that $\langle h_i,r_i\rangle\geq\norm{h_i}_1/3=\widetilde{M}_i/3$.

To compare total masses, average the upper cut inequality in
\Cref{lem:gradient_estimation} over a uniformly random subset $S\subseteq V$.
Each edge crosses this cut with probability $1/2$, so
\[
 \frac12\langle h_i,r_i\rangle
 \leq\frac{1+\gamma}{2}\langle w_i,r_i\rangle+B.
\]
This averaging uses the simultaneous all-cut guarantee and incurs no additional failure
probability. Since $\widetilde M_i\geq M_i$ and
$M_i\geq10B$ by \eqref{eq:continuing-mass-bounds},
\[
 \langle w_i,r_i\rangle
 \geq\frac{\widetilde M_i/3-2B}{1+\gamma}
 \geq\frac{(1/3-1/5)M_i}{1+\gamma}
 \geq\frac{M_i}{20},
\]
where $\gamma\leq1/4$.

By the partial estimator collection in \Cref{alg:main}, we have \[w_{i+1}=a_{i+1}\odotc w=a_i\odotc(\one_{\binom{n}{2}}-r_i)\odotc w=w_i\odotc(\one_{\binom{n}{2}}-r_i).\] 
Thus $M_{i+1}=M_i-\langle w_i,r_i\rangle\leq19M_i/20.$ 
Since $M_1=O(n^2U)$, and every continuing outer round satisfies $M_i > M_{\mathrm{stop}} - 2\tau_2$, at most $T_1=O(\log(n^2U))=O(\log(nU))$ outer rounds can occur.
\end{proof}

Next we prove the utility guarantee of \Cref{alg:main}.
This essentially assembles the per-invocation guarantees of Lemma~\ref{lem:gradient_estimation} across the entire algorithm.

\begin{lemma}\label{lem:utility}
Let $n\geq2$, $U\geq1$, $0<\gamma\leq1/4$, $0<\eps\leq1$, and $0<\delta,\beta<1/10$.
Choose all parameters as in \Cref{alg:main}, in particular
$\eps_{\mathrm{GE}}=\eps/(4T_1\sqrt{2T_2\log(2T_1/\delta)})$,
$\delta_{\mathrm{GE}}=\delta/(2T_1T_2)$, and $M_{\mathrm{stop}}=C(nU\log\frac{nU}{\delta\beta\gamma\eps}+\frac{n}{\gamma\varepsilon}(\log\frac{nU}{\delta\beta\gamma\eps})^6)
\geq10(B+\tau_2/\gamma)$, for a sufficiently large universal constant $C$.
Given a nonnegative weighted $n$-vertex graph $G$ with edge weights bounded by $U$, \Cref{alg:main} outputs a
nonnegative weighted graph $\widetilde{G}$ satisfying, 
with probability at least $1-\beta$, simultaneously
for every $S\subseteq V$,
\begin{align*}
 \left|w_G(S,V\setminus S)-w_{\widetilde{G}}(S,V\setminus S)\right|
 \leq \gamma w_G(S,V\setminus S)  + O\left(nU\log\frac{nU}{\delta\beta\gamma\eps}+\frac{n}{\gamma\varepsilon}
 \left(\log\frac{nU}{\delta\beta\gamma\eps}\right)^7\right).
\end{align*}
\end{lemma}
\begin{proof}
Assume that the good event $\mathcal E$ defined in the proof of Lemma~\ref{lem:outer_round} holds. This event occurs with probability at least $1 - T_1T_2\beta_{\mathrm{GE}}-T_1\beta_2$. By setting $\beta_2=\beta/2T_1$ and $\beta_{\mathrm{GE}}=\beta/2T_1T_2$, it occurs with probability at least $1-\beta$.

By Lemma~\ref{lem:gradient_estimation}, for every outer iteration with released partial estimate $h_i$ and mask $r_i$, and every cut $S$, \[(1-\gamma)\langle w_i,c_S\odotc r_i\rangle-B\leq \langle h_i,c_S\odotc r_i\rangle\leq(1+\gamma)\langle w_i,c_S\odotc r_i\rangle+B.\]
This is equivalent to,
\begin{equation}\label{eq:one round utility}
    (1-\gamma)\langle w_i\odotc r_i,c_S\rangle-B\leq \langle h_i\odotc r_i,c_S\rangle\leq(1+\gamma)\langle w_i\odotc r_i,c_S\rangle+B.
\end{equation}

Assume after $T\leq T_1$ outer rounds, $\widetilde{M}_{T+1}\leq M_{\mathrm{stop}}$. 
Since $a_{i+1}=a_i\odotc(\one_{\binom{n}{2}}-r_i)$ and $a_1=\one_{\binom{n}{2}}$, we have   \[\sum_{i=1}^{T}w_i\odotc r_i=\sum_{i=1}^{T}(a_i\odotc r_i)\odotc w=(a_1-a_{T+1})\odotc w=w-w_{T+1},\]where $\norm{w_{T+1}}_1\leq M_{\mathrm{stop}}$. Combining with the fact $\widetilde{w}=\sum_{i=1}^T h_i\odotc r_i$, summing up \eqref{eq:one round utility} over all the $T\leq T_1$ rounds,
\[(1-\gamma)\langle w-w_{T+1},c_S\rangle-T_1B\leq \langle \widetilde{w},c_S\rangle\leq(1+\gamma)\langle w-w_{T+1},c_S\rangle+T_1B.\]
Since $\norm{w_{T+1}}_1\leq M_{\mathrm{stop}}$, we can conclude that \[\left|w_G(S,V\setminus S)-w_{\widetilde{G}}(S,V\setminus S)\right|
 \leq \gamma w_G(S,V\setminus S)  + O(T_1B+M_{\mathrm{stop}}).\]
By Lemma~\ref{lem:gradient_privacy} and the choices of $\eps_{\mathrm{GE}}, \del_{\mathrm{GE}}, \beta_{\mathrm{GE}}, T_1, T_2$, we have
\[
\lambda=\Theta\left(\frac{1}{\eps_{1}}\log\frac{n}{\beta_{\mathrm{GE}}\del_{\mathrm{GE}}}\right)=O\left(\frac{4KT_1\sqrt{T_2}}{\eps}\log^{\frac{5}{2}}\frac{2nT_1^2T_2^2}{\beta\del}\right)\leq
 \frac{\left(\log(nU/(\delta\beta\gamma\eps))\right)^{C_1}}{\gamma\varepsilon}
\]
for some universal constant $C_1$.

More explicitly, put $\mathcal L=\log(nU/(\delta\beta\gamma\eps))$. Then $T_1,K=O(\mathcal L)$, $T_2=O(\mathcal L/\gamma^2)$, and every logarithm in the oracle parameters is $O(\mathcal L)$. Substitution gives
\[
 \eps_{\mathrm{GE}}^{-1}=O\left(\frac{\mathcal L^2}{\gamma\eps}\right),
 \quad \eps_1^{-1}=O\left(\frac{\mathcal L^4}{\gamma\eps}\right),
 \quad \lambda,\tau_1=O\left(\frac{\mathcal L^5}{\gamma\eps}\right),
 \quad \tau_2=O\left(\frac{\mathcal L^2}{\eps}\right).
\]
Consequently, $B=O(nU+n\mathcal L^6/(\gamma\eps))$ and $T_1B=O(nU\mathcal L+n\mathcal L^7/(\gamma\eps))$. These bounds imply that the public threshold $M_{\mathrm{stop}}=C(nU\mathcal L+n\mathcal L^6/(\gamma\eps))$ satisfies $M_{\mathrm{stop}}\geq10(B+\tau_2/\gamma)$ for a sufficiently large fixed $C$, uniformly over the stated parameter range.
Thus 
plugging $\lambda$ and $\tau_1,\tau_2$ defined in \Cref{alg:main}, \Cref{alg:gradient_aggregation}, we have $T_1B+M_{\mathrm{stop}}\leq C'(nU\mathcal L+n\mathcal L^7/(\gamma\varepsilon))$ for a universal constant $C'$ larger than the stopping-threshold constant $C$.
\end{proof}

Now combining Lemma~\ref{lem:preprocess}, Lemma~\ref{lem:main-privacy} and Lemma~\ref{lem:utility}, we can finally give the proof of \cref{thm:main}.

\begin{proof}[Proof of \cref{thm:main}]
Given an $n$-vertex input graph $G$, we first invoke Algorithm~\ref{alg:preprocess} with $\eps_0=\eps/2,\beta_0=\beta/2$ and obtain a preprocessed graph $G'$ with edge weight $w'$ bounded by $U=\frac{8}{\eps_0\gamma}\log\frac{n}{\beta_0}$. By Lemma~\ref{lem:preprocess}, with probability at least $1-\beta/2$, for every cut $S$,
\[
 |Y(S,V\setminus S)+w'(S,V\setminus S)-w(S,V\setminus S)|\leq\frac{\gamma}{2}w(S,V\setminus S).
\]

Then we run \Cref{alg:main} on $w'$ with parameters $\gamma/2,\eps/2, \del,\beta/2$ and obtain an estimate $\widetilde{G}'$. By Lemma~\ref{lem:main-privacy}, \Cref{alg:main} is $(\eps/2, \del)$-DP.

By Basic Composition (Lemma~\ref{lem:basic-composition}), the total algorithm is $(\eps, \del)$-DP. By Lemma~\ref{lem:utility} and $U=\frac{8}{\eps_0\gamma}\log\frac{n}{\beta_0}$ defined in \Cref{sec:preprocess}, 
with probability at least $1-\beta/2$, for every cut $S$, 
\begin{align*}
    \left|w'(S,V\setminus S)-w_{\widetilde{G}'}(S,V\setminus S)\right|
 &\leq \frac{\gamma}{2} w'(S,V\setminus S)  + O\left(nU\log\frac{nU}{\delta\beta\gamma\eps}+\frac{n}{\gamma\varepsilon}
 \left(\log\frac{nU}{\delta\beta\gamma\eps}\right)^7\right)\\
 &\leq\frac{\gamma}{2} w'(S,V\setminus S)  + 
 O\left(\frac{n}{\gamma\varepsilon}
 \left(\log(n/(\delta\beta\gamma\eps))\right)^{7}\right).
\end{align*}

Finally, we release $\widetilde{G}=\widetilde{G}'+Y$. With probability at least $1-\beta$, for every cut $S$, 
\begin{align*}
    \left|w_{\widetilde{G}}(S,V\setminus S)-w(S,V\setminus S)\right|&\leq\left|w_{\widetilde{G}'}(S,V\setminus S)-w'(S,V\setminus S)\right|+\left|Y(S,V\setminus S)+w'(S,V\setminus S)-w(S,V\setminus S)\right|\\
    &\leq\frac{\gamma}{2} w'(S,V\setminus S)  + O\left(\frac{n}{\gamma\varepsilon}
 \left(\log(n/(\delta\beta\gamma\eps))\right)^{7}\right)+\frac{\gamma}{2}w(S,V\setminus S)\\
 &\leq\gamma w(S,V\setminus S)+O\left(\frac{n}{\gamma\varepsilon}
 \left(\log(n/(\delta\beta\gamma\eps))\right)^{7}\right).
\end{align*}
The last inequality comes from the fact $w'\leq w$.

We now show that \Cref{alg:main} has a polynomial-time implementation preserving the stated privacy and utility guarantees.\footnote{With additional work, one obtains $\widetilde O_{\eps,\del,\gamma,\beta}(n^8)$ time using the cutting-plane algorithm of Lee, Sidford, and Wong~\cite{lee2015faster}, as in~\cite[Appendix~A]{EKKL20}. After the clipping described below, the expected cost per optimization is uniformly $\widetilde O_{\eps,\del,\gamma,\beta}(n^6)$. Independent restarts with public time limits add only a logarithmic factor and bound the running time on every execution. If a call exhausts its allotted restarts, use $I_n$ as the covariance matrix and continue; reserve half of the total-variation error budget below for the event that this occurs in any call and halve the covariance tolerance $\mu$ accordingly.}

For this implementation, immediately after generating each noisy mass estimate $\widetilde M_i$ in \Cref{alg:main}, we cap it at the public bound $\binom n2 U+2\tau_2$. This is post-processing of each private mass release, so the privacy analysis is unchanged. On the utility good event, the uncapped estimate is at most $M_i+2\tau_2\leq\binom n2 U+2\tau_2$, so clipping does not change the execution. On every execution, it ensures $\norm{h_i}_1\leq\binom n2 U+2\tau_2$ and hence $\norm{D^\pm}_1\leq\norm{h_i}_1+(1+\gamma)\norm{w_i}_1=O(n^2U+\tau_2)$ for every oracle call.

By \Cref{alg:main} and Lemma~\ref{lem:utility}, we have $T_1 = O(\log(nU))$, $T_2=O(\log n/\gamma^2)$. In each call to \Cref{alg:gradient_aggregation}, note that it calls \Cref{alg:gradient_estimation} at most $n^2$ times. Thus there are at most $n^2T_1T_2=\widetilde{O}_{\eps,\del,\gamma,\beta}(n^2)$ SDP calls.

Let $X^*$ be the maximizer of $\langle D,q(X)\rangle+\lambda\log\det X$ over $\mathcal X$. In the affine space of symmetric matrices with unit diagonal, of dimension $\binom n2$, the feasible region contains the Frobenius ball of radius $1/2$ centered at $I_n$ and is contained in the ball of radius $n$ centered there. Separation and objective-gradient evaluation reduce to eigenvalue computation and matrix inversion, respectively, and take polynomial time. The objective range is at most $\norm{D}_1+\lambda n\log n$. The standard deterministic ellipsoid method therefore computes a feasible approximate maximizer in time polynomial in $n$, $\log(nU+\tau_2+\lambda)$, and the logarithm of the inverse objective precision. For $0<\mu\leq1/2$, we require
\[
 \norm{(X^*)^{-1/2}(X^*-X)(X^*)^{-1/2}}_F\leq\mu.
\]
Indeed, all feasible eigenvalues lie in $[1/n,n]$, so the negative objective is $\lambda/n^2$-strongly convex in Frobenius norm; a feasible solution with objective gap at most $\lambda\mu^2/(2n^4)$ suffices for this covariance bound.

To account for this approximation, let $R=(X^*)^{-1/2}X(X^*)^{-1/2}$ and write its eigenvalues as $1+a_i$. Taking expectations of the Gaussian log-density ratio, as computed in the proof of \Cref{lem:gaussian-privacy}, gives
\begin{align*}
 &\KL(\mathcal N(0,X)^{\otimes k}\|\mathcal N(0,X^*)^{\otimes k})\\
 &\qquad=\frac{k}{2}\bigl(\tr R-n-\log\det R\bigr)
 =\frac{k}{2}\sum_{i=1}^n\bigl(a_i-\log(1+a_i)\bigr)\\
 &\qquad\leq\frac{k}{2}\sum_{i=1}^n a_i^2
 =\frac{k}{2}\norm{R-I}_F^2\leq\frac{k\mu^2}{2}.
\end{align*}
Here $|a_i|\leq\norm{R-I}_F\leq\mu\leq1/2$, and we used $t-\log(1+t)\leq t^2$ for $|t|\leq1/2$.
Pinsker's inequality bounds the total variation distance of each Gaussian batch by $O(\sqrt{k}\mu)$. Conditional on the batch and a common history, validation and all subsequent updates use the same rules. There are at most $N=n^2KT_1T_2$ batches on every execution. Adaptive coupling therefore bounds the total variation distance between the two output laws by $cN\sqrt{k}\mu$, for a universal constant $c$.
For a bounded call with target parameters $(\eps,\delta,\beta)$, choose its internal parameters using $(\eps,\delta/2,\beta/2)$ and take
\[
 \mu=\min\left\{\frac12,\frac{\nu}{cN\sqrt{k}}\right\},
 \qquad
 \nu=\min\left\{\frac{\beta}{2},\frac{\delta}{2(1+e^\eps)}\right\}.
\]
Total variation distance at most $\nu$ increases the utility failure probability by at most $\nu$ and the additive privacy parameter by at most $(1+e^\eps)\nu$. Thus the approximate mechanism has the required $(\eps,\delta)$ privacy and failure probability at most $\beta$; the constant-factor changes to $\delta,\beta$ are absorbed in the logarithmic error bound. This precision has $\log(1/\mu)=O(\log(nU/(\delta\beta\gamma\eps)))$, and the parameter bounds above imply $\log(nU+\tau_2+\lambda)=O(\log(nU/(\delta\beta\gamma\eps)))$. Each optimization therefore takes polynomial time, and there are at most $n^2T_1T_2$ such calls, proving the polynomial-time claim. This completes the proof of Theorem~\ref{thm:main}.
\end{proof}

\section{AI Disclosure}
The first proofs of our main privacy theorems were developed independently by multiple groups using AI systems, including an initially more complex proof using GPT-5.5 and later simpler proofs using GPT-5.6 and Claude 4.6 in early July 2026.
The current proof is chosen among  different algorithms and different proof strategies. We have subsequently also used GPT-6 Astra and Claude Opus 5.5 to refine the proofs, polish the exposition, and generate the Lean formalization. 
The authors have verified, significantly simplified, and rewritten the algorithms and proofs to improve readability and presentation. 
In particular, the authors have restructured the algorithm, provided intuition and explanations, and abstracted the use of the success-charged composition (\Cref{lem:pay-what-you-get}), which originates from works including \cite{gupta2010differentially,kaplan2021sparse,durfee2019practical,cohen2023target}, in the privacy analysis. The authors take full responsibility for the correctness of all results and the final presentation of this paper.

\bibliographystyle{alpha}
\bibliography{references}

\newpage
\appendix
\section{Success-Charged Composition}\label{sec:pay-what-you-get}
In this section, we prove the composition lemma that only charges for successes (Lemma~\ref{lem:pay-what-you-get}). This lemma ensures that, in an adaptive sequence of private mechanisms, we only need to charge the privacy loss when the subroutine returns a successful output, while incurring only a small privacy loss when it fails. 
Importantly, at each step we incur only a fraction of the privacy loss, determined by the amount of information revealed by the output.

\begin{proof}[Proof of Lemma~\ref{lem:pay-what-you-get}]
Fix adjacent inputs $w,w'$. Write $P_{<t}^b$ for the law of the history preceding step $t$ under input $b$, and define
\[
 \mu_{t-1}=P_{<t}^0+P_{<t}^1,
 \qquad p_{t-1}^b=\frac{dP_{<t}^b}{d\mu_{t-1}}.
\]
Common histories are those with $p_{t-1}^0p_{t-1}^1>0$. On these histories set $Q_t^b=N_t^b$; on $\{p_{t-1}^b>0,\ p_{t-1}^{1-b}=0\}$ set $Q_t^0=Q_t^1=P_t^b$; elsewhere set both kernels to the point mass at $\bot$. Extend $\ell_t$ by zero outside common histories.

These definitions give, for every history $h$,
\begin{equation}
 e^{-\ell_t(h)}Q_t^1(\cdot\mid h)
 \leq Q_t^0(\cdot\mid h)
 \leq e^{\ell_t(h)}Q_t^1(\cdot\mid h),
 \label{eq:pwg-kernel-ratio}
\end{equation}
and, for $P_{<t}^b$-almost every history $h$,
\begin{equation}
    d_{\mathrm{TV}}
 \bigl(P_t^b(\cdot\mid h),Q_t^b(\cdot\mid h)\bigr)
 \leq\delta'.
 \label{eq:kenerl-tv-distance}
\end{equation}

Let $P^b$ and $ Q^b$ denote the full distributions induced by the conditional distributions $ P_t^b$ and $Q_t^b$, respectively. 
By triangle inequality and \eqref{eq:kenerl-tv-distance} we have,
\begin{align*}
 d_{\mathrm{TV}}
 ( P_{\leq t}^b, Q_{\leq t}^b)
 &\leq
 d_{\mathrm{TV}}( P_{<t}^b, Q_{<t}^b)+
 \int d_{\mathrm{TV}}
 \bigl(P_t^b(\cdot\mid h),Q_t^b(\cdot\mid h)\bigr)
 \, P_{<t}^b(dh)\\
 &\leq
 d_{\mathrm{TV}}( P_{<t}^b, Q_{<t}^b)+\delta'.
\end{align*}
Induction over all $k$ time steps yields
\begin{equation}
 d_{\mathrm{TV}}( P^b, Q^b)\leq k\delta'.
 \label{eq:pwg-tv}
\end{equation}

If $\ell^*=0$,~\eqref{eq:pwg-kernel-ratio} gives
$ Q^0= Q^1$, so~\eqref{eq:pwg-tv} yields
$ P^0(A)\leq P^1(A)+2k\del'$
and its reverse, which imply the stated guarantee.
Assume henceforth that $\ell^*>0$.

Along a transcript, define
\[
 C_t=\sum_{j=1}^t
 \ell_j(y_{<j})\mathbf 1\{y_j\neq\bot\},
 \qquad
 Z_t=\log
 \frac{dQ_t^0(\cdot\mid y_{<t})}
      {dQ_t^1(\cdot\mid y_{<t})}(y_t).
\]

$C_t$ represents the cumulative privacy charge of successful calls up to a specific time step $t$.
$Z_t$ is the log-likelihood ratio at output $y_t$, conditioned on the history $y_{<t}$, under the auxiliary distributions $Q_t^0$ and $Q_t^1$.

Under $ Q^0$, we define
\[
 W_t=
 \exp\left(
 \frac{1}{4\ell^*}\left(\sum_{j=1}^tZ_j-4C_t\right)
 \right),
 \qquad W_0=1.
\]
$W_t$ serves as a measure of how much the accumulated privacy loss exceeds the cumulative privacy charge that includes \emph{only successful calls}. 
Our proof will demonstrate that this sequence is a supermartingale, and thus will not grow too large with high probability. 

Observe that $W_t$ can be written as 
\[ W_t = W_{t-1}\exp\left(\frac{Z_t - 4\ell_t \mathbf 1\{y_t\neq\bot\}}{4\ell^*}\right).\]

Thus, to prove the supermartingale property $\mathbb E_{Q^0}[W_t | y_{<t}]\leq W_{t-1},$
it suffices to prove:
\begin{equation}\label{eq:one-step-moment}
    \mathbb E_{Q^0}\left[\exp\left(\frac{1}{4\ell^*}(Z_t-4\ell_t\mathbf 1\{y_t\neq\bot\})\right)\mid y_{<t}\right]\leq1.
\end{equation}

Conditioned on $y_{<t}$, by \eqref{eq:pwg-kernel-ratio}, $Z_t\leq\ell_t(y_{<t})$. For simplicity, write $\ell_t:=\ell_t(y_{<t}).$ For $y_t\ne\bot$, by the inequality $e^{-x}\leq1-\frac{x}{2}$ which holds for $0\leq x\leq1$,
\[\exp\left(\frac{1}{4\ell^*}(Z_t-4\ell_t\mathbf 1\{y_t\neq\bot\})\right)\leq \exp\left(-\frac{3\ell_t}{4\ell^*}\right)\leq1-\frac{3\ell_t}{8\ell^*}.\]

If $Q_t^0(\bot)=0$, then $y_t\ne\bot$ almost surely under $Q^0$, so the preceding bound proves \eqref{eq:one-step-moment}. Henceforth assume $Q_t^0(\bot)>0$. By \eqref{eq:pwg-kernel-ratio}, $Q_t^1(\bot)>0$ as well, and we define $F_t:=\frac{Q_t^1(\bot)}{Q_t^0(\bot)}$.

For $y_t=\bot$,
\[\exp\left(\frac{1}{4\ell^*}(Z_t-4\ell_t\mathbf 1\{y_t\neq\bot\})\right)=\exp\left(\frac{Z_t}{4\ell^*}\right)=\left(\frac{1}{F_t}\right)^{1/{4\ell^*}}.\]

If $Q_t^0(\bot)\leq Q_t^1(\bot)$, $F_t\geq1$. Since $\frac{1}{4\ell^*}\geq1$,
\begin{align*}
    &\mathbb E_{Q^0}\left[\exp\left(\frac{1}{4\ell^*}(Z_t-4\ell_t\mathbf 1\{y_t\neq\bot\})\right)\mid y_{<t}\right]\\&\leq \left(1-Q^0_t(\bot)\right)\cdot \left(1 - \frac{3\ell_t}{8\ell^*}\right)+Q^0_t(\bot)\cdot\left(\frac{1}{F_t}\right)^{1/{4\ell^*}}\leq1.
\end{align*}

The more interesting case is when $Q_t^0(\bot)> Q_t^1(\bot)$.
By \eqref{eq:pwg-kernel-ratio}, 
\[1-Q^1_t(\bot)\leq e^{\ell_t}(1-Q^0_t(\bot)).\]
Rearranging, 
\begin{equation*}
Q^0_t(\bot)-Q^1_t(\bot)\leq(e^{\ell_t}-1)(1-Q^0_t(\bot)).
\end{equation*}
Since $\frac{1}{4\ell^*}\geq1$ and the inequality $1-x^{\theta}\leq\theta(1-x)$ which holds for $\theta\geq1, 0\leq x\leq1$, 
\begin{equation*}
    1-F_t^{1/{4\ell^*}}\leq\frac{1}{4\ell^*}\left(1-F_t\right).
\end{equation*}
Additionally, by \eqref{eq:pwg-kernel-ratio}, $Q^0_t(\bot)\leq e^{\ell_t}Q^1_t(\bot)$, thus $1/F_t\leq e^{\ell_t}.$
Therefore,
\begin{align*}
    &\mathbb E_{Q^0}\left[\exp\left(\frac{1}{4\ell^*}(Z_t-4\ell_t\mathbf 1\{y_t\neq\bot\})\right)\mid y_{<t}\right]\\
    &\leq\left(1-Q^0_t(\bot)\right)\left(1-\frac{3\ell_t}{8\ell^*}\right)+Q^0_t(\bot)\left(\frac{1}{F_t}\right)^{1/{4\ell^*}}\\
    &=\left(1-Q^0_t(\bot)\right)\left(1-\frac{3\ell_t}{8\ell^*}\right)+Q^0_t(\bot)+Q^0_t(\bot)\left(\frac{1}{F_t}\right)^{1/{4\ell^*}}\left(1-F_t^{1/{4\ell^*}}\right)\\
    &\leq\left(1-Q^0_t(\bot)\right)\left(1-\frac{3\ell_t}{8\ell^*}\right)+Q^0_t(\bot)+\frac{e^{\ell_t/4\ell^*}}{4\ell^*}\left(Q_t^0(\bot)-Q^1_t(\bot)\right)\\
    &\leq\left(1-Q^0_t(\bot)\right)\left(1-\frac{3\ell_t}{8\ell^*}\right)+Q^0_t(\bot)+\frac{e^{\ell_t/4\ell^*}}{4\ell^*}\left(e^{\ell_t}-1\right)\left(1-Q^0_t(\bot)\right)\\
    &\leq1.
\end{align*}
The last inequality comes from \[\frac{e^{\ell_t/4\ell^*}}{4\ell^*}(e^{\ell_t}-1)\leq\frac{e^{\ell_t/4\ell^*}}{4\ell^*}\ell_te^{\ell_t}\leq\frac{3\ell_t}{8\ell^*},\]
since $e^{\ell_t/4\ell^*+\ell_t}\leq e^{0.35}\leq3/2.$
Therefore, the claim \eqref{eq:one-step-moment} is proved, and thus,
\[\mathbb E_{ Q^0}
 [W_t\mid y_{<t}]
 \leq W_{t-1}.\]
By $W_0=1$ and induction this further gives 
\begin{equation}\label{eq:expectation-w_k}
    \mathbb E_{ Q^0}[W_k]\leq1.
\end{equation}
By the chain rule for Radon--Nikodym derivatives, the log-likelihood ratio of the entire transcript is
\[
 \Lambda:=
 \log\frac{dQ^0}{dQ^1}(y_{\leq k})
 =\sum_{t=1}^kZ_t.
\]
By $\eps=4L + 4\ell^*\log(1/\del'')$, 
on the event $\{C_k\leq L\}\cap\{\Lambda>\varepsilon\}$,
\[
 W_k=\exp\left(\frac{1}{4\ell^*}(\Lambda-4C_k)\right)
 >\exp\left(\frac{1}{4\ell^*}(\varepsilon-4L)\right)=1/\del''.
\]

The first and third assumptions of \Cref{lem:pay-what-you-get} give absolute continuity $N_t^b\ll P_t^0,P_t^1$ on common histories. Induction therefore gives $Q_{<t}^b\ll P_{<t}^0,P_{<t}^1$, so the auxiliary process stays on common histories almost surely. The assumed bound on successful-call charges thus gives $Q^b(C_k\leq L)=1$.
Consequently, Markov's inequality and \eqref{eq:expectation-w_k} imply
\begin{equation}
  Q^0
 \bigl(\{\Lambda>\varepsilon\}\bigr)
 = Q^0
 \bigl(\{C_k\leq L\}\cap\{\Lambda>\varepsilon\}\bigr)
 \leq Q^0(W_k\geq1/\del'')
 \leq\frac{\mathbb E_{ Q^0}[W_k]}{1/\del''}\leq\del''.
 \label{eq:pwg-tail}
\end{equation}

Since $ Q^0\leq e^\varepsilon  Q^1$ on
$\{\Lambda\leq\varepsilon\}$, for any event $A$, we have
\begin{align*}
  P^0(A)
 &\leq Q^0(A)+k\del'\\
 &\leq Q^0(A\cap\{\Lambda\leq\varepsilon\})+Q^0(\{\Lambda>\varepsilon\})+k\del'\\
 &\leq
 e^\varepsilon Q^1(A)+\del''+k\del'\\
 &\leq
 e^\varepsilon P^1(A)
 +\del''+(1+e^\varepsilon)k\del'.
\end{align*}
Here the first and fourth lines use~\eqref{eq:pwg-tv}, and the third~
uses~\eqref{eq:pwg-tail} and the definition of $\Lambda$. Exchanging the two inputs gives the reverse
inequality with the same parameters. Since the adjacent pair was
arbitrary, the full mechanism is
$(\varepsilon,\delta)$-differentially private.
\end{proof}

\section{A Gaussian Fractional Witness}\label{sec:Gaussian_approximate}
We first recall the following facts on the concentration of Chi-squared distribution and some auxiliary tools that will be used in our analysis.
\begin{lemma}[Chi-squared Tail Bound~\cite{laurent2000adaptive}]\label{lem:chi_square_tail}
Let $Y_1, \dots, Y_k$ be independent standard normal random variables, i.e., $Y_i \sim \mathcal{N}(0, 1)$. The random variable $Q = \sum_{i=1}^k Y_i^2$ follows a Chi-squared distribution with $k$ degrees of freedom, denoted as $Q \sim \chi_k^2$. For any such $Q$, there exists a universal constant $c > 0$ such that
\[
 \Prb\left[\frac{Q}{k} \notin \left[\frac{1}{2}, 2\right]\right] \leq 2e^{-ck}.
\]
\end{lemma}

A \emph{degree-two polynomial in independent standard Gaussian random variables} (often referred to as a Gaussian chaos of degree two) is a random variable $W$ that can be expressed in the form $W = Z^\top A Z + b^\top Z + c$, where $Z = (Z_1, \dots, Z_m)$ is a vector of independent standard random variables $\mathcal{N}(0, 1)$, $A$ is a real symmetric matrix, $b$ is a real vector, and $c$ is a real scalar. 

\begin{lemma}[Gaussian Hypercontractivity for Degree-Two Polynomials~\cite{Janson97}]\label{lem:hypercontractivity}
Let $W$ be a degree-two polynomial in independent standard Gaussian random variables. Then its higher moments can be bounded by its second moment. Specifically, for the $4$-th moment, we have:
\[
    \|W\|_4 \le 3\|W\|_2,
\]
where $\|W\|_p = (\E[|W|^p])^{1/p}$ denotes the $L_p$ norm of the random variable.
\end{lemma}

\begin{lemma}[Log-Convexity of $L_p$ Norms]\label{lem:holder_interpolation}
For any random variable $W$ and $1 \le p < r < q \le \infty$, the $L_p$ norms satisfy the interpolation inequality $\|W\|_r \le \|W\|_p^\theta \|W\|_q^{1-\theta}$, where $\theta \in (0, 1)$ is determined by $\frac{1}{r} = \frac{\theta}{p} + \frac{1-\theta}{q}$. 
In particular, choosing $p=1, r=2$, and $q=4$ yields $\theta = 1/3$, which gives:
\[
    \|W\|_2 \le \|W\|_1^{1/3} \|W\|_4^{2/3}.
\]
\end{lemma}

Before giving the proofs of Lemma~\ref{lem:chi} and Lemma~\ref{lem:chaos}, we first note that, for any $e = \{u,v\}$, $\overline g(u,v)$ is indeed an unbiased estimation of $q_e(X)$:

\begin{proposition}[Unbiased Estimator of Fractional Query]\label{prop:unbiased_estimator}
Let $X \in \mathcal{X}$ be a valid covariance matrix and let $Z \sim \mathcal{N}(0, X)$. For any pair of vertices $e = \{u,v\}$, we have
\[
    \E\left[\frac{(Z_u - Z_v)^2}{4}\right] = q_e(X).
\]
Consequently, for $\overline{g}(u,v) = \frac{1}{k} \sum_{j=1}^k \frac{(Z^{(j)}_u - Z^{(j)}_v)^2}{4}$ where $Z^{(j)}$ are drawn independently from $\mathcal{N}(0, X)$, it holds that $\E[\overline{g}(u,v)] = q_e(X)$.
\end{proposition}

\begin{proof}
Let $b_e = \mathbf{e}_u - \mathbf{e}_v$. The difference $Z_u - Z_v$ can be expressed as the inner product $b_e^\top Z$. Since $\E[Z Z^\top] = X$, we have:
\[
    \E[(Z_u - Z_v)^2] = \E[(b_e^\top Z)(Z^\top b_e)] = b_e^\top \E[Z Z^\top] b_e = b_e^\top X b_e.
\]
Recall from \cref{eq:qe} that $q_e(X) = \frac{1}{4} b_e^\top X b_e$, which yields $\E\left[\frac{(Z_u - Z_v)^2}{4}\right] = q_e(X)$. The second statement directly follows from the linearity of expectation over the $k$ independent samples.
\end{proof}

Now, we are ready to present the proofs for Lemma~\ref{lem:chi} and Lemma~\ref{lem:chaos}.
\begin{proof}[Proof of Lemma~\ref{lem:chi}]
For any fixed edge $e = \{u, v\}$, let $b_e = \mathbf{e}_u - \mathbf{e}_v$. Since $Z^{(j)} \sim \mathcal{N}(0, X)$, the difference $Z^{(j)}_u - Z^{(j)}_v = b_e^\top Z^{(j)}$ is a Gaussian random variable with mean $0$ and variance $b_e^\top X b_e = 4q_e(X)$.
Since $X \succeq (1/n)I_n$ implies $q_e(X) \geq 1/(2n) > 0$, we can define the normalized random variable $Y_j := \frac{Z^{(j)}_u - Z^{(j)}_v}{2\sqrt{q_e(X)}}$ such that $Y_j \sim \mathcal{N}(0, 1)$. 

By the definition in \Cref{alg:gradient_estimation}, $\overline{g}(u,v) = \frac{1}{k} \sum_{j=1}^k \frac{(Z^{(j)}_u - Z^{(j)}_v)^2}{4}$. We can rewrite the ratio of our estimate to the true value as:
\[
    \frac{\overline{g}(u,v)}{q_e(X)} = \frac{1}{k}\sum_{j=1}^k \left( \frac{Z^{(j)}_u - Z^{(j)}_v}{2\sqrt{q_e(X)}} \right)^2 = \frac{1}{k}\sum_{j=1}^k Y_j^2.
\]
By definition, $\sum_{j=1}^k Y_j^2$ follows a Chi-squared distribution $\chi_k^2$. Applying the Chi-squared tail bound (Lemma~\ref{lem:chi_square_tail}), there exists a universal constant $c > 0$ such that:
\[
    \Prb\left[ \frac{\overline{g}(u,v)}{q_e(X)} \notin \left[\frac{1}{2}, 2\right] \right] \le 2e^{-ck}.
\]
For a sufficiently large universal constant $C$, the assumption $k \ge C \log(n / (\beta_\chi \del_1))$ implies $k \ge \frac{1}{c} \log\left(\frac{2n^2}{\min(\beta_\chi, \del_1)}\right)$, so this probability is at most $\frac{\min(\beta_\chi, \del_1)}{n^2}$. Thus, taking a union bound, this concentration holds simultaneously for all ${n\choose 2}$ pairs with probability at least $1 - \min(\beta_\chi, \del_1)$.

On this high-probability event, we have $\overline{g}(u,v) / 2 \in [q_e(X)/4, q_e(X)]$ simultaneously for all pairs. Conditioned on this, since $X \in \mathcal{X}$ implies $q_e(X) \le 1$, the clipping operation $$\hat{g}(u,v) = \min\{\overline{g}(u,v)/2, 1\} \in [q_e(X)/4, q_e(X)],$$ which completes the proof.
\end{proof}
\begin{proof}[Proof of Lemma~\ref{lem:chaos}]
Let $\mu = \sum_{e=\{u,v\} \in \binom{n}{2}} D_{uv} q_e(X)$ be the true linear component of the objective $F(D)$. First, we analyze the estimator before clipping. Define the unclipped fractional witness as $\hat{g}^\circ(u,v) = \overline{g}(u,v)/2$. By the definition of $\overline{g}(u,v)$, we have:
\[
    \langle D, \hat{g}^\circ \rangle = \frac{1}{2k} \sum_{j=1}^k \sum_{e=\{u,v\} \in \binom{n}{2}} D_{uv} \frac{(Z^{(j)}_u - Z^{(j)}_v)^2}{4}.
\]
Since $Z^{(j)} \sim \mathcal{N}(0, X)$, $\langle D, \hat{g}^\circ \rangle$ is a degree-two polynomial in independent standard Gaussian random variables (a Gaussian chaos of degree two). From Proposition~\ref{prop:unbiased_estimator}, we see that $\E[\overline{g}(u,v)] = q_e(X)$, so we have $\E[\langle D, \hat{g}^\circ \rangle] = \mu/2$.

Let $W = \langle D, \hat{g}^\circ \rangle - \mu/2$ be the centered random variable, and let $\sigma^2 = \E[W^2]$ be its variance. If $\sigma=0$, then $W=0$ almost surely; on the good event in Lemma~\ref{lem:chi}, clipping is inactive, so the desired conditional probability is $1$.
Henceforth assume $\sigma>0$. Gaussian hypercontractivity for degree-two polynomials (Lemma~\ref{lem:hypercontractivity}) gives $\|W\|_4 \le 3\|W\|_2 = 3\sigma$.
Using the log-convexity of $L_p$ norms (H{\"o}lder's inequality, Lemma~\ref{lem:holder_interpolation}), we have:
\[
    \sigma = \|W\|_2 \le \|W\|_1^{1/3} \|W\|_4^{2/3} \le (\E|W|)^{1/3} (3\sigma)^{2/3}.
\]
Rearranging this inequality yields $\E|W| \ge \sigma/9$. Since $\E[W] = 0$, then $\E[W_+] = \E[W_-] = \E|W|/2$ and thus $\E[W_+] \ge \sigma/18$. Applying the Cauchy-Schwarz inequality to the positive part $W_+$:
\[
    (\E[W_+])^2 = (\E[W \cdot \ind_{W \ge 0}])^2 \le \Prb[W \ge 0] \E[W_+^2] \le \Prb[W \ge 0] \E[W^2] = \Prb[W \ge 0] \sigma^2.
\]
Substituting $\E[W_+] \ge \sigma/18$ gives $\Prb[W \ge 0] \ge (\sigma/18)^2 / \sigma^2 = 1/324$. Therefore, the unclipped estimator satisfies $\Prb[\langle D, \hat{g}^\circ \rangle \ge \mu/2] \ge 1/324$.

Finally, we connect this back to the clipped estimator $\hat{g}$ by conditioning on the high-probability event $\mathcal{E}_{\mathrm{good}}$ from Lemma~\ref{lem:chi}. On $\mathcal{E}_{\mathrm{good}}$, we are guaranteed that $\overline{g}(u,v)/2 \le q_e(X) \le 1$ for all pairs, meaning $\hat{g}(u,v) = \hat{g}^\circ(u,v)$ deterministically on $\mathcal{E}_{\mathrm{good}}$.

Let $\xi = \Prb[\neg\mathcal{E}_{\mathrm{good}}]$ be the failure probability of Lemma~\ref{lem:chi}. Its proof gives $\xi \le n^2e^{-ck}$. Taking the universal constant $C$ in $k\geq C\log(n/(\beta_\chi\del_1))$ sufficiently large ensures $\xi \le 1/648$, so we can bound the joint probability:
\[
    \Prb\left[\langle D, \hat{g} \rangle \ge \frac{\mu}{2} \text{ and } \mathcal{E}_{\mathrm{good}}\right] = \Prb\left[W \ge 0 \text{ and } \mathcal{E}_{\mathrm{good}}\right] \ge \Prb[W \ge 0] - \Prb[\neg\mathcal{E}_{\mathrm{good}}] \ge \frac{1}{324} - \xi \ge \frac{1}{648}.
\]
This yields a conditional probability of:
\[
    \Prb\left[\langle D, \hat{g} \rangle \ge \frac{\mu}{2} \;\middle|\; \mathcal{E}_{\mathrm{good}}\right] = \frac{\Prb[\langle D, \hat{g} \rangle \ge \mu/2 \text{ and } \mathcal{E}_{\mathrm{good}}]}{\Prb[\mathcal{E}_{\mathrm{good}}]} \ge \Prb\left[\langle D, \hat{g} \rangle \ge \frac{\mu}{2} \text{ and } \mathcal{E}_{\mathrm{good}}\right] \ge \frac{1}{648}.
\]
Setting $c_0 = 1/648$ concludes the proof.
\end{proof}

\section{Applications}\label{app:applications}

Here we discuss applications of Theorem~\ref{thm:main} to several important private graph problems. Except for correlation clustering, each application runs a nonprivate algorithm on the released graph and transfers its guarantee back to the input graph, following prior private graph-release work~\cite{arora2019differentially,BEK21,aamand2025breaking}. In particular, \cite{aamand2025breaking} stated the Max-Cut, Max-$k$-Cut, and minimum-bisection consequences of their mixed relative-additive release; our results for these problems use the same transfer with \cref{thm:main} in place of their release, and for Max-$k$-Cut we additionally remove their factor $k$ using the random block coloring of Liu~\cite{liu2022better}. Throughout this section, let $C_A$ be a sufficiently large universal constant, and abbreviate
\begin{equation}\label{eq:application-additive-error}
 A_\eta
 :=\frac{C_A n}{\eta\varepsilon}
 \left(\log\frac{n}{\delta\beta\eta\varepsilon}\right)^7 .
\end{equation}
Thus, from \cref{thm:main}, for any input non-negative weighted graph $G$, our $(\eps,\delta)$-differentially private algorithm outputs a synthetic graph $H$ such that with probability at least $1-\beta$:
\begin{equation}\label{eq:application-cut-event}
 \left|w_H(S,V\setminus S)-w_G(S,V\setminus S)\right|
 \leq \eta w_G(S,V\setminus S)+A_\eta
 \qquad(S\subseteq V).
\end{equation}
If an application uses a constant number of private mechanisms, we divide
$(\varepsilon, \delta,\beta)$ among them by constant factors; this only changes the universal
constants in \eqref{eq:application-additive-error}.  In particular,
$A_\eta=\widetilde O_{\eta,\varepsilon}(n)$ when $\delta$ and $\beta$ are inverse polynomial in
$n$. For the cut-release mechanisms in the high-probability corollaries below, we take $\beta=n^{-10}$; other algorithmic failure probabilities are accounted for separately.

For edge-indexed vectors $x,f$, write $x(f)=\langle x,f\rangle$;
for an edge set $A$, write $x(A)=\sum_{e\in A}x_e$.

\subsection{Correlation Clustering}\label{sec:correlation-clustering}

We first introduce an application for the standard unweighted complete-graph min-disagree correlation clustering
problem~\cite{BansalBlumChawla04}.  An instance is represented by its positive-edge graph $G=(V,E^+)$; every non-edge is
negative, and we denote their set by $E^-$.
For a partition $\Pi$ of $V$, let $F_\Pi$ be the set of pairs separated by $\Pi$ and
write the cost function as
\[
 \operatorname{cc}_G(\Pi)
 =|E^+\cap F_\Pi|+|E^-\setminus F_\Pi|.
\]
That is, $\operatorname{cc}_G(\Pi)$ counts the number of edges outside any cluster in $\Pi$ and the number of non-edges inside clusters in $\Pi$. Write $\operatorname{OPT}_G:=\min_\Pi\operatorname{cc}_G(\Pi)$.
Changing one sign is exactly edge adjacency of the positive-edge graph.  

\subsubsection{Related Work on Private Correlation Clustering}

The study of private correlation clustering was initiated by Bun, Eli\'a\v{s}, and
Kulkarni~\cite{BEK21}.  Their framework splits the signed input into its positive and negative
graphs, privately releases both, and runs a nonprivate correlation-clustering algorithm on the
resulting synthetic signed graph.  For complete unweighted instances they use the
Gupta--Roth--Ullman release~\cite{GRU12}; for weighted or incomplete instances they use the
synthetic-graph release of Eli\'a\v{s}, Kapralov, Kulkarni, and Lee~\cite{EKKL20}.  Their complete-unweighted analysis
gave cost at most $2.06\operatorname{OPT}_G+\widetilde O(n^{7/4}/\varepsilon)$.
The exponent $7/4$ came from bounding the error of a clustering with $k$ blocks by $k$ times the
cut distance and then merging small blocks to balance this term against the cost of merging.
Liu~\cite{liu2022better} observed, using a random bipartition of the blocks, that the factor $k$
is unnecessary.
Combined with the state-of-the-art purely additive graph release algorithm~\cite{EKKL20}, this yields an $\widetilde O(n^{3/2})$ additive error\footnote{While there exist mixed-error graph release algorithms with substantially smaller additive terms (e.g., $\tilde{O}(n^{13/12 +o(1)})$ in \cite{fan2026private}), they inherently introduce a relative error component. The BEK-Liu reduction instead assumes a purely additive cut-distance bound; applying it directly to a mixed-error release would incur a relative error proportional to the total dense graph mass $\Theta(n^2)$, thereby destroying the approximation guarantee.}
(for general weighted instances Liu obtained an $O(\log n)$ multiplicative approximation with
density-sensitive additive error).  In particular, updating the unweighted BEK--Liu route with the
current best nonprivate factor $1.387$~\cite{garcia2026approximate} gives the same
multiplicative factor with roughly $\widetilde O(n^{3/2})$ additive error.

Putting the approximation ratio aside, the prior best additive error for efficient private correlation clustering (unweighted, complete-graph, min-disagree) is due to
\cite{CohenAddadEtAlPrivateCC}. Their direct private algorithm (i.e., not based on running non-private approximation algorithms on a synthetic graph) achieves
cost 
$O(1)\operatorname{OPT}_G
 +O\left(\frac{n\log^2 n\log(1/\delta)}{\varepsilon^2}\right)$,
which is optimal up to logarithmic factors in its additive term, but its multiplicative constant
is at least in the thousands, far from
 the best nonprivate factor.  Thus the prior frontier had two incomparable points:
the factor $1.387$ with about $n^{3/2}$ additive error via graph release, and nearly linear
additive error with a large constant factor via the direct algorithm.  Our goal is to
obtain both guarantees simultaneously. Next, we explain the several difficulties one encounters on the way, and how we deal with them.
\subsubsection{Barriers in Prior Methods and Our Technical Overview}
We note that the BEK--Liu reduction cannot simply be instantiated with \cref{thm:main}.  Their transfer lemma
assumes a purely \emph{additive} cut-distance bound.  Our guarantee has an additional relative
term.  For a fractional positive graph $z\in[0,1]^{\binom V2}$ and a partition $\Pi$, with $F_\Pi$
denoting the separated pairs, the min-disagree cost and its rewriting are
\begin{equation}\label{eq:cc-objective-identity}
 \operatorname{cc}_z(\Pi)
 =z(F_\Pi)+(\one-z)(\tbinom V2\setminus F_\Pi)
 =2z(F_\Pi)+\sum_{C\in\Pi}\binom{|C|}{2}-z(\tbinom V2),
\end{equation}
the second equality using $\sum_{C\in\Pi}\binom{|C|}{2}=\binom n2-|F_\Pi|$; the cost agrees with
$\operatorname{cc}_G$ when $z$ is the $0/1$ positive-edge indicator.
Following Liu~\cite{liu2022better}, randomly bipartitioning the blocks relates $z(F_\Pi)$ to an ordinary cut, and randomly
bipartitioning the vertices does the same for the total mass.  Applied directly to the positive
graph, however, \cref{thm:main} permits an error $\gamma|E^+|+O(A)$.  This is
$\Theta(\gamma n^2)$ on a dense instance even when $\operatorname{OPT}_G=0$, so it cannot be
absorbed into the desired relative approximation.  Releasing the positive and negative graphs
separately only replaces $|E^+|$ by their combined mass, again $\Theta(n^2)$.\footnote{
Nonprivately, an additive term proportional to the total edge weight likewise appears when cut
sparsifiers are used for correlation clustering~\cite{AhnEtAl21CCStreams}; Assadi, Khanna, and
Putterman~\cite{AssadiKhannaPutterman25} avoid it by preserving the total weight exactly, which their
reconstruction obtains from the exact number of edges together with a spectral sparsifier.
}

To overcome this $\Theta(n^2)$ relative-error barrier, our core intuition is to bypass running the release mechanism directly on the original dense graph $G$, and instead apply it exclusively to a \emph{residual} graph. Specifically, we first compute a public baseline clustering $\Pi_0$ using the near-linear-additive private algorithm of \cite{CohenAddadEtAlPrivateCC}. While this baseline has a large constant approximation ratio, its total disagreement mass is bounded by $r=\operatorname{cc}_G(\Pi_0)\leq c_{\rm base}\operatorname{OPT}_G+ \widetilde{O}(n)$, which is \textbf{proportional} to $\operatorname{OPT}_G$ albeit with an extra $\widetilde{O}(n)$ additive error. We then construct a residual graph representing solely the mistakes made by $\Pi_0$ (i.e., positive pairs separated by $\Pi_0$ and negative pairs joined by $\Pi_0$). By running our cut-release mechanism (which builds Theorem~\ref{thm:main}) on this low-mass residual graph rather than the full instance, the relative error $\gamma$ is assessed against $r$ rather than the dense graph mass $\Theta(n^2)$. Consequently, the relative error becomes $\gamma r \le (\gamma c_{\rm base})\operatorname{OPT}_G + \widetilde{O}(n)$, which absorbs the massive dense-graph error into the multiplicative approximation factor.

However, implementing the above strategy first needs to resolve an output-model mismatch. The release in \cref{thm:main} is an arbitrary
nonnegative weighted graph, while the $1.387$ theorem of
\cite{garcia2026approximate} takes a complete unweighted signed graph.
Trying to insert the released fractional weights into the 
cluster-LP objective is not supported by \cite{garcia2026approximate}.
Likewise, clipping weights to $[0,1]$ afterward is unsafe, because cut accuracy controls signed
sums of errors rather than the total clipped excess.  Also the postprocessing in \cite{BEK21}
does not apply out of the box, as it was designed for a
purely additive error.

To resolve this, we specialize our fractional multiplicative-weights mechanism to binary inputs and enforce a strict unit coordinate cap inside each MWU round. Crucially, simple post-hoc clipping is unsafe; instead, we integrate this cap by projecting the candidate distribution onto a bounded convex polytope during the update process. The resulting capped disagreement release then combines with the public baseline signs to define a fractional positive-edge graph $p\in[0,1]^{\binom V2}$. Independent Bernoulli rounding produces a simple unweighted graph while preserving all clustering objectives up to an arbitrarily small fixed relative error and an $O(n\log n)$ additive error. Only at this point does the instance match the standard input model, allowing us to invoke an arbitrary non-private approximation algorithm as a black box.

The following two subsections formalize this capped fractional release and the subsequent black-box reduction.

\subsubsection{A Capped Unit-Bounded Release}

We first need a bounded-output specialization of Theorem~\ref{thm:main}. The coordinate cap lets us turn the fractional release into an unweighted graph without entering the harder weighted correlation-clustering problem. Note that the residuals used in Section~\ref{sec:black-box-reduction-correlation-clustering} are binary, but in the following proof we only assume each coordinate of the input is bounded in $[0,1]$, so we state the stronger form. For $S\subseteq V$ and some $D\subseteq {V\choose 2}$, write
\[
 \delta_D(S)=\{uv\in D:|\{u,v\}\cap S|=1\}.
\]

The following lemma will be useful:

\begin{lemma}[Generalized Pythagorean Inequality for KL Divergence~\cite{Csiszar75}]\label{lem:pythagorean}
Let $\mathcal{K}$ be a closed convex set of probability distributions, and let $\overline{p}$ be an arbitrary strictly positive probability distribution. If $p'$ is the projection of $\overline{p}$ onto $\mathcal{K}$, defined as:
\[
    p' = \argmin_{z \in \mathcal{K}} \KL(z \| \overline{p}),
\]
then for any target distribution $p^* \in \mathcal{K}$, it holds that:
\[
    \KL(p^* \| \overline{p}) \ge \KL(p^* \| p') + \KL(p' \| \overline{p}).
\]
In particular, since $\KL(p' \| \overline{p}) \ge 0$, this implies the distance-reduction property:
\[
    \KL(p^* \| p') \le \KL(p^* \| \overline{p}).
\]
\end{lemma}

\begin{lemma}[Capped unit-bounded cut release]\label{lem:capped-unit-release}
Let $D\subseteq\binom V2$ be public and let $x\in[0,1]^D$. For $0<\gamma\leq1/4$, $0<\varepsilon\leq1$, and $0<\delta,\beta_{\rm cap}<1/10$, there is a polynomial-time $(\varepsilon, \delta)$-DP mechanism under adjacency defined in Definition~\ref{def:DP} on $x$ that outputs $q\in[0,1]^D$ such that, except with probability $\beta_{\rm cap}$, simultaneously for every $S\subseteq V$,
\begin{equation*}
\left|q(\delta_D(S))-x(\delta_D(S))\right| \leq \gamma x(\delta_D(S))+A_{\rm cap},
\end{equation*}
where
\[
 A_{\rm cap} \leq O\left(\frac{n}{\gamma\varepsilon}
 \left(\log\frac{n}{\delta\beta_{\rm cap}\gamma\varepsilon}\right)^7\right).
\]
\end{lemma}

\begin{proof}
If $D=\emptyset$, output the empty vector. Otherwise, we naturally embed $x$ into $\mathbb{R}^{\binom{n}{2}}$ by padding it with zeros outside $D$. To ensure that the mechanism outputs synthetic weights strictly within $D$, we run \Cref{alg:main} on this padded input but strictly limit the algorithmic updates to the public support $D$ by initializing the outer active mask as $a_1 = \mathbf{1}_D$ (rather than $\mathbf{1}_{\binom{n}{2}}$). Since the input edge weights are already bounded by $U=1$, we omit the heavy-coordinate preprocessing step (Section~\ref{sec:preprocess}). We execute the algorithm with relative parameter $\gamma' = \gamma$ and an overall utility failure probability of $\beta_{\rm cap}$.

For the adaptive oracle calls in this variant, the utility failure probability in
Lemma~\ref{lem:gradient_estimation} can be counted locally. Conditional only on
the history before a call to \Cref{alg:gradient_estimation} whose regularized objective is at least $4\tau_1$,
Lemmas~\ref{lem:chi} and~\ref{lem:chaos} and the independent Laplace test give
acceptance probability at least $c_0/4$ per trial. Thus the probability of returning
$\bot$ is at most $(1-c_0/4)^K$. A union bound over at most $n^2$ calls, together
with the per-trial tail bounds for accepted witnesses, gives failure probability
\[
 n^2(1-c_0/4)^K+n^2K(\beta_1+\beta_\chi)\leq\beta_{\mathrm{GE}}
\]
for a sufficiently large constant in $K=\Theta(\log(n/\beta_{\mathrm{GE}}))$.
This supplies the oracle utility event used below without conditioning on a good
event for the entire execution.

To enforce the coordinate cap without violating privacy, we make two modifications that only affect the public \textit{post-processing} of the noisy masses and the candidate weights. Both are crucial: a lower-biased mass ensures the normalized true histogram remains feasible for the capped simplex, while projection prevents the exponential-weights updates from concentrating above the unit cap.

First, at outer round $i$, let $a_i$ be the active mask, $w_i=a_i\odotc x$, and $M_i=\norm{w_i}_1$. In place of the upper-biased noisy mass, we compute a lower-biased private estimate:
\begin{equation*}
 \widehat M_i = \operatorname{clip}_{[0,a_i(\one)]}(M_i + L_i - \tau_2), \qquad L_i\sim\Lap(1/\eps_2).
\end{equation*}
On the high-probability event $|L_i|\leq\tau_2$, we have
\begin{equation}\label{eq:lower-biased-mass-good}
 M_i-2\tau_2 \leq \widehat M_i \leq M_i.
\end{equation}
For this variant, take $M_{\mathrm{stop}}$ to be ten times the stopping threshold in
\Cref{alg:main} with $U=1$ and relative parameter $\gamma'$, so that
$M_{\mathrm{stop}}\geq100(B+\tau_2/\gamma')$, and use
$\widehat M_i$ in place of $\widetilde M_i$ throughout the round, halting when
$\widehat M_i\leq M_{\mathrm{stop}}$. In a continuing round,
$M_i\geq\widehat M_i>M_{\mathrm{stop}}$, so
$\widehat M_i/M_i\geq1-2\tau_2/M_i\geq1-\gamma'/50$ and $B/M_i\leq1/100$.

Second, all candidate updates in this round are constrained to satisfy $0\leq h_i(e)\leq a_i(e)$ and $\norm{h_i}_1=\widehat M_i$. Work on the public active support $D_i=\{e\in D:a_i(e)>0\}$, extending distributions by zero elsewhere. Equivalently, for $\widehat M_i>0$, we write $h_i = \widehat M_i p$ and maintain the distribution $p$ within the convex polytope:
\[
 \mathcal K_i = \left\{p\in\mathbb R_{\geq0}^{D_i} : p(\one)=1,\quad p_e\leq\frac{a_i(e)}{\widehat M_i} \text{ for every } e\in D_i\right\}.
\]
Clipping ensures $\widehat M_i\leq a_i(\one)$ on every execution, so $\mathcal K_i$ is non-empty and the initialization $p_e = a_i(e)/a_i(\one)$ is feasible. After the ordinary multiplicative-weights update with feature $f\in[0,1]^D$ yields a normalized distribution $\overline p$, we set
\begin{equation*}
p' = \arg\min_{z\in\mathcal K_i}\KL(z\|\overline p).
\end{equation*}
This projection has the form $p'_e=\min\{a_i(e)/\widehat M_i,c\overline p_e\}$, where $c>0$ makes $\sum_e p'_e=1$; sorting the breakpoints computes it in polynomial time. All candidate distributions remain positive on $D_i$.
The true target distribution $p^* = w_i/M_i$ remains feasible in $\mathcal K_i$ on the event \eqref{eq:lower-biased-mass-good}, because $$p^*_e = \frac{a_i(e)x_e}{M_i} \leq \frac{a_i(e)}{M_i} \leq \frac{a_i(e)}{\widehat M_i}.$$
The Bregman Pythagorean inequality (Lemma~\ref{lem:pythagorean}), applied on $D_i$,
guarantees that
\[
    D_{\mathrm{KL}}(p^* \| p')
    \le D_{\mathrm{KL}}(p^* \| \overline p).
\]
Thus projection cannot increase the KL potential, and the one-step
potential-decrease calculation in the proof of Lemma~\ref{lem:inner_round} remains valid
for each multiplicative update followed by projection.
For the current candidate distribution $p$ and a winning violating
feature $f$, we have $p(f) \ge 1/3$.
Writing $t=\widehat M_i/M_i\in[1-\gamma'/50,1]$, its normalized margin is
\begin{align*}
 p(f)-p^*(f)&\geq\left(1-\frac{t}{1+\gamma'}\right)p(f)
 \geq\frac{\gamma'}{3(1+\gamma')} &&\text{for overestimation},\\
 p^*(f)-p(f)&\geq\left(\frac{t}{1-\gamma'}-1\right)p(f)
 \geq\frac{49\gamma'}{150(1-\gamma')} &&\text{for underestimation}.
\end{align*}
Both are at least $\gamma'/[4(1+\gamma')]$. Initialization still gives $\KL(p^*\|p)\leq\log n$ since $M_i>100B\geq100n$. Thus the same learning rate and explicit $T_2$ in \Cref{alg:main} apply with $\gamma'$ in place of $\gamma$.

When the oracle returns an accurate mask $r_i$, we release the partial estimate $h_i\odotc r_i$ and update $a_{i+1}=a_i\odotc(\one_D-r_i)$ as before. The new cap implies coordinatewise $h_i\odotc r_i \leq a_i\odotc r_i = a_i-a_{i+1}$. Hence, the final synthesized graph $q = \sum_i h_i\odotc r_i$ inherently lies in $[0,1]^D$. 

When $r_i$ is selected, $\langle h_i,r_i\rangle\geq\widehat M_i/3$.
The cut-averaging argument in \Cref{lem:outer_round} therefore gives
\[
 \langle w_i,r_i\rangle
 \geq\frac{\widehat M_i/3-2B}{1+\gamma'}
 \geq\frac{(1-\gamma'/50)/3-1/50}{1+\gamma'}M_i
 \geq\frac{M_i}{20}.
\]
Thus the same outer contraction and telescoping cut-accuracy argument applies.
At termination, \eqref{eq:lower-biased-mass-good} bounds the discarded mass by
$M_{\mathrm{stop}}+2\tau_2$. With $U=1$ and $\gamma'=\gamma$, the parameter
calculation in \Cref{lem:utility} yields
\[
 T_1B+M_{\mathrm{stop}}+2\tau_2
 =O\left(\frac{n}{\gamma\varepsilon}
 \left(\log\frac{n}{\delta\beta_{\rm cap}\gamma\varepsilon}\right)^7\right).
\]
Finally, conditional on every common public history, the modified mass release is $\eps_2$-DP, and the projection and candidate updates are functions of the public transcript. Lemma~\ref{lem:gradient_privacy} applies uniformly to each resulting fixed candidate and mask. With the same iteration caps, the adaptive-composition argument of Lemma~\ref{lem:main-privacy} therefore proves $(\varepsilon,\delta)$-DP for this variant.
\end{proof}

\subsubsection{Black-Box Reduction to an Unweighted Instance}\label{sec:black-box-reduction-correlation-clustering}

\begin{theorem}[Black-box private correlation-clustering reduction]
\label{thm:cc-black-box-reduction}
Let $\rho\geq1$ and $p_0>0$ be constants, and let $\mathcal A$ be a polynomial-time randomized algorithm such that, on every unweighted complete-graph min-disagree instance $Q$,
\[
 \Prb\left[\operatorname{cc}_Q(\mathcal A(Q)) \leq \rho\operatorname{OPT}_Q\right] \geq p_0.
\]
Then, for every fixed $\xi>0$, $\varepsilon>0$, and $\delta=n^{-10}$, there is a polynomial-time $(\varepsilon, \delta)$-DP algorithm that, on every complete-graph min-disagree instance $G$, with high probability, outputs a partition $\widehat\Pi$ satisfying
\[
 \operatorname{cc}_G(\widehat\Pi) \leq (\rho+\xi)\operatorname{OPT}_G+\widetilde O_{\rho,\xi,\varepsilon}(n).
\]
\end{theorem}

\begin{proof}
Let $\chi_e=\ind[e\in E^+]$ be the positive-edge incidence vector of the input instance $G$. Without loss of generality, we assume $n \ge 2$, $0 < \xi \leq 1$, and $\varepsilon \le 1$. 

We allocate a constant fraction of the privacy budget to run the private constant-factor algorithm of~\cite{CohenAddadEtAlPrivateCC}. This returns a public baseline partition $\Pi_0$ such that, with probability at least $1-n^{-2}$,
\begin{equation}\label{eq:cc-baseline}
 r := \operatorname{cc}_G(\Pi_0) \leq c_{\rm base}\operatorname{OPT}_G+A_0, \qquad A_0= \widetilde{O}(n),
\end{equation}
for a universal constant $c_{\rm base}$. Let $D_+$ be the pairs separated by $\Pi_0$ and $D_-=\tbinom V2\setminus D_+$. We define the binary disagreement indicator
\[
 x_e=\begin{cases}
  \ind[e\in E^+],& e\in D_+,\\
  \ind[e\notin E^+],& e\in D_-.
 \end{cases}
\]
By definition, its total mass is exactly $x(\tbinom V2)=r$.

Setting $\beta_{\rm cap}=\beta_{\rm rnd}=n^{-10}/10$, we invoke Lemma~\ref{lem:capped-unit-release} once on $x$, consuming another constant fraction of the privacy budget. With relative parameter $\gamma$ and failure probability $\beta_{\rm cap}$, this yields a fractional vector $h\in[0,1]^{\binom V2}$ governed by the additive bound $A_{\rm cap}$. We then define a continuous fractional positive-edge graph $p\in[0,1]^{\binom V2}$ by setting $p_e=h_e$ for $e\in D_+$, and $p_e=1-h_e$ for $e\in D_-$.

We now compare clustering objectives evaluated on $p$ versus $G$. Let $d=h-x$. Following Liu~\cite{liu2022better}, for an arbitrary partition $\mathcal Q$, we color every block of $\mathcal Q$ independently and uniformly red or blue, defining a cut $S$ as the union of the red blocks. Every pair in $F_{\mathcal Q}$ crosses this random cut with probability $1/2$, while pairs within the same block never do. Thus, conditioned on the all-cut preservation event from Lemma~\ref{lem:capped-unit-release}, we have:
\begin{equation}\label{eq:cc-random-block-cut}
 |d(F_{\mathcal Q})| \leq \gamma x(F_{\mathcal Q})+2A_{\rm cap}.
\end{equation}
This holds because $|\E_S d(\delta(S))| = |d(F_{\mathcal Q})|/2$, which is bounded by $\E_S|d(\delta(S))| \leq \gamma x(F_{\mathcal Q})/2 + A_{\rm cap}$. Applying this to the discrete partition (singletons) yields the same bound for the total mass $d(\tbinom V2)$. Using the identity for $\operatorname{cc}_z$ from \eqref{eq:cc-objective-identity}, and substituting $p-\chi=d$ on $D_+$ and $p-\chi=-d$ on $D_-$, a direct algebraic expansion reveals:
\begin{equation*}
\operatorname{cc}_p(\Pi)-\operatorname{cc}_G(\Pi) = 2d(F_\Pi)+2d(D_+)-4d(F_\Pi\cup D_+)+d(\tbinom V2).
\end{equation*}
The four target sets $F_\Pi$, $D_+=F_{\Pi_0}$, $F_\Pi\cup D_+$, and $\tbinom V2$ correspond to the separated pairs of $\Pi$, $\Pi_0$, their common refinement $\Pi\wedge\Pi_0$, and the discrete partition, respectively. Since each has an $x$-mass of at most $r$, applying \eqref{eq:cc-random-block-cut} individually bounds each term $|d(\cdot)|$ by $\gamma r+2A_{\rm cap}$. Summing them with coefficients $2,2,4,1$ gives, simultaneously for every partition $\Pi$:
\begin{equation*}
 |\operatorname{cc}_p(\Pi)-\operatorname{cc}_G(\Pi)| \leq O(\gamma r+A_{\rm cap}).
\end{equation*}
Substituting \eqref{eq:cc-baseline} and using $\operatorname{OPT}_G\leq\operatorname{cc}_G(\Pi)$, we bound this difference by:
\begin{equation}\label{eq:cc-relative-transfer}
 a\operatorname{cc}_G(\Pi)+b, \qquad \text{where} \quad a=O(c_{\rm base}\gamma), \quad b=O(\gamma A_0+A_{\rm cap}).
\end{equation}

To obtain an unweighted graph $J$, we independently instantiate every pair $e$ as an edge with probability $p_e$, as in the rounding steps of~\cite{BEK21,AssadiKhannaPutterman25}. For any fixed partition $\Pi$, $\operatorname{cc}_J(\Pi)$ is a sum of independent Bernoulli variables with mean $\operatorname{cc}_p(\Pi)$. For any fixed $0<\eta_{\rm rnd}\leq1$, applying a two-sided Bernstein-Chernoff bound with $t=n\log n+\log(2/\beta_{\rm rnd})$ and taking a union bound over all $\le n^n$ partitions ensures that with probability at least $1-\beta_{\rm rnd}$, simultaneously for every $\Pi$:
\begin{equation}\label{eq:cc-bernoulli-rounding}
 |\operatorname{cc}_J(\Pi)-\operatorname{cc}_p(\Pi)| \leq \eta_{\rm rnd}\operatorname{cc}_p(\Pi)+ \frac{3}{\eta_{\rm rnd}}\left(n\log n+\log\frac{2}{\beta_{\rm rnd}}\right).
\end{equation}
Combining \eqref{eq:cc-relative-transfer} and \eqref{eq:cc-bernoulli-rounding} ensures that $J$ acts as a uniform surrogate for $G$:
\begin{equation}\label{eq:cc-uniform-surrogate}
 |\operatorname{cc}_J(\Pi)-\operatorname{cc}_G(\Pi)| \leq \theta\operatorname{cc}_G(\Pi)+B \qquad\text{for every }\Pi,
\end{equation}
where $\theta=a+\eta_{\rm rnd}(1+a)$ and $B=(1+\eta_{\rm rnd})b+\frac{3}{\eta_{\rm rnd}}(n\log n+\log\frac{2}{\beta_{\rm rnd}})$.

Finally, we independently invoke the non-private approximation algorithm $\mathcal A$ on the unweighted graph $J$ for $O(p_0^{-1}\log n)$ times, obtaining the partition $\widehat\Pi$ with the minimum $\operatorname{cc}_J$ cost. This succeeds with probability at least $1-n^{-10}$. If $\Pi^*$ is the optimal partition for $G$, \eqref{eq:cc-uniform-surrogate} combined with the $\rho$-approximation guarantee gives:
\begin{align*}
 \operatorname{cc}_G(\widehat\Pi) &\leq \frac{\operatorname{cc}_J(\widehat\Pi)+B}{1-\theta} \leq \frac{\rho\operatorname{cc}_J(\Pi^*)+B}{1-\theta} \\
 &\leq \rho\frac{1+\theta}{1-\theta}\operatorname{OPT}_G +\frac{\rho+1}{1-\theta}B.
\end{align*}
We choose $\eta_{\rm rnd}$ and $\gamma$ sufficiently small such that $\theta\leq \xi/(2\rho+\xi)$ (e.g., small constant multiples of $\xi/\rho$ and $\xi/(c_{\rm base}\rho)$). This guarantees that the multiplicative factor $\rho\frac{1+\theta}{1-\theta} \leq \rho+\xi$. The remaining additive term scales as $n$ multiplied by a fixed polynomial in $\rho/\xi$, $1/\varepsilon$, $\log n$, and $\log(1/\delta)$.

Privacy follows from adaptive composition (Lemma~\ref{lem:basic-composition}): the sets $D_+$ and $D_-$ are derived from the baseline transcript, and altering a single input sign changes exactly one coordinate of $x$. The Bernoulli rounding and non-private approximation are strictly post-processing steps (Lemma~\ref{lem:post-processing}). A union bound over the baseline, capped release, rounding, and approximation gives overall failure probability $O(n^{-2})$, proving the high-probability claim.
\end{proof}
\subsubsection{Proof of Corollary~\ref{cor:cc-intro}}
Finally, we are now ready to present the proof for our result on private correlation clustering, which we restate here for the ease of reading.
\privatecorrelationclustering*
\begin{proof}
Set $\tau=1/4000$ and run the deterministic non-private approximation algorithm of \cite{garcia2026approximate} with approximation slack $\tau$. It returns a clustering of cost at most $(1.3865+\tau)\operatorname{OPT}_Q$, so the black-box hypothesis holds with $p_0=1$.

We can now invoke the black-box reduction from Theorem~\ref{thm:cc-black-box-reduction} by setting $\rho=1.3865+\tau$ and $\xi=\tau$. Since $\rho + \xi = 1.3865 + 2\tau = 1.387$, the mechanism outputs a partition $\widehat\Pi$ satisfying:
\[
 \operatorname{cc}_G(\widehat\Pi) \leq 1.387\operatorname{OPT}_G+\widetilde O_{\varepsilon}(n)
\]
with high probability. This completes the proof.
\end{proof}

\subsection{Correlation Clustering with At Most $O(1)$ Clusters}
In the standard settings we discussed in Section~\ref{sec:correlation-clustering}, there is no constraint on the number of clusters. For a positive integer $k$, let $\operatorname{OPT}_{G,k}$ be the minimum disagreement cost among
partitions of $V$ into at most $k$ nonempty blocks.  Giotis and
Guruswami~\cite{GiotisGuruswami06} gave a PTAS for this problem when $k$ is fixed.  Restricting the
black-box reduction in \cref{thm:cc-black-box-reduction} to the same feasible family gives the following consequence.

\begin{corollary}[Private fixed-$k$ correlation clustering]
\label{cor:cc-fixed-k}
For every fixed $k\geq2$ and $\xi>0$, constant $\varepsilon>0$, and $\delta=n^{-10}$, there is a
polynomial-time $(\varepsilon, \delta)$-edge-DP algorithm for complete unweighted
Min-Disagree with at most $k$ clusters which, with high probability, outputs a partition
$\widehat\Pi$ satisfying
\[
 \operatorname{cc}_G(\widehat\Pi)
 \leq (1+\xi)\operatorname{OPT}_{G,k}
   +\widetilde O_{\xi,\varepsilon}(n).
\]
\end{corollary}

\begin{proof}
Apply the proof of \cref{thm:cc-black-box-reduction} to partitions with at most $k$ blocks.  The
PTAS of \cite[Section~4]{GiotisGuruswami06}, run with accuracy $\xi/2$, returns with high
probability a partition of $J$ of cost at most
$(1+\xi/2)\operatorname{OPT}_{J,k}$.  The baseline remains unrestricted, but
\eqref{eq:cc-baseline} gives
\[
 r\leq c_{\rm base}\operatorname{OPT}_G+A_0
   \leq c_{\rm base}\operatorname{OPT}_{G,k}+A_0.
\]
The uniform comparison inequalities hold for the restricted family.  Applying the reduction with
$\rho=1+\xi/2$ and transfer slack $\xi/2$ proves the claim.  The running time is polynomial for
fixed $k$ and $\xi$.
\end{proof}

For comparison, the fixed-$k$ theorem in \cite[Section~5.4, Theorem~29]{BEK21} was pure DP and had the
same PTAS-type multiplicative guarantee but additive error
$\widetilde O_\varepsilon(k n^{3/2})$; Liu's random block coloring~\cite{liu2022better} removes the factor $k$ from this transfer.  Thus
\cref{cor:cc-fixed-k} improves the polynomial-time additive dependence from $n^{3/2}$ to nearly
linear, but under $(\eps, \delta)$-DP.

\subsection{Max-Cut and Max-$k$-Cut}

For a partition $\Pi$ of $V$, let
$\operatorname{cut}_G(\Pi)=w_G(F_\Pi)$ be the weight of pairs in different blocks.  Thus Max-Cut
is the case $|\Pi|\leq2$, and Max-$k$-Cut restricts to $|\Pi|\leq k$.  Define
\[
 \operatorname{OPT}^{\operatorname{Max-}k\operatorname{-Cut}}_G
 :=\max_{|\Pi|\leq k}\operatorname{cut}_G(\Pi).
\]
Abbreviate the case $k=2$ by $\operatorname{OPT}^{\operatorname{Max-Cut}}_G$.

\begin{theorem}[Private Max-$k$-Cut]
\label{thm:private-max-k-cut}
Suppose a polynomial-time nonprivate algorithm, on every nonnegative weighted graph $Q$, outputs
an at-most-$k$ partition $\Pi$ and, for some $0<\rho_k\leq1$, satisfies
\[
 \E[\operatorname{cut}_Q(\Pi)]
 \geq\rho_k\operatorname{OPT}^{\operatorname{Max-}k\operatorname{-Cut}}_Q.
\]
For every fixed $\xi>0$, constant $\varepsilon>0$, and
$\delta=n^{-10}$, there is a polynomial-time $(\varepsilon, \delta)$-edge-DP algorithm which, with
high probability, outputs an at-most-$k$ partition $\widehat\Pi$ satisfying
\[
 \operatorname{cut}_G(\widehat\Pi)
 \geq (\rho_k-\xi)\operatorname{OPT}^{\operatorname{Max-}k\operatorname{-Cut}}_G
   -\widetilde O_{\xi,\varepsilon}(n).
\]
The additive term is independent of $k$.  In particular, for Max-Cut, the algorithm of
Goemans and Williamson~\cite{GoemansWilliamson95} gives
\[
 \operatorname{cut}_G(\widehat\Pi)
 \geq (0.878-\xi)\operatorname{OPT}^{\operatorname{Max-Cut}}_G
   -\widetilde O_{\xi,\varepsilon}(n).
\]
For $k\geq3$, one may use the standard SDP ratio $\rho_k$ of Frieze and
Jerrum~\cite{FriezeJerrum97}.
\end{theorem}

\begin{proof}
Release $H$ using \cref{thm:main}.  We first lift \eqref{eq:application-cut-event} from cuts to
arbitrary partitions without paying for the number of blocks, using the random block coloring of
Liu~\cite{liu2022better}.  Independently color every block of
$\Pi$ red or blue, and let $S$ be the union of the red blocks.  Every pair in $F_\Pi$ crosses the
resulting cut with probability $1/2$.  Writing $d=H-G$ gives
\begin{equation}
    \begin{aligned}
 |\operatorname{cut}_H(\Pi)-\operatorname{cut}_G(\Pi)|
 &=2|\E_S d(S,V\setminus S)| \\
 &\leq2\E_S|d(S,V\setminus S)| \\
 &\leq\eta\operatorname{cut}_G(\Pi)+2A_\eta .
\end{aligned}
\label{eq:max-k-cut-transfer}
\end{equation}

The cited SDP algorithms give their ratios in expectation.  We may assume
$0<\xi<\rho_k$, since otherwise the theorem is vacuous.  Since the returned value lies between
zero and the optimum on $H$, a standard bounded-variable argument shows that one rounding trial
achieves ratio at least $\overline\rho_k:=\rho_k-\xi/3$ with probability
$\Omega_{\rho_k,\xi}(1)$.  Repeat the rounding
$O_{\rho_k,\xi}(\log n)$ times on the public graph $H$ and retain the best partition.  With high
probability, the retained partition $\widehat\Pi$ has value at least
$\overline\rho_k\operatorname{OPT}^{\operatorname{Max-}k\operatorname{-Cut}}_H$.

Let $\Pi^*$ be optimal for $G$.  On this rounding event, applying
\eqref{eq:max-k-cut-transfer} twice yields
\begin{align*}
 \operatorname{cut}_G(\widehat\Pi)
 &\geq\frac{\operatorname{cut}_H(\widehat\Pi)-2A_\eta}{1+\eta}\\
 &\geq\overline\rho_k\frac{1-\eta}{1+\eta}\operatorname{cut}_G(\Pi^*)
   -\frac{2(\overline\rho_k+1)}{1+\eta}A_\eta.
\end{align*}
Choose $\eta=\min\{1/4,\xi/(3\rho_k)\}$; then the displayed multiplicative factor is at least
$\rho_k-\xi$.  The theorem
follows by postprocessing.
As in Liu's correlation-clustering analysis~\cite{liu2022better}, random block coloring, rather than
summing the $k$ block-boundary errors, removes the factor $k$ in the additive term of the
Max-$k$-Cut bound of~\cite{aamand2025breaking}.
\end{proof}

\begin{remark}[Approximate-DP Max-Cut lower bound]\label{rem:maxcut-tightness}
A direct-sum adaptation of the privacy comparison in~\cite{dharangutte2026price}
shows that the linear additive dependence on $n$ is necessary even for unweighted
Max-Cut under approximate differential privacy. Specifically, let $n\geq8$, $0\leq\varepsilon\leq1$,
$0\leq\delta,\beta\leq1/100$, and $0.868\leq\rho\leq1$.
If an $(\varepsilon,\delta)$-edge-DP algorithm $\mathcal M$ returns a cut of value
at least $\rho\operatorname{OPT}_G-B$, where $B\geq0$, with probability at least
$1-\beta$ on every unweighted $n$-vertex graph, then
$B\geq\lfloor n/8\rfloor/100=\Omega(n)$.

To see this, partition $8m$ vertices into $m=\lfloor n/8\rfloor$ public blocks
of size eight, and put one independently uniform random edge in each block;
any remaining vertices are isolated. Every such graph has optimum $m$.
Fix a block and the edges in all other blocks. Let $G_0$ leave this block empty,
and let $G_e=G_0+e$ for each of its $28$ possible edges $e$.
Let $E_e$ be the event that the output cut does not cut $e$.
Differential privacy gives
\[
 \Prb[\mathcal M(G_e)\in E_e]
 \geq e^{-\varepsilon}\bigl(\Prb[\mathcal M(G_0)\in E_e]-\delta\bigr).
\]
Every cut separates at most $16$ of the $28$ possible edges in the block, so
averaging over $e$ bounds its expected contribution under $\mathcal M(G_e)$ by
\[
 q:=1-e^{-\varepsilon}(3/7-\delta)
 \leq1-e^{-1}(3/7-1/100)<0.847.
\]
This holds for every choice of the other blocks. Averaging over all input edges
and summing over blocks therefore gives some graph $G$ with
$\E[\operatorname{cut}_G(\mathcal M(G))]\leq qm$.
The assumed utility guarantee and nonnegativity of cut values give
$\E[\operatorname{cut}_G(\mathcal M(G))]\geq(1-\beta)\rho m-B$.
Consequently,
\[
 B\geq\bigl((1-\beta)\rho-q\bigr)m
 >\bigl(0.99\cdot0.868-0.847\bigr)m>m/100.
\]
In particular, this applies to the ratio $0.878-\xi$ for $0<\xi\leq1/100$,
establishing the claimed tightness up to polylogarithmic factors in that range.
\end{remark}

We compare our result on private Max-Cut approximation with previous works. Arora and Upadhyay~\cite[Theorem~5]{arora2019differentially} gave an explicit
polynomial-time private Max-Cut algorithm with the Goemans--Williamson factor and worst-case
$\widetilde O(n^{3/2})$ additive error for fixed privacy parameters.
For weighted inputs,
\cite{aamand2025breaking} applied the same route to their mixed relative-additive release and
stated  additive error $n^{5/4+o(1)}$ for Max-Cut with the
Goemans--Williamson factor up to arbitrarily small slack, and $k\,n^{5/4+o(1)}$ for Max-$k$-Cut;
their proofs use the same transfer as above, with their release in place of \cref{thm:main}.
For unweighted inputs,
applying the $n^{13/12+o(1)}$ all-cut release of
\cite[Theorem~1]{fan2026private} improves that generic route to $n^{13/12+o(1)}$ for both Max-Cut
and Max-$k$-Cut.  The pure-DP Max-Cut algorithm of Chandra, Dinitz, Fan, and
Zou~\cite[Section~5.3 of the arXiv version]{chandra2026differentially} has
$O(n\log n)$ error for constant privacy, but
its running time invokes an exact Max-Cut algorithm and is exponential.  Thus
\cref{thm:private-max-k-cut} improves the preceding polynomial-time release-based bound to nearly
linear additive error; for Max-Cut it obtains the Goemans--Williamson factor up to an arbitrarily
small fixed slack. 
\subsection{Minimum Bisection and Balanced Cuts}

The same release also transfers minimization over any public family of feasible cuts.  This
captures exact bisections, cardinality-constrained cuts, and bicriteria balanced-separator
algorithms.

\begin{theorem}[Private bicriteria cut minimization]
\label{thm:private-constrained-cut}
Fix $0<\eta\leq1/4$, $0<\varepsilon\leq1$, $0<\delta,\beta<1/10$,
$\rho\geq1$, and $0\leq\tau<1$.  Let
$\emptyset\ne\mathcal F,\mathcal F'\subseteq2^V$ be public families and
define
\[
 \operatorname{OPT}_{G,\mathcal F}
 :=\min_{S\in\mathcal F}w_G(S,V\setminus S).
\]
Suppose a polynomial-time nonprivate algorithm, on every nonnegative weighted graph, returns
$\widehat S\in\mathcal F'$ with probability at least $1-\tau$ and, on that event,
\[
 w(\widehat S,V\setminus\widehat S)
 \leq \rho\min_{S\in\mathcal F}w(S,V\setminus S).
\]
Then there is a polynomial-time $(\varepsilon, \delta)$-edge-DP algorithm which, with probability
at least $1-\beta-\tau$, returns $\widehat S\in\mathcal F'$ satisfying
\[
 w_G(\widehat S,V\setminus\widehat S)
 \leq \rho\frac{1+\eta}{1-\eta}\operatorname{OPT}_{G,\mathcal F}
   +\frac{\rho+1}{1-\eta}A_\eta.
\]
Here $A_\eta$ is the quantity in \eqref{eq:application-additive-error} for the displayed
$(\varepsilon, \delta,\beta)$.
Consequently, for every fixed $\xi>0$ and $\rho\geq1$, choosing
$\eta=\min\{1/4,\xi/(2\rho+\xi)\}$ gives
$(\rho+\xi)\operatorname{OPT}_{G,\mathcal F}
+\widetilde O_{\xi,\varepsilon}(\rho^{8}n)$ for constant $\varepsilon$ and
inverse-polynomial $\delta,\beta$.
\end{theorem}

\begin{proof}
Run \cref{thm:main}, apply the nonprivate algorithm to $H$, and let $S^*$ minimize the cut in
$G$ over $\mathcal F$. On the intersection of the cut-release accuracy event \eqref{eq:application-cut-event} and the success event of the nonprivate algorithm, which has probability at least $1-\beta-\tau$ by a union bound,
\begin{align*}
 (1-\eta)w_G(\widehat S,V\setminus\widehat S)-A_\eta
 &\leq w_H(\widehat S,V\setminus\widehat S)\\
 &\leq \rho w_H(S^*,V\setminus S^*)\\
 &\leq \rho(1+\eta)\operatorname{OPT}_{G,\mathcal F}+\rho A_\eta.
\end{align*}
Rearranging proves the desired bound.
\end{proof}

Define $\operatorname{OPT}_{\rm bis}(G)$ as the minimum cut weight over partitions whose side
sizes differ by at most one.  For even $n$, take $\mathcal F'=\mathcal F$ to be the exact
bisections, set $\tau=0$, and use the deterministic
$\rho=O(\log n)$ algorithm of R{\"a}cke~\cite[Section~3, Lemma~9 and the argument
following it]{Raecke08}.  Running \cref{thm:private-constrained-cut} with any fixed $\eta$ gives
\[
 w_G(\widehat S,V\setminus\widehat S)
 \leq O(\log n)\operatorname{OPT}_{\rm bis}(G)+\widetilde O_\varepsilon(n).
\]
For odd $n$, after releasing $H$, add the same public isolated zero-weight vertex $v_0$ to both
$G$ and $H$.  The cut-release event is unchanged because every augmented cut has the value of its
restriction to $V$.  Compute an exact bisection of the augmented graph and remove $v_0$; this
gives the same guarantee for parts of sizes $\lfloor n/2\rfloor$ and $\lceil n/2\rceil$.

For a weighted graph $Q$ on vertex set $U$, write
$\phi_Q(S)=w_Q(S,U\setminus S)/\min\{|S|,|U\setminus S|\}$ for the expansion of $S$.
Call $S\subseteq U$ $b$-balanced if both $S$ and $U\setminus S$ have size at least $b|U|$.
For a fixed target balance $0<b\leq1/4$, let $\mathcal F_b$ contain the $b$-balanced cuts.
Arora, Rao, and
Vazirani~\cite[Corollary~2]{AroraRaoVazirani09} show that there is a constant
$b'=b'(b)>0$ and a randomized polynomial-time $O(\sqrt{\log n})$
pseudo-approximation that outputs a cut in $\mathcal F_{b'}$ while comparing its expansion with
the best cut in $\mathcal F_b$.  Their corollary is presented for unweighted graphs.  For nonnegative capacities
$c_e$, use the same SDP with objective
$W=\frac14\sum_e c_e\|v_i-v_j\|_2^2$, but keep the region-growing edge length
$\ell_e=\frac14\|v_i-v_j\|_2^2$ without a capacity factor.  In the proof of Corollary~2, replace
$|E_s|$ by the crossing capacity $c(E_s)$.  The identical calculation gives
$\int_0^\Delta c(E_s)\,ds\leq\sum_e c_e\ell_e=W$, proving the weighted version needed for the
released graph $H$.  Repeating the rounding $O(\log n)$ times and obtaining the best feasible cut makes
$\tau\leq n^{-10}$; if all trials fail the public balance check, output an arbitrary cut in
$\mathcal F_{b'}$.
Because both denominators are within constant factors of $n$, this is also a bicriteria
$\rho=O_b(\sqrt{\log n})$ guarantee for cut weight.  Taking
$(\mathcal F,\mathcal F')=(\mathcal F_b,\mathcal F_{b'})$ in
\cref{thm:private-constrained-cut} gives a $b'$-balanced cut with
\[
 w_G(\widehat S,V\setminus\widehat S)
 \leq O_b(\sqrt{\log n})\operatorname{OPT}_{G,\mathcal F_b}
   +\widetilde O_{b,\varepsilon}(n).
\]
Dividing by the smaller output-side cardinality gives the edge-expansion guarantee
\[
 \phi_G(\widehat S)
 \leq O_b(\sqrt{\log n})\min_{S\in\mathcal F_b}\phi_G(S)
 +\widetilde O_{b,\varepsilon}(1).
\]
If uniform sparsity is defined by
$\psi_G(S)=w_G(S,V\setminus S)/(|S|(n-|S|))$, then dividing the displayed inequality by the
output denominator gives
\[
 \psi_G(\widehat S)
 \leq O_b(\sqrt{\log n})\min_{S\in\mathcal F_b}\psi_G(S)
 +\widetilde{O}_{b,\varepsilon}(1/(b'(1-b')n)).
\]
Arora and Upadhyay~\cite[Theorem~5]{arora2019differentially} previously combined their private
sparsifier with the algorithm of~\cite{AroraRaoVazirani09} for uniform sparsest cut without a balance
constraint, with additive error $\widetilde O(1/\sqrt n)$ for fixed privacy parameters.

\cite{aamand2025breaking} stated the same R{\"a}cke-based reduction for minimum
bisection with their release, obtaining additive error $n^{5/4+o(1)}$ for weighted graphs.
For unweighted inputs, substituting the release of \cite[Theorem~1]{fan2026private} in the same
reductions gives the preceding release-based additive term
$n^{13/12+o(1)}$ (polylogarithmic approximation factors are absorbed in $n^{o(1)}$).  The new
release reduces that term to $n^{1+o(1)}$ for exact minimum bisection and for the bicriteria
balanced-cut consequence above.

\subsection{Hierarchical Clustering}

Dasgupta~\cite{Dasgupta16} defined the cost of a hierarchical clustering tree $T$ on a
nonnegative weighted similarity graph $G$ by
\[
 \operatorname{hcost}_G(T)
 =\sum_{\{u,v\}}w_G(u,v)\,|T[u\vee v]|,
\]
where $T[u\vee v]$ is the smallest subtree containing leaves $u$ and $v$, and
$|T[u\vee v]|$ denotes its number of leaves.  For $A\subseteq V$ and $S\subset A$, use
$\phi_{G[A]}(S)$ for the expansion defined above on the induced graph $G[A]$.
Define $\operatorname{OPT}^{\rm HC}_G:=\min_T\operatorname{hcost}_G(T)$.

We first present a modified \textit{bicriteria} recursion statement needed later. Its proof adapts the level-set charging argument of Charikar and Chatziafratis~\cite{CharikarChatziafratis17} (see also~\cite{CohenAddadEtAl19HC}). Their original recursive-sparsest-cut theorem allows for arbitrarily unbalanced cuts, which in the worst case would lead to an $O(n)$ tree depth and cause our accumulated additive privacy error to explode. To bound the tree depth to $O(\log n)$, the lemma below enforces a hard limit on the split ratio, while comparing its expansion cost against the optimal cut under a balance requirement specified by $b$. Balanced recursions were also analyzed by~\cite{AssadiEtAl22HCStreams}, building on~\cite{CharikarChatziafratis17}; we state the bicriteria form needed for the pseudo-approximation of~\cite{AroraRaoVazirani09}.

\begin{lemma}[Bicriteria balanced recursion]\label{lem:hc-bicriteria-recursion}
Fix constants $0<b\leq1/3$ and $0<b'\leq1/3$, and let $a\geq1$.  Suppose an algorithm, on every
nonnegative weighted graph with vertex set $A$ of size at least $1/b$, returns a $b'$-balanced
cut $S$ satisfying
\[
 \phi_{G[A]}(S)\leq a\min\{\phi_{G[A]}(R):R\text{ is $b$-balanced}\}.
\]
Recursively applying this algorithm, and splitting smaller clusters as evenly as possible,
produces a hierarchical clustering tree $T$ with
\[
 \operatorname{hcost}_G(T)\leq O_b(a)\operatorname{OPT}^{\rm HC}_G.
\]
\end{lemma}

\begin{proof}
Fix an optimal tree $T^*$ for $G$.  For $t\geq1$, let $\mathcal P_t$ be the partition into the
maximal clusters of $T^*$ having size at most $t$, and let $E_t$ be the pairs whose endpoints lie
in different blocks of $\mathcal P_t$.  Set $E_0=\tbinom V2$.  Counting the contribution of each
pair according to the size of its least common ancestor gives
\begin{equation}\label{eq:hc-opt-levels}
 \operatorname{hcost}_G(T^*)=\sum_{t=0}^{n-1}w_G(E_t).
\end{equation}

Consider a recursive cluster $A$ of size $r$ that the algorithm splits into $S$ and
$A\setminus S$, and put $s=\min\{|S|,|A\setminus S|\}$.  Intersect the blocks of
$\mathcal P_{\lfloor br\rfloor}$ with $A$.  When $r\geq1/b$, these pieces have size at most $br$
and can be grouped greedily into two unions, each of size at least $br$;
the assumption $b\leq1/3$ ensures that the complement of the first union is also that large.
Every edge of this candidate cut belongs to
$E_{\lfloor br\rfloor}\cap\tbinom A2$.  Therefore the oracle guarantee implies
\[
 \phi_{G[A]}(S)
 \leq \frac{a}{br}\,w_G(E_{\lfloor br\rfloor}\cap\tbinom A2),
\]
and the contribution of this split to the hierarchical cost is at most
\begin{equation}\label{eq:hc-one-split-charge}
 r\,w_G(S,A\setminus S)
 \leq \frac{a}{b}s\,w_G(E_{\lfloor br\rfloor}\cap\tbinom A2).
\end{equation}
If $2\leq r<1/b$, split $A$ as evenly as possible.  Then $s\geq1$, the split is
$b'$-balanced because $b'\leq1/3$, and, using $a\geq1$ and
$E_{\lfloor br\rfloor}=E_0$, the same inequality follows from
\[
 r\,w_G(S,A\setminus S)
 \leq r\,w_G(E_0\cap\tbinom A2)
 \leq \frac{a}{b}s\,w_G(E_0\cap\tbinom A2).
\]
For every integer $t\in\{r-s,\ldots,r-1\}$, monotonicity of the level partitions gives
\[
 w_G(E_{\lfloor bt\rfloor}\cap\tbinom A2)
 \geq w_G(E_{\lfloor br\rfloor}\cap\tbinom A2).
\]
For a fixed $t$, the recursive clusters satisfying $r-s\leq t<r$ are disjoint: they are exactly
some of the minimal clusters in the output tree having size greater than $t$.  Summing
\eqref{eq:hc-one-split-charge} first over all recursive clusters and then over $t$ thus gives
\[
 \operatorname{hcost}_G(T)
 \leq \frac{a}{b}\sum_{t=0}^{n-1}w_G(E_{\lfloor bt\rfloor})
 \leq O_b(a)\sum_{j=0}^{n-1}w_G(E_j).
\]
Together with \eqref{eq:hc-opt-levels}, this proves the lemma.
\end{proof}

\begin{theorem}[Private hierarchical clustering]
\label{thm:private-hierarchical-clustering}
For every $0<\eta\leq1/4$, constant $0<\varepsilon\leq1$, and $\delta=n^{-10}$, there is a
polynomial-time $(\varepsilon, \delta)$-edge-DP algorithm that, with high probability, outputs a
$b'$-balanced hierarchical clustering tree $\widehat T$, for a universal constant $b'>0$,
satisfying
\[
 \operatorname{hcost}_G(\widehat T)
 \leq O(\sqrt{\log n})\operatorname{OPT}^{\rm HC}_G
  +O_\eta(A_\eta n\log^{3/2}n).
\]
In particular, for fixed $\eta$ the additive term is
$\widetilde O_{\eta,\varepsilon}(n^2)$.
\end{theorem}

\begin{proof}
We first show that all constant-balanced trees are simultaneously preserved by the release,
analogously to the cut-sparsifier result of~\cite{AssadiEtAl22HCStreams}.  For a set
$X\subseteq V$, abbreviate $c_G(X)=w_G(X,V\setminus X)$.  If an internal cluster $A$ of $T$ is
split into children $B,C$, then
\[
 w_G(B,C)=\frac12\bigl(c_G(B)+c_G(C)-c_G(A)\bigr).
\]
We write $A\to(B,C)$ for this internal split: $A$ is the leaf set below an internal node of $T$,
and $B,C$ are the leaf sets below its two children.
Substituting this identity in
$\operatorname{hcost}_G(T)=\sum_{A\to(B,C)}|A|w_G(B,C)$ and regrouping the coefficient of each
cut gives the decomposition of~\cite{AgarwalKhannaLiPatil22} (see
also~\cite[Section~5.1]{ImolaEtAl23})
\begin{equation}\label{eq:hc-cut-decomposition}
 \operatorname{hcost}_G(T)
 =\frac12\sum_{A\to(B,C)}
    \bigl(|C|c_G(B)+|B|c_G(C)\bigr)
  +\frac12\sum_{v\in V}c_G(\{v\}).
\end{equation}
All coefficients in \eqref{eq:hc-cut-decomposition} are nonnegative.  Applying
\eqref{eq:application-cut-event} term by term therefore gives
\begin{equation}\label{eq:hc-uniform-transfer}
 |\operatorname{hcost}_H(T)-\operatorname{hcost}_G(T)|
 \leq \eta\operatorname{hcost}_G(T)
 +\frac{A_\eta}{2}\left(n+\sum_{A\text{ internal}}|A|\right).
\end{equation}
If every split of $T$ is $c$-balanced for a fixed $c>0$, every root-to-leaf path has length at
most $D_c=\lceil\log_{1/(1-c)}n\rceil$.  Combining \eqref{eq:hc-uniform-transfer} with the fact
that charging each internal cluster once to every leaf below it gives
$\sum_A|A|\leq nD_c$, we obtain, simultaneously for every $c$-balanced $T$,
\begin{equation}\label{eq:hc-balanced-transfer}
 |\operatorname{hcost}_H(T)-\operatorname{hcost}_G(T)|
 \leq\eta\operatorname{hcost}_G(T)+B_c,
 \qquad B_c=O_c(A_\eta n\log n).
\end{equation}

Fix $b=1/4$.  By \cite[Corollary~2]{AroraRaoVazirani09} and the weighted-capacity extension
explained above, there is a constant $b_0=b_0(b)>0$ and a randomized polynomial-time algorithm
that returns a $b_0$-balanced cut whose expansion is within
$a_{|A|}=O(\sqrt{\log |A|})$ of the best $b$-balanced cut on each recursive cluster $A$.
Set $b'=\min\{b_0,b\}$.  On clusters $A$ with $2\leq|A|<1/b$, split as evenly as possible, as in
\cref{lem:hc-bicriteria-recursion}.  On clusters with $1/b\leq|A|$ below a sufficiently large
absolute constant, brute-force the optimal $b$-balanced cut; this takes constant time and has
approximation factor one.  On all larger clusters, amplify every ARV call to failure probability
at most $n^{-12}$ by
independent repetition, retaining the best public feasible cut and using an arbitrary
$b'$-balanced cut if all trials fail the public balance check.  A union bound over the at most
$n-1$ recursive calls shows that all calls succeed with high probability, and
$a_{|A|}\leq a_n=O(\sqrt{\log n})$ uniformly.  Recursively apply this procedure to the induced
subgraphs of $H$.  By \cref{lem:hc-bicriteria-recursion}, its output $\widehat T$
satisfies
\[
 \operatorname{hcost}_H(\widehat T)
 \leq O(a_n)\operatorname{OPT}^{\rm HC}_H.
\]

Let $T^{\rm bal}$ be obtained by recursively taking an exact minimum-expansion $b$-balanced cut
on $G$.  Applying \cref{lem:hc-bicriteria-recursion} with $a=1$ shows
$\operatorname{hcost}_G(T^{\rm bal})=O(\operatorname{OPT}^{\rm HC}_G)$ (cf.~\cite{CharikarChatziafratis17,AssadiEtAl22HCStreams} for exact minimum $1/3$-balanced cuts).  Both
$\widehat T$ and $T^{\rm bal}$ have constant-balanced splits, so
\eqref{eq:hc-balanced-transfer} gives
\begin{align*}
 \operatorname{hcost}_G(\widehat T)
 &\leq\frac{\operatorname{hcost}_H(\widehat T)+B_{b'}}{1-\eta}\\
 &\leq\frac{O(a_n)\operatorname{hcost}_H(T^{\rm bal})+B_{b'}}{1-\eta}\\
 &\leq O(a_n)\frac{1+\eta}{1-\eta}\operatorname{hcost}_G(T^{\rm bal})
   +O((a_nB_b+B_{b'})/(1-\eta))\\
 &\leq O(\sqrt{\log n})\operatorname{OPT}^{\rm HC}_G
   +O_\eta(A_\eta n\log^{3/2}n).
\end{align*}
\end{proof}

Let us compare this result to prior works.
Imola, Epasto, Mahdian, Cohen-Addad, and Mirrokni~\cite[Section~5.1, Theorem~2]{ImolaEtAl23}
proved the following black-box polynomial-time $(\varepsilon, \delta)$-edge-DP guarantee: given a
nonprivate $a_n$-approximation, their reduction has multiplicative factor $(1+o(1))a_n$ and
additive error
\[
 O\left(\frac{n^{2.5}\log^2 n\log^2(1/\delta)}{\varepsilon}\right).
\]
Instantiating $a_n=O(\sqrt{\log n})$ gives
\[
 O(\sqrt{\log n})\operatorname{OPT}^{\rm HC}_G
 +\widetilde O(n^{2.5}/\varepsilon).
\]
They also proved an $\Omega(n^2/\varepsilon)$ additive lower bound for pure edge DP
\cite[Section~4, Theorem~1]{ImolaEtAl23}; since its hard instances have optimum $O(n/\varepsilon)$,
it applies to every multiplicative factor $o(n)$.  They further gave an exponential-time algorithm with
$O(n^2\log n/\varepsilon)$ additive error \cite[Section~5.2, Lemma~3]{ImolaEtAl23}.  Thus
\cref{thm:private-hierarchical-clustering} improves the polynomial-time additive dependence from
$n^{2.5}$ to nearly quadratic while maintaining the standard nonprivate multiplicative factor.
For unweighted inputs, applying the balanced-tree transfer above to the prior release of~\cite{fan2026private} also gives a derived $n^{25/12+o(1)}$ additive bound; our release improves this to nearly quadratic.
Deng, Gao, Upadhyay, Wang, and Zhou~\cite{DengEtAl25} obtain a
pure-DP $O(\log^{3/2}n/\varepsilon)$ multiplicative guarantee under weight privacy for connected
graphs whose topology is public and whose edges all have weight at least one.  That result uses a
different privacy model and is incomparable to the edge-private guarantee above.  On the lower-bound
side, \cite{DengEtAl25} extend the pure-DP $\Omega(n^2/\varepsilon)$ bound to weight
privacy on the complete graph, and \cite{ImolaPaghRetschmeier26} prove
an approximate-DP lower bound that gives $\Omega(n^2/\varepsilon)$ additive error on expander
topologies for algorithms without multiplicative loss.  Both bounds apply under the adjacency of
\cref{def:DP}.

\section{Lower Bound for Private Cut Release with Relative Error}
\label{sec:lower_bound}
As a counterpart to the algorithmic upper bound in \cref{thm:main}, we show that,
for $n^{-1/2}\leq\gamma\leq1/4$, the $\widetilde{O}(n/(\eps\gamma))$ additive error
is information-theoretically optimal up to logarithmic factors. For smaller
$\gamma$, our lower bound is $\Omega(n^{3/2}/\eps)$. We establish this lower bound
via the discrepancy-based reconstruction argument of~\cite{EKKL20}.
Their argument builds on the reconstruction framework of Muthukrishnan and
Nikolov~\cite{MN12} and a variant of the graph discrepancy result of
Bollob{\'a}s and Scott~\cite{BS06}.

We restate the discrepancy lemma in graph notation, which says that if two graphs differ in many edges, then at least one cut will be significantly discrepant in the number of crossing edges. 

\begin{lemma}[{\cite[Lemma~5.2]{EKKL20}}]
\label{lem:ekkl-discrepancy}
There is an absolute constant $0<\kappa\leq1$ such that the following holds.
Let $n\geq2$, $1\leq d\leq n/2$, and $\sigma\in[0,1]$.
Let $G,G'$ be simple unweighted graphs on the same $n$ vertices, with every
vertex having degree in $[d/2,2d]$ in each graph. If
$|E(G)\mathbin{\triangle}E(G')|\geq\sigma nd$, then
\[
 \max_{\substack{S,T\subseteq V\\S\cap T=\emptyset}}
       |w_G(S,T)-w_{G'}(S,T)|\geq\kappa\sigma n\sqrt d.
\]
\end{lemma}

At a high level, the lower bound proof proceeds as follows.
We start with a private bitstring $x \in \{0,1\}^m$, and construct a graph where an edge is present if and only if the corresponding bit in the bitstring is $1$. 
Each edge has weight $1/\varepsilon$. 
We run a private mechanism to construct a synthetic graph, and ``decode'' that graph to obtain a bitstring $x'$ which ought to resemble $x$. 
(Group) privacy implies that $x$ and $x'$ will differ in at least $\Omega(m)$ entries on average. 
Such a difference translates into a large discrepancy between two disjoint vertex sets (\Cref{lem:ekkl-discrepancy}), which is thus at odds with utility for an appropriate class of graphs. 
The construction and argument are simpler  when $\gamma = \Omega(1)$, where a more direct utility analysis holds on a family of star-like graphs.

\begin{theorem}[Lower Bound for Relative Cut Release]
\label{thm:lower_bound}
Let $n\geq33$, $\varepsilon\in(0,1)$, $0\leq\delta\leq\varepsilon/100$,
$\beta\in (0,1/40]$, and $\gamma \in [0, 1)$.
Suppose that an $(\varepsilon,\delta)$-differentially private mechanism
$\mathcal M$, on every nonnegative weighted graph $G$ on a fixed set $V$
of $n$ vertices, outputs a nonnegative weighted graph $H$ such that,
with probability at least $1-\beta$, simultaneously for all $S\subseteq V$,
\[
 \left|w_G(S, V\setminus S) - w_H(S, V\setminus S)\right| \leq \gamma w_G(S, V\setminus S) + A.
\]
Then the additive error must satisfy
\[
 A=\Omega\left(
   (1-\gamma)\cdot\min\left\{\frac{n^{3/2}}{\varepsilon},
                    \frac{n}{\varepsilon\gamma}\right\}
 \right),
\]
where the implicit constant is absolute and $1/\gamma$ is interpreted as
$+\infty$ when $\gamma=0$.
\end{theorem}

\begin{proof}
Fix a mechanism $\mathcal M$ satisfying the stated privacy and utility guarantees.
We first describe the construction of a family of graphs which will be used to prove this lower bound.

We fix an even integer $d$ such that $2 \leq d \leq n/2$.
Recall that $K_{d,d}$ is the complete bipartite graph on $2d$ vertices, with $d^2$ edges. 
We consider $b = \frac{n}{2d}$ disjoint copies of $K_{d,d}$:\footnote{In this proof, for simplicity, we assume $\frac{n}{2d}$ and $1/\varepsilon$ are integers. The general cases follow similarly.} for each copy, decompose the edges (arbitrarily) into $d$ perfect matchings. 
For each copy, $d/2$ of these matchings will be present in every graph (the \emph{fixed} edges). 
The remaining $d^2/2$ edges (the \emph{variable} edges) in this copy will have their presence determined by a private bitstring $x$.
Specifically, for $m=  b \cdot \frac{d^2}{2} = \frac{nd}{4}$, we consider $x \in \{0,1\}^m$, where each coordinate corresponds to one variable edge $e$.
That edge is present if and only if $x_e = 1$. 
This defines an unweighted graph corresponding to $x$, $F_x$. 
We assign a weight $t = 1/\eps$ to each edge, creating a weighted graph $G_x$.
Observe that, by construction, each vertex of $F_x$ has degree in $[d/2, d]$, and the total edge weight of $G_x$ is at most $2tm$.

Changing one bit of $x$ changes exactly
one edge weight by $t$. Given any graph $H$, we define a post-processing operation $\mathcal{D}$ that returns the first $y\in\{0,1\}^m$ in a fixed ordering such that for every $S\subseteq V$,
\begin{equation}
\label{eq:lb_compatible_decoder}
 |w_{G_y}(S,V\setminus S)-w_{H}(S,V\setminus S)|\leq\gamma w_{G_y}(S,V\setminus S)+A,
\end{equation}
and outputs the zero vector if no such $y$ exists.

By group privacy, $x\mapsto\mathcal M(G_x)$ is
$(t\varepsilon,\Delta)$-private if we change one bit of $x$, where
\[
 \Delta=\delta\sum_{j=0}^{t-1}e^{j\varepsilon}
 \leq et\delta\leq e/100,
\]
by $0\leq\delta\leq\varepsilon/100$.
Let $X$ be uniform on $\{0,1\}^m$ and let $\widehat X$ be the decoded string $\mathcal{D}(\mathcal{M}(G_X))$.
Fix a coordinate $i\in[m]$ and a value $z\in\{0,1\}^{m-1}$ of the other
coordinates $X_{-i}$. For $b\in\{0,1\}$, let
\[
 p_b(z):=\Prb[\widehat X_i\ne b\mid X_i=b,\ X_{-i}=z].
\]
By post-processing (\Cref{lem:post-processing}), $x\mapsto\mathcal{D}(\mathcal{M}(G_x))$ is also
$(t\varepsilon,\Delta)$-private. Applying privacy to the two inputs with
other coordinates $z$ gives
\[
 1-p_0(z)\leq e^{t\varepsilon}p_1(z)+\Delta,
 \qquad
 1-p_1(z)\leq e^{t\varepsilon}p_0(z)+\Delta.
\]
Adding these inequalities yields $(p_0(z)+p_1(z))/2\geq \frac{1-\Delta}{1+e^{t\varepsilon}} =: q$.
Since $X$ is uniform, conditional on $X_{-i}=z$ the bit $X_i$ takes each
value with probability $1/2$. Thus
\[
 \Prb[\widehat X_i\ne X_i\mid X_{-i}=z]
 =\frac{p_0(z)+p_1(z)}2\geq q.
\]
Averaging over $z$ gives $\Prb[\widehat X_i\ne X_i]\geq q$ for every
$i\in[m]$. The Hamming distance counts the coordinates decoded incorrectly,
so linearity of expectation gives
\begin{equation}
\label{eq:lb_private_reconstruction}
 \mathbb E\left[ \|X-\widehat X\|_1\right]
 =\sum_{i=1}^m\Prb[\widehat X_i\ne X_i]\geq qm.
\end{equation}

Let $\mathcal E$ be the utility success event of $\mathcal{M}$ for the true graph $G_X$.
It has probability at least $1-\beta$.
On $\mathcal E$, $G_X$ and $H = \mathcal{M}(G_X)$ satisfy \eqref{eq:lb_compatible_decoder}
by the definition of the success event. The decoder's feasible set is therefore
nonempty, so it returns a feasible index $\widehat X = \mathcal{D}(\mathcal{M}(G_X))$ rather than using its fallback.
Consequently, $\widehat X$ also satisfies \eqref{eq:lb_compatible_decoder}.

By $t\varepsilon\leq1$, $\Delta\leq e/100$, and $\beta\leq1/40$, we have $q-\beta\geq\frac15$. Since $\|X-\widehat X\|_1\leq m$ on every output, \eqref{eq:lb_private_reconstruction} implies
\begin{equation}
\label{eq:lb_success_reconstruction}
 \mathbb E[\|X-\widehat X\|_1\,\mathbf{1}_{\mathcal E}] = 
  \mathbb E[\|X-\widehat X\|_1]
  -
  \mathbb E[\|X-\widehat X\|_1\,\mathbf{1}_{\mathcal E^c}]
 \geq qm - \beta m\geq m/5.
\end{equation}

For simplicity, write $c_G(S)=w_G(S,V\setminus S)$. Define
\[
 d_{ST}(G,G')
 :=\max_{\substack{S,T\subseteq V\\S\cap T=\emptyset}}
       |w_G(S,T)-w_{G'}(S,T)|.
\]

Fix $x,y\in\{0,1\}^m$. Since $|E(F_x)\mathbin{\triangle}E(F_y)|=\|x-y\|_1$, we apply Lemma~\ref{lem:ekkl-discrepancy} to the unweighted graphs $F_x,F_y$
on their $n$ vertices, with
$\sigma=\|x-y\|_1/(nd)\leq1/4$.
Scaling by the weights $t$ gives
\begin{equation}
\label{eq:lb_discrepancy}
 d_{ST}(G_x,G_y)\geq\kappa t\sigma n\sqrt d=\frac{\kappa t}{\sqrt d}\|x-y\|_1.
\end{equation}

To compare this discrepancy with the released cuts, for disjoint $S,T$ note that
\[
 w_G(S,T)=\frac{c_G(S)+c_G(T)-c_G(S\cup T)}2,
 \qquad
 c_G(S)+c_G(T)+c_G(S\cup T)\leq2\sum_e w_{G}(e).
\]
We thus decompose $d_{ST}(H,G_y)$ into these three cuts, and apply the utility guarantees of \eqref{eq:lb_compatible_decoder} to each:
\begin{align*}
|w_H(S,T)-w_{G_y}(S,T)|
&\le \frac12 \left(\sum_{U\in\{S,T,S\cup T\}}
    |c_H(U)-c_{G_y}(U)|\right) \\
&\le \frac{1}{2}\left( 
    \sum_{U\in\{S,T,S\cup T\}} \left(\gamma c_{G_y}(U) + A\right)\right) \\
&\le \gamma \sum_e w_{G_y}(e) + \frac{3A}{2}.
\end{align*}

Since $\sum_e w_{G_y}(e)\leq2tm$, we have that
\[
 d_{ST}(H,G_y)\leq\gamma \sum_e w_{G_y}(e)+\frac32 A
 \leq2\gamma tm+\frac32 A.
\]
Therefore on $\mathcal E$, the triangle inequality gives
$d_{ST}(G_X,G_{\widehat X})\leq4\gamma tm+3A$.
Combining this with \eqref{eq:lb_success_reconstruction} and
\eqref{eq:lb_discrepancy}, we obtain
\[
 \frac{\kappa tm}{5\sqrt d}\leq\frac{\kappa t}{\sqrt d}\mathbb E[\|X-\widehat X\|_1\mathbf{1}_{\mathcal E}]
 \leq\mathbb E[d_{ST}(G_X,G_{\widehat X})\mathbf{1}_{\mathcal E}]
 \leq4\gamma tm+3A.
\]
Thus we can conclude,
\begin{equation}
\label{eq:lb_master_inequality}
 A\geq tm\left(\frac{\kappa}{15\sqrt d}-\frac43\gamma\right).
\end{equation}
First suppose $0\leq\gamma\leq\kappa/80$ and choose
\[
 d=2\left\lfloor
       \min\left\{\frac n4,\frac{\kappa^2}{3200\gamma^2}\right\}
     \right\rfloor\in[4,n/2].
\]
Moreover,
$\gamma\sqrt d\leq\kappa/40$ and
$d=\Theta(\min\{n,\gamma^{-2}\})$.
Using these facts in combination with \eqref{eq:lb_master_inequality},
\[
 A\geq\frac{\kappa tm}{30\sqrt d}
   =\frac{\kappa t n\sqrt d}{120}
   =\Omega\left(
       \frac n\varepsilon\cdot\min\left\{\sqrt n,\frac1\gamma\right\}
     \right),
\]
proving the desired bound for this case. 

For $\gamma>\kappa/80 = \Omega(1)$, we have that $\min\{\sqrt{n},1/\gamma\} = O(1)$, and thus it suffices to prove a simpler $\Omega(n/\varepsilon)$ lower bound. 
To this end, we consider an entirely different family of graphs, which is constructed from the star graph on $n$ nodes. 
Specifically, consider the set of graphs with one fixed center and $n-1$ leaves.
Letting $x \in \{0,1\}^{n-1}$ be a private bitstring, unweighted graph $F_x$ has an edge from the center to node $i$ if and only if $x_i = 1$, and the weighted graph $G_x$ has weight $1/\varepsilon$ on every present edge. 
Use the same compatibility decoder $\mathcal{D}$ defined by \eqref{eq:lb_compatible_decoder} restricted to this
family, and define $X,\widehat X,\mathcal E$ as above. A single-bit change
again changes only one edge weight by $t$. The preceding group-privacy and
reconstruction argument therefore applies with $m=n-1$, giving
\eqref{eq:lb_success_reconstruction} for this family as well.
\par\nopagebreak[4]
On $\mathcal E$, let $U_+$ be the leaves with $X_i=1,\widehat X_i=0$ and
$U_-$ those with $X_i=0,\widehat X_i=1$.
The cut $U_+$ has values $t|U_+|$ and $0$ in the true and decoded graphs.
Using these cut values and the utility guarantee \eqref{eq:lb_compatible_decoder} of $H$ with respect to the true graph and the reconstructed graph, respectively, we get
\[(1-\gamma)t|U_+|-A\leq c_H(U_+)\leq (1+\gamma) 0 + A.\]
A symmetric argument on $U_-$ gives $(1-\gamma)t|U_-|\leq2A$ as well.
Thus \[\|X-\widehat X\|_1=|U_+|+|U_-|\leq4A/((1-\gamma)t)\] on $\mathcal E$.
Combining this with \eqref{eq:lb_success_reconstruction} yields
\begin{equation*}
A\geq\frac{(1-\gamma)t(n-1)}{20}
   =\Omega\left(\frac{(1-\gamma)n}{\eps}\right),
\end{equation*}
as desired.
\end{proof}

\end{document}